\pdfoutput=1
\RequirePackage[l2tabu, orthodox]{nag}
\RequirePackage{ifpdf}
\RequirePackage[usenames,dvipsnames]{xcolor}
\RequirePackage{tikz}
\RequirePackage{hyperref}
\let\href\undefined
\documentclass[a4paper,nohyper]{JHEP3}
\usepackage{hyperref}
\usepackage[utf8]{inputenc}
\usepackage{mathtools, amscd, dsfont, amsthm, amssymb,array}
\usepackage{graphicx}
\usepackage[toc,page]{appendix}
\usepackage{etoolbox}
\usepackage{cite}
\mathtoolsset{showonlyrefs=false}
\usetikzlibrary{arrows,calc,shapes.misc,decorations.markings,snakes}
\tikzset{cross/.style={cross out, draw=black, minimum size=2*(#1-\pgflinewidth), inner sep=0pt,
        outer sep=0pt},cross/.default={3pt},
    gluon/.style={decorate, decoration={coil,aspect=0.9,segment length=5pt, amplitude=3pt}}}

\hypersetup{colorlinks=true, urlcolor=black, linkcolor=black, citecolor=[rgb]{0.15,0.35,0.65}}

\usepackage{axodraw}

\def\be{\begin{equation}}
\def\ee{\end{equation}}
\newcommand{\beq}{\begin{equation}}
\newcommand{\eeq}{\end{equation}}
\newcommand{\bea}{\begin{align}}
\newcommand{\eea}{\end{align}}
\def\bsp#1\esp{\begin{split}#1\end{split}}

\newcommand{\cA}{\begin{cal}A\end{cal}}

\newcommand{\cE}{\begin{cal}E\end{cal}}
\newcommand{\cF}{\begin{cal}F\end{cal}}
\newcommand{\cG}{\begin{cal}G\end{cal}}
\newcommand{\cH}{\begin{cal}H\end{cal}}

\newcommand{\cL}{\begin{cal}L\end{cal}}

\newcommand{\cN}{\begin{cal}N\end{cal}}
\newcommand{\cO}{\begin{cal}O\end{cal}}
\newcommand{\cP}{\begin{cal}P\end{cal}}

\newcommand{\cR}{\begin{cal}R\end{cal}}

\newcommand{\cW}{\begin{cal}W\end{cal}}

\newcommand{\fa}{\mathfrak{a}}
\newcommand{\fb}{\mathfrak{b}}
\newcommand{\fc}{\mathfrak{c}}
\newcommand{\fd}{\mathfrak{d}}
\newcommand{\fe}{\mathfrak{e}}
\newcommand{\ff}{\mathfrak{f}}

\newcommand{\zb}{{\bar z}}

\newcommand{\brc}[1]{\{#1\}}
\newcommand{\nn}{\nonumber}
\newcommand{\<}{{\langle}}
\renewcommand{\>}{{\rangle}}

\title{The seven-gluon amplitude in multi-Regge kinematics\\
beyond leading logarithmic accuracy}

\author{Vittorio Del Duca$^{a}$\footnote{On leave from Istituto Nazionale di Fisica Nucleare,
Laboratori Nazionali di Frascati, Italy.},
Stefan Druc$^b$,
James Drummond$^b$,
Claude Duhr$^{c,d}$,
Falko Dulat$^e$,
Robin Marzucca$^d$,
Georgios Papathanasiou$^f$,
Bram Verbeek$^d$\\
${}^a$ Institute for Theoretical Physics, ETH Z\"urich, 8093 Z\"urich, Switzerland.\\
${}^b$ School of Physics \& Astronomy, University of Southampton, \\
\phantom{${}^b$} Highfield, Southampton, SO17 1BJ, United Kingdom.\\
${}^c$ Theoretical Physics Department, CERN, CH-1211 Geneva 23, Switzerland.\\
${}^d$ Center for Cosmology, Particle Physics and Phenomenology (CP3),\\
\phantom{{}$^d$} Universit\'e catholique de Louvain,\\
\phantom{{}$^d$} Chemin du Cyclotron 2, 1348 Louvain-La-Neuve, Belgium.\\
${}^e$ SLAC National Accelerator Laboratory, Stanford University, \\
\phantom{${}^e$} Stanford, CA 94309, USA.\\
${}^f$ DESY Theory Group, DESY Hamburg, Notkestra{\ss}e 85, D-22607 Hamburg, Germany.
}

\preprint{CERN-TH-2018-016, CP3-18-06\\
    DESY 18-009, SLAC-PUB-17220}

\abstract{
We present an all-loop dispersion integral, well-defined to arbitrary logarithmic accuracy,
describing the multi-Regge limit of the $2\to5$ amplitude in planar $\cN=4$ super Yang-Mills theory.
It follows from factorization, dual conformal symmetry and consistency with soft limits, and
specifically holds in the region where the energies of all produced particles have been analytically
continued. After promoting the known symbol of the 2-loop $N$-particle MHV amplitude in this region
to a function, we specialize to $N=7$, and extract from it the next-to-leading order (NLO)
correction to the BFKL central emission vertex, namely the building block of the dispersion integral
that had not yet appeared in the well-studied six-gluon case. As an application of our results, we explicitly compute the seven-gluon amplitude at next-to-leading logarithmic accuracy through 5 loops for the MHV case, and through 3 and 4 loops for the two independent NMHV helicity configurations, respectively.}


\newcommand{\contourOne}{
\begin{tikzpicture}{scale=1.0}
        \draw[->,thick] (0,3) -- (6,3) node [right] {$\Re(\nu)$};
        \draw[->,thick] (3,0) -- (3,6) node [above] {$\Im(\nu)$};
        \draw (4.5,3) node[cross] (A) {};
        \node [below of=A, node distance=10pt] {$\pi \Gamma$};
        \draw (1.5,3) node[cross] (mA) {};
        \node [below of=mA, node distance=10pt] {$-\pi \Gamma$};

        \draw[thick,RoyalBlue,yshift=2, decoration={
                markings,
                mark = at position 0.23 with {\arrow{>}},
                mark = at position 0.4 with {\arrow{>}},
                mark = at position 0.6 with {\arrow{>}},
                mark = at position 0.8 with {\arrow{>}}},
                postaction={decorate}]
        (0,3) -- (1.2, 3) arc (-180:0:.3) -- (4.2, 3) arc (180:0:.3) -- (6,3);

\end{tikzpicture}}
\newcommand{\contourTwo}{
\begin{tikzpicture}{scale=1.0}
        \draw[->,thick] (0,3) -- (6,3) node [right] {$\Re(\nu_1)$};
        \draw[->,thick] (3,0) -- (3,6) node [above] {$\Im(\nu_1)$};
        \draw (4.5,3) node[cross] (A) {};
        \node [below of=A, node distance=10pt] {$\pi \Gamma$};
        \draw (1.5,3) node[cross] (mA) {};
        \node [below of=mA, node distance=10pt] {$\nu_2$};

        \draw[thick,RoyalBlue,yshift=2, decoration={
                markings,
                mark = at position 0.23 with {\arrow{>}},
                mark = at position 0.4 with {\arrow{>}},
                mark = at position 0.6 with {\arrow{>}},
                mark = at position 0.8 with {\arrow{>}}},
                postaction={decorate}]
        (0,3) -- (1.2, 3) arc (-180:0:.3) -- (4.2, 3) arc (180:0:.3) -- (6,3);

\end{tikzpicture}}
\newcommand{\contourThree}{
\begin{tikzpicture}{scale=1.0}
        \draw[->,thick] (0,3) -- (6,3) node [right] {$\Re(\nu_2)$};
        \draw[->,thick] (3,0) -- (3,6) node [above] {$\Im(\nu_2)$};
        \draw (4.5,3) node[cross] (A) {};
        \node [below of=A, node distance=10pt] {$\nu_1$};
        \draw (1.5,3) node[cross] (mA) {};
        \node [below of=mA, node distance=10pt] {$-\pi \Gamma$};

        \draw[thick,RoyalBlue,yshift=2, decoration={
                markings,
                mark = at position 0.23 with {\arrow{>}},
                mark = at position 0.4 with {\arrow{>}},
                mark = at position 0.6 with {\arrow{>}},
                mark = at position 0.8 with {\arrow{>}}},
                postaction={decorate}]
        (0,3) -- (1.2, 3) arc (-180:0:.3) -- (4.2, 3) arc (180:0:.3) -- (6,3);

\end{tikzpicture}}

\newcommand{\contourRSevenNuOne}{
\begin{tikzpicture}{scale=1.0}
        \draw[->,thick] (0,3) -- (6,3) node [right] {$\Re(\nu_1)$};
        \draw[->,thick] (3,0) -- (3,6) node [above] {$\Im(\nu_1)$};
        \draw (4.5,3) node[cross] (A) {};
        \node [below of=A, node distance=10pt] {$\pi \Gamma$};
        \draw (1.5,3) node[cross] (mA) {};
        \node [below of=mA, node distance=10pt] {$\nu_2$};

        \draw[thick,RoyalBlue,yshift=2, decoration={
                markings,
                mark = at position 0.13 with {\arrow{>}},
                mark = at position 0.25 with {\arrow{>}},
                mark = at position 0.5 with {\arrow{>}},
                mark = at position 0.9 with {\arrow{>}}},
                postaction={decorate}]
        (-0.5,2) arc(180:90:1) -- (1.2, 3) arc (-180:0:.3) -- (4.2, 3) arc (-180:0:.3) -- (5.5,3) arc (90:0:1);
\end{tikzpicture}}
\newcommand{\contourRSevenNuTwo}{
\begin{tikzpicture}{scale=1.0}
        \draw[->,thick] (0,3) -- (6,3) node [right] {$\Re(\nu_2)$};
        \draw[->,thick] (3,0) -- (3,6) node [above] {$\Im(\nu_2)$};
        \draw (4.5,3) node[cross] (A) {};
        \node [below of=A, node distance=10pt] {$\nu_1$};
        \draw (1.5,3) node[cross] (mA) {};
        \node [below of=mA, node distance=10pt] {$-\pi\Gamma$};

        \draw[thick,RoyalBlue,yshift=2, decoration={
                markings,
                mark = at position 0.13 with {\arrow{>}},
                mark = at position 0.25 with {\arrow{>}},
                mark = at position 0.5 with {\arrow{>}},
                mark = at position 0.9 with {\arrow{>}}},
                postaction={decorate}]
        (-0.5,4) arc(-180:-90:1) -- (1.2, 3) arc (180:0:.3) -- (4.2, 3) arc (180:0:.3) -- (5.5,3) arc (-90:0:1);

\end{tikzpicture}}
\newcommand{\bfklViz}{
    \begin{tikzpicture}{scale=1.0}
\begin{scope}[thick,decoration={markings, mark=at position 0.6 with {\arrow{>}}}]
        \draw[postaction={decorate}] (0,6) node[left]{$p_2$} -- (6,6) node[right]{$p_3$};
        \draw[postaction={decorate}] (0,0) node[left]{$p_1$} -- (6,0) node[right]{$p_7$};
        \draw[gluon] (1.8,6) -- (2,5.0); 
        \draw[gluon] (1.8,0) -- (2,1.0);
        \draw[->](4.5,5.0) -- (5.5,5.0) node[right]{$p_4$};
        \draw[->](4.5,3.0) -- (5.5,3.0) node[right]{$p_5$};
        \draw[->](4.5,1.0) -- (5.5,1.0) node[right]{$p_6$};
    \end{scope}
    \draw (2,1.0) -- (2,5.0);
    \draw (4,1.0) -- (4,5.0);
    \draw[black,fill=CadetBlue!60] (3,5.0) ellipse (1.5cm and 0.4cm);
    \draw[black,fill=CadetBlue!60] (3,1.0) ellipse (1.5cm and 0.4cm);
    \node at (2.5,5.0){$\chi_{\nu_1}$};
    \node at (2.5,1.0){$\chi_{\nu_2}$};
    \draw[black,fill=RedOrange!40] (3,3) ellipse (1.5cm and 0.6cm);
    \node at (2.5,3.0){$C_{\nu_1,\nu_2}$};
    \node[anchor=mid] at (2.5,4.0){$\omega_{\nu_1}$};
    \node[anchor=mid] at (3.5,4.0){$\tau_1$};
    \node[anchor=mid] at (2.5,2.0){$\omega_{\nu_2}$};
    \node[anchor=mid] at (3.5,2.0){$\tau_2$};
    \draw[dashed,thick] (3,.2) -- (3,5.8);
    \node[anchor=mid] at (5.0,4.0){$z_1$};
    \node[anchor=mid] at (5.0,2.0){$z_2$};
\end{tikzpicture}}

\begin{document}

\maketitle
\tableofcontents
\hypersetup{linkcolor=[rgb]{0.15,0.35,0.65}}

\section{Introduction}
From {phenomenological studies} to implications for quantum gravity, the description of scattering at high energies, or in the (multi-)Regge limit, is a subject with rich history and impact on many branches of theoretical physics. In the simplest case of $2\to2$ scattering, where the limit corresponds to a center-of-mass energy squared $s$ being much larger  than the momentum transfer $|t|$, the necessity for the respective amplitude to behave as
\be\label{eq:A4Regge}
\cA(s,t)\sim s^{\alpha(t)}\,,
\ee
so as to agree with experimental results of the time, was a key ingredient in the construction of the Veneziano amplitude, marking the birth of string theory~\cite{DiVecchia:2007vd}.

With the establishment of QCD as the theory of strong interactions, the same behavior~\eqref{eq:A4Regge} was obtained from perturbative QCD by resumming contributions from an infinite number of Feynman diagrams, each of which contains large logarithms in $s$, within the Balitsky-Fadin-Kuraev-Lipatov (BFKL) framework \cite{Kuraev:1976ge,Kuraev:1977fs,Balitsky:1978ic}. The leading logarithmic approximation (LLA) amounts to only keeping the largest power of $\log s$ at fixed order, with the obvious generalization to N(ext-to)$^k$LLA. This procedure gives rise to the concept of an effective particle with $t$-dependent angular momentum $\alpha(t)$, the reggeized gluon, as well as its bound states. Due to their interpretation in the complex angular momentum plane, the contributions of the reggeized gluon and its bound states to the amplitude are also known as Regge pole and cut respectively. For pedagogical introductions to Regge theory and BFKL see for example \cite{BROWER1974257,Collins:1977jy,Forshaw:1997dc,DelDuca:1995hf}.

Multi-Regge kinematics (MRK) and the BFKL dynamics describing them, have also been of paramount
importance for computing amplitudes in arguably the simplest gauge theory, $\cN=4$ super Yang-Mills
theory \cite{Brink:1976bc,Gliozzi:1976qd} in the planar limit \cite{tHooft:1973alw}, which is the
focus of this article. While a conjecture for the all-loop $N$-particle amplitude of the planar
theory \cite{Anastasiou:2003kj,Bern:2005iz} known as the BDS ansatz accurately describes the $N=4,5$
cases, the first evidence that this is no longer true for $N\ge 6$ came from the inconsistency of
the ansatz with the BFKL approach \cite{Bartels:2008ce}. Due to the duality between amplitudes and
null polygonal Wilson loops
\cite{Alday:2007hr,Drummond:2007aua,Brandhuber:2007yx,Drummond:2007cf,Drummond:2007au,Bern:2008ap,Drummond:2008aq}
and the associated dual conformal
symmetry~\cite{Drummond:2007au,Drummond:2006rz,Bern:2006ew,Bern:2007ct,Alday:2007he}, the correction
to the ansatz, namely the BDS-normalized amplitude, has to be a function of conformal cross ratios of the Mandelstam invariants.

In the same vein as in ref.~\cite{Bartels:2008ce}, dispersion integrals yielding the BDS-normalized
amplitude for $N=6$ gluons in MRK were derived in refs.~\cite{Bartels:2008sc, Lipatov:2010qf,Fadin:2011we,Lipatov:2012gk,Dixon:2014iba}. In Fourier-Mellin space, they factorize into universal building blocks known as the adjoint BFKL eigenvalue and impact factor, which may be determined perturbatively from first principles, or extracted~\cite{Lipatov:2010qg,Lipatov:2010ad,Bartels:2010tx} from fixed-order expressions for the amplitude that have been obtained by other means~\cite{DelDuca:2009au,DelDuca:2010zg,Goncharov:2010jf}. A leading-order strong-coupling analysis is also possible \cite{Bartels:2010ej,Bartels:2013dja}, but even more remarkably, the building blocks in question can also be obtained to all loops \cite{Basso:2014pla} by means of analytic continuation from a collinear limit where the dynamics is governed by an integrable flux tube \cite{Alday:2010ku,Gaiotto:2010fk,Gaiotto:2011dt,Sever:2011da,Basso:2013vsa,Basso:2013aha,Basso:2014koa,Basso:2014jfa,Basso:2014nra,Basso:2014hfa,Basso:2015rta,Basso:2015uxa}, see also \cite{Bartels:2011xy,Hatsuda:2014oza,Drummond:2015jea}. These developments render the MRK as one of the best sources of `boundary data' \cite{Dixon:2012yy,Pennington:2012zj,Drummond:2015jea,Broedel:2015nfp} for determining the six-gluon amplitude in general kinematics through five loops, by exploiting its analytic structure with the help of the bootstrap method \cite{Dixon:2011pw,Dixon:2013eka,Dixon:2014iba,Dixon:2014voa,Dixon:2015iva,Caron-Huot:2016owq}.

It would be of course very exciting if higher-point amplitudes in MRK could also be computed to all loops. While the amplitude bootstrap method in general kinematics remarkably turns out to require no nontrivial boundary data through 4 loops~\cite{Drummond:2014ffa,Dixon:2016nkn} for $N=7$,
in its present formulation, it is unclear how this generalizes for higher point amplitudes because their analytic structure is significantly more complicated \cite{ArkaniHamed:2012nw,Golden:2013xva,CaronHuot:2012ab,Prlina:2017tvx} and in general is not known.
In MRK however the analytic structure of amplitudes simplifies significantly.
Indeed, in \cite{DelDuca:2016lad} we argued that all amplitudes in MRK at the leading log approximation are described by single-valued polylogarithms associated to the moduli space of $(N-2)$ points on a Riemann sphere.
Thus the study of MRK is an important stepping stone for extending to general kinematics.

With this goal in mind, in this paper we turn our attention to the simplest object beyond the solved six-gluon case, the $2\to5$ amplitude in MRK. While predictions for the latter to LLA have been worked out in
\cite{Bartels:2011ge,Bartels:2013jna,Bartels:2014jya}, see also \cite{DelDuca:2016lad} for an
extension to $N$-gluons, a major obstacle for their generalization to
arbitrary logarithmic accuracy, is that the dispersion integral yielding the Regge cut contribution
diverges when considering the N$^{\ell-1}$LLA term at $\ell$ loops. This phenomenon is related to the fact
that within the BFKL approach, the dual conformal invariance of the theory is not maintained at the
intermediate steps of the calculation, and implies that some terms in the Regge pole contribution
develop unphysical poles. As discussed in \cite{Bartels:2011ge}, it is possible to shift these terms from the pole to the cut contribution by modifying
their definitions. The necessity of this step suggests that there may be a certain degree of arbitrariness in separating the pole and cut contributions in a conformally invariant theory.

Instead of the BFKL approach, in our analysis we will draw from the eikonal framework of \cite{Caron-Huot:2013fea}, where the two incoming high-energy gluons are approximated by straight Wilson lines. Within this framework, only a Regge cut contribution, respecting high-energy factorization, dual conformal symmetry and consistency with soft limits, is necessary to describe the amplitude in MRK. This procedure also provides a natural regularization for the Regge cut at finite coupling, and allows for its straightforward expansion at weak coupling. In this manner, we obtain the first significant result of this paper, namely a dispersion integral describing the $2\to5$ amplitude in MRK to arbitrary logarithmic accuracy. {After reviewing some background knowledge on the kinematical dependence and behavior of amplitudes in the limit in section \ref{sec:MRKreview}, we present this result in section \ref{sec:BFKLFinite}}.

The $2\to5$ dispersion integral contains a new building block compared to the six-gluon case, the
BFKL central emission vertex, to date known to leading order~\cite{Bartels:2011ge}.  The second significant result of this paper is the extraction of the NLO correction to the central emission vertex, from the known NLLA contribution to the 2-loop $7$-particle MHV amplitude~\cite{Bargheer:2015djt}, see also refs.~\cite{Prygarin:2011gd,Bartels:2011ge} for earlier work on the LLA contribution.

More precisely, since the aforementioned NLLA contribution had been previously determined \cite{Bargheer:2015djt} at level of the symbol \cite{Goncharov:2010jf}, in section~\ref{sec:SymbolsToFunctions} we show how to uniquely promote it to a function, based on information we derive on the leading discontinuity of the amplitude, together with its expected behavior under soft limits, and single-valuedness properties of the function space in which it ``lives''. As a bonus, from this result we in fact also obtain the function-level 2-loop MHV amplitude for \emph{any} number of gluons in MRK, since the latter has been shown to decompose into $6$- and $7$-gluon building blocks in momentum space~\cite{Bargheer:2015djt}.

Then, in section~\ref{sec:CNLO} we present the final expression for the NLO correction to the BFKL central emission vertex, and detail our approach for obtaining it, by translating the momentum-space expression for the amplitude, to the Fourier-Mellin space of the dispersion integral. Using the same approach, we similarly extract the next-to-next-to-leading order (NNLO) correction to the central emission vertex, up to transcendental constants, from the 3-loop MHV heptagon symbol~\cite{Drummond:2014ffa}, which will be discussed in future work.

With the $2\to5$ dispersion integral and the NLO BFKL central emission vertex at hand, we then move on to produce from them the third significant result of this paper, namely new predictions for the seven-gluon amplitude in MRK to NLLA at higher loops. In particuler, we obtain explicit expressions in momentum space through 5 loops for the MHV case, and through 3 and 4 loops for the two independent NMHV helicity configurations, respectively. This is achieved by direct evaluation of the dispersion integral, using a combination of nested sum algorithms and convolution methods, that we describe in section \ref{sec:HigherLoopPredictions}.

{Finally, section \ref{sec:Conclusions} contains our conclusions, and discusses possible future directions of inquiry. We also include two appendices,  the first one describing the class of functions that is relevant for seven-gluon scattering in MRK, and the second one containing explicit 2-loop results for the respective amplitude in momentum space, that are too long to present in the main text. Finally, we attach our higher-loop results as ancillary files, accompanying the submission on this article on the \texttt{arXiv}.}

\section{ $\mathcal{N}=4$ amplitudes in the multi-Regge limit}\label{sec:MRKreview}

\subsection{Multi-Regge kinematics}
\label{sec:kinematics}

Let us start by recalling the precise definition of the limit and the kinematic region we will be considering, mostly following the conventions of~\cite{DelDuca:2016lad}. We will be keeping the number of particles $N$ general, since in the next section we will first be reviewing the $N=6$ case, before focusing on $N=7$ for the rest of the paper.

In $1+2\to 3+\ldots+N$ scattering, and in conventions where all momenta are outgoing, multi-Regge kinematics are defined as the limit where the final-state gluons are strongly ordered in rapidity, or equivalently in lightcone $+$-coordinates,
\beq\label{eq:MRK_def}
p_3^+\gg p_4^+\gg \ldots p_{N-1}^+\gg p_N^+\,, \qquad |{\bf p}_3|\simeq\ldots \simeq|{\bf p}_N|\,,
\eeq
where the lightcone and complex transverse coordinates are defined as
\be
p^\pm\equiv p^0\pm p^z\,,\quad {\bf p}_k \equiv p_{k\bot} = p_k^x + i p_k^y\,,
\ee
and without loss of generality we have chosen a frame where the initial-state gluons lie on the $z$-axis, such that ${\bf p}_1={\bf p}_2=0$.

The limit implies that the incoming gluons are only mildly deflected in the scattering process, so that helicity is conserved along their trajectory. Thus different helicity configurations are only distinguished by the helicities $h_1,\ldots,h_{N-4}$  of the produced gluons, for which we define the BDS-normalized ratio
\beq\label{eq:Rhhh_definition}
\cR_{h_1\ldots h_{N-4}} \equiv \frac{A_N(-,+,h_1,\ldots,h_{N-4},+,-)}{A_N^{\textrm{BDS}}(-,+,h_1,\ldots,h_{N-4},+,-)}\,.
\eeq
Here $A_N^{\textrm{BDS}}$ denotes the tree-level amplitude of the given helicity, times the same scalar factor that is present in the BDS amplitude for the MHV case, encoding the universal infrared-divergent structure of amplitudes in $\cN=4$ super Yang-Mills theory.

Due to their equivalence to bosonic Wilson loops, the MHV and $\overline{\text{MHV}}$ ratios are equal to each other, and are also alternatively described by their logarithm, the remainder function,
\be\label{eq:RN}
\cR_{\underset{N-4}{\underbrace{+\ldots+}}}=\cR_{\underset{N-4}{\underbrace{-\ldots-}}}\equiv e^{R_N}\,.
\ee
Unless otherwise stated, $\cR$ and $R_N$ will denote the BDS-normalized amplitude or remainder
function specifically in the multi-Regge limit, and we refrain from using additional superscripts
denoting this, in order to avoid cluttering our expressions.

As a consequence of the dual conformal invariance of the theory, the dependence of $\cR$ on the kinematics only enters through conformal cross ratios of the Mandelstam invariants,
\be\label{eq:Ucrossratios}
U_{ij}\equiv\frac{x^2_{i+1j}x^2_{ij+1}}{x^2_{ij}x^2_{i+1j+1}}\,,\quad p_i=x_i-x_{i-1}\,,
\ee
with indices cyclically identified, $i+N\simeq i$. {As shown in \cite{Bartels:2012gq,Bartels:2014mka}, in the following convenient choice of $3N-15$ algebraically independent cross ratios, the multi-Regge limit \eqref{eq:MRK_def} becomes}
\be\label{eq:UcrossratiosMRK}
u_{1i}\equiv U_{i+1i+4}\to 1\,,\quad  u_{2i}\equiv U_{Ni+2}\to 0\,,\quad u_{3i}\equiv U_{1i+3}\to 0\,,
\ee
with
\be
\tilde u_{2i}\equiv\frac{u_{2i}}{1-u_{1i}}\equiv \frac{1}{|1-z_i|^2}\,,\quad\tilde u_{3i}\equiv\frac{u_{3i}}{1-u_{1i}}\equiv \frac{|z_i|^2}{|1-z_i|^2}\,,\quad i=1,\ldots,N-5\,,
\ee
kept finite. In the right-hand side, the surviving transverse cross ratios describing the limit are chosen slightly differently compared to another widely used convention in the literature,
\be
z_i=-w_i=\frac{({\bf x}_1 -{\bf x}_{i+3})\,({\bf x}_{i+2} -{\bf x}_{i+1})}{({\bf x}_1 -{\bf x}_{i+1})\,({\bf x}_{i+2} -{\bf x}_{i+3})}\,,
\ee
so as to prevent a proliferation of minus signs in what follows. We have also expressed the
transverse cross ratios above in terms of transverse dual points ${\bf x}_i$,
\be
{\bf p}_{i+3}\equiv {\bf x}_{i+2}-{\bf x}_{i+1}\,,\quad i=0,\ldots,N-4\,,
\ee
in order to illustrate that the kinematics is invariant under $SL(2,\mathbb{C})$ transformations, comprised of translations,  dilatations and special conformal transformations.

The ${\bf x}_i$ correspond to $N-2$ points on $\mathbb{CP}^1$, or equivalently the Riemann sphere $\mathbb{C}\cup\{\infty\}$. This implies that we may pick a gauge by exploiting the $SL(2,\mathbb{C})$ symmetry to set three of the coordinates to specific values, and parametrize our kinematics equally well in terms of the remaining $N-5$ coordinates ${\bf x}_i$. Along with the $z_i$, a very convenient set of coordinates we will be using, also known as \emph{simplicial MRK coordinates}, amounts to the choice
\be\label{eq:def_rho}
({\bf x}_1\,\ldots,{\bf x}_{N-2})\to(1,0,\rho_1,\ldots,\rho_{N-5},\infty)\,.
\ee
From the last two equations, it is clear that the two sets of coordinates are related by
\begin{equation}\label{ztorho}
z_i=\frac{(1-\rho_{i+1})(\rho_i-\rho_{i-1})}{(1-\rho_{i-1})(\rho_i-\rho_{i+1})}\,,
\quad
\rho_0=0\,,\rho_{N-4}=\infty\,.
\end{equation}

Apart from the finite cross ratios, we may also define the $N-5$ independent variables that become small in the limit as
\be
\tau_i\equiv\sqrt{u_{2i}u_{3i}}\,,\quad i=1,\ldots, N-5\,,
\ee
where a symmetric choice was made, relying on the fact that $u_{2i}$ and $u_{3i}$ vanish at the same rate for each $i$. {Given that the variables $z_i$ in \eqref{ztorho} are complex, we are thus left with $2(N-5)$ finite real parameters that survive in the limit, which together with the small variables  of the last equation match the total number of $3N-15$ independent cross ratios.}

In the Euclidean region, where all Mandelstam invariants are spacelike, the BDS amplitude accurately
describes the multi-Regge limit \cite{DelDuca:2008jg}, and thus $\cR\to1$ there. In other words,
while $\cR$ develops large logarithms in the variables $\tau_i$ defined above, their coefficient
functions will vanish in the limit in the Euclidean region.

{In order to obtain a nontrivial result, we first need to analytically continue the energy components
of some of the particles to opposite sign, with the different choices for doing so giving rise to
different so-called Mandelstam regions. In this paper we will be focusing on the regions where all produced particles have their energy components flip sign under analytic continuation, which amounts \cite{Bartels:2014mka,Bargheer:2015djt} to transforming a single cross-ratio as follows,
\be\label{eq:Ucontinuation}
\tilde u\equiv U_{2,N-1}\to e^{-2\pi i}U_{2,N-1}\,.
\ee
After analytic continuation, the ratio $\cR$ will thus be a polynomial of large logarithms $\log\tau_i$, whose coefficients will have residual kinematical dependence on the $z_i$.
The case $N=6$ has been extensively studied in the literature.
In the context of this paper we will only consider the case $N=7$, and we leave the extension to more legs to future work.
}

\subsection{Symmetries and soft limits}\label{sec:SymmetriesLimits}
Let us now discuss the relevant discrete symmetries of multi-Regge kinematics. Parity $P$ corresponds to spatial reflection, or more correctly to the exchange of the two factors comprising the Lorentz group, $SO(3,1)\simeq SL(2)\times SL(2)$. As such, it can be shown (for example by reflecting along the $x$-axis) that it reverses particle helicities, and amounts to complex conjugation for the variables parametrizing the limit,
\be\label{eq:Parityzrho}
P: z\leftrightarrow \bar z\quad \text{and}\quad \rho_i\leftrightarrow\bar \rho_i \,,
\ee
leaving all cross ratios $U_{ij}$ invariant. Thus at the level of the amplitude it acts as
\be\label{eq:ParitycR}
P\cR_{h_1,\ldots,h_{N-4}}(\tau_1,{z}_1,\ldots,\tau_{N-5},{z}_{N-5}) =
\cR_{-h_1,\ldots,-h_{N-4}}(\tau_1,{z}_1,\ldots,\tau_{N-5},{z}_{N-5}) \,.
\ee
Due to \eqref{eq:RN}, we observe that parity is a symmetry of BDS-normalized MHV amplitudes.

Next, the target-projectile transformation $F$ is a particular flip symmetry of the dihedral symmetry respected by MHV amplitudes and superamplitudes that has a natural action in the Multi-Regge limit. As its name suggests, it amounts to exchanging the two initial-state gluons and reversing their orientation, or more generally $p_i\to -p_{N+3-i}$. At the level of cross ratios and the variables describing the limit, this translates to
\be\label{eq:TargetProjectileUzrho}
F:
\begin{cases}
u_{1i}\leftrightarrow u_{1N-4-i}\,,\,u_{2i}\leftrightarrow u_{3N-4-i}\quad \text{and}\quad \tau_i\leftrightarrow \tau_{N-4-i}\,,\\
z_i\leftrightarrow 1/{z_{N-4-i}}\quad \text{and}\quad \rho_i\leftrightarrow 1/\rho_{N-4-i}\,,
\end{cases}
\ee
respectively. The reversal of the orientation has the same effect on particle helicities, such that the action of target-projectile transformation on the amplitude becomes
\be\label{eq:TargetProjectilecR}
F\cR_{h_1,\ldots,h_{N-4}}\left(\tau_{1},{z_{1}},\ldots,\tau_{N-5},{z_{N-5}}\right) =
\cR_{-h_{N-4},\ldots,-h_1}\left(\tau_{1},{z_{1}},\ldots,\tau_{N-5},{z_{N-5}}\right) \,.
\ee
In particular for MHV amplitudes again \eqref{eq:RN} reveals again that the target-projectile transformation is
a symmetry of the terms with a symmetric combination $\prod_i \log^k\tau_i$ of large logarithms.

Finally, for our purposes it will also be necessary to review the behavior of the amplitude when
taking soft limits. As a consequence of the fact that the BDS amplitude captures all {soft/collinear} divergences of the amplitudes, the BDS-normalized ratio \eqref{eq:Rhhh_definition} is finite in the limit where the momentum of any of the produced gluons becomes soft, and reduces to the same quantity with one leg less.

More concretely, in terms of the transverse cross ratios in MRK, the limits and the behavior of $\cR$ in them are described by{
\begin{gather}
z_1\to 0: \cR_{h_1h_2\ldots}(\tau_1,z_1,\tau_2,z_2,\ldots)\to \cR_{h_2\ldots}(\tau_2,z_2,\ldots)\,,\label{eq:wsoftlimits}\\
\nonumber\\
z_i\to 0\,,\, z_{i-1}z_{i}\,\,\text{fixed}:\cR_{\ldots h_{i}\ldots}(\ldots,\tau_{i-1},z_{i-1},\tau_{i},z_{i},\ldots)\to \cR_{\ldots \hat h_{i}\ldots}(\ldots,\tau_{i-1}\tau_{i},-z_{i-1}z_{i},\ldots)\,,\nonumber\\
\nonumber\\
z_{N-5}\to \infty : \cR_{\ldots h_{N-5}h_{N-4}}(\ldots,\tau_{N-6},z_{N-6},\tau_{N-5},z_{N-5})\to \cR_{\ldots h_{N-5}}(\ldots,\tau_{N-6},z_{N-6})\,,\nonumber
\end{gather}}
with $2\le i\le N-5$ on the second line, and where for compactness we only included the dependence on the relevant helicities, with the hat indicating that the corresponding helicity is absent. The same limits can be equivalently formulated in terms of the $\rho_i$ variables, and correspond to
\be\label{eq:vsoftlimits}
\rho_{i-1}=\rho_{i}\,,i=1,\ldots,N-4\,,\quad\text{with fixed }\,\rho_0=0, \rho_{N-4}=\infty\,.
\ee
For the MHV ratio, of course all symmetries and soft limits discussed here also carry over to its logarithm, the remainder function \eqref{eq:RN}.

\subsection{Single-valued multiple polylogarithms}\label{sec:SVMPL}

$N$-particle amplitudes in MRK are expected to be described by single-valued iterated integrals on the space
of configurations of $N-2$ points on the Riemann sphere. This implies that they do not have branch cuts in the $z_i$ plane, only isolated
singularities when at least one of their symbol letters vanishes. These functions can always be expressed in terms of polylogarithms, {defined by the iterated integral
\beq
G(a_1,\ldots,a_n; z) = \int_0^z\frac{dt}{t-a_1}G(a_2,\ldots,a_n;t)\,,
\eeq
and the recursion starts with $G(;z)=1$. In the special case where all the $a_i$ are zero, we define
\beq
G(\underbrace{0,\ldots,0}_{n\textrm{ times}};z)= \frac{1}{n!}\log^nz\,.
\eeq
{As we argued in our previous work \cite{DelDuca:2016lad}, for $N$-particle amplitudes in the multi-Regge limit} the
 singularities, i.e., the symbol entries, are described by the cluster algebra $A_{N-5}\times A_{N-5}$, where the two copies of $A_{N-5}$ are
 conjugate to each other. } We refer to this class of functions as \emph{(single-valued) $A_{N-5}\times A_{N-5}$ polylogarithms}.

Focusing on the case of a single variable $z$ and its complex conjugate for simplicity, i.e. the case $A_1\times A_1$, if $F(z,\bar z)$ is a single-valued multiple polylogarithm (SVMPL), its \emph{holomorphic} part $F^h(z)$ is defined to be the strict limit $\bar z\to 0$ with  $z$ fixed, and with any divergent logarithms $\log\bar z$ also removed. Namely,
\be\label{eq:HolomorphicPart}
F^h(z)=F(z,0)\Big|_{\log\bar z\to0}\,,
\ee
with the obvious generalization to the antiholomorphic part, or to more variables.
We refer to the holomorphic part of single-valued $A_{N-5}\times A_{N-5}$ polylogarithms as \emph{$A_{N-5}$ polylogarithms}.

A key property of SVMPLs, is that they can be uniquely reconstructed from the knowledge of their
(anti)-holomorphic part with the help of the single-valued map $\bf{s}$,
\be\label{eq:SingleValuedMap}
F(z,\bar z) = F^h(z)\Big|_{G(\vec a;z)\to {\bf s}(G(\vec a;z))}\,.
\ee
This immediately follows from the fact that
\be
{\bf s}(G(\vec a;z))\Big|_{\bar{z}\to0,\,\log\bar z\to0} = G(\vec a;z)\,.
\ee
The single-valued map ${\bf s}$ was algorithmically constructed in \cite{BrownSVMPLs,Brown_Notes} for arbitrary periods, and explicitly specialized to multiple polylogarithms in \cite{DelDuca:2016lad}.
As an example we give its action on weight 1 and 2 polylogarithms below, \begin{align}
{\bf{s}}\left(G_{a}(z)\right)\equiv\cG_{a}(z)&=G_{a}(z)+G_{\bar a}(\bar z),\label{eq:ExampleSVMPL}\\
{\bf{s}}\left(G_{a,b}(z)\right)\equiv\cG_{a,b}(z)&= G_{a,b}(z)+G_{\bar{b},\bar{a}}(\bar{z})+G_{b}(a) G_{\bar{a}}(\bar{z})+G_{\bar{b}}(\bar{a}) G_{\bar{a}}(\bar{z})\nonumber\\
&-G_{a}(b) G_{\bar{b}}(\bar{z})+G_{a}(z) G_{\bar{b}}(\bar{z})-G_{\bar{a}}(\bar{b}) G_{\bar{b}}(\bar{z})\,,
\end{align}
{where here and in what follows we adopt the convention}
\be
G(a_1,\ldots a_n;z)=G_{a_1,\ldots a_n}(z)\,,
\ee
{and similarly for $\cG$, for compactness.}

The single-valued map provides an important isomorphism between the algebra of the multiple
polylogarithms and that of the SVMPLs. It furthermore plays a crucial role in the reconstruction
of single-valued results from the dispersive representation of the amplitude in MRK,
which provides a straightforward way to obtain the (anti)-holomorphic part of the amplitude through
a Fourier-Mellin integral.

\section{The BFKL equation at finite coupling}\label{sec:BFKLFinite}
In this section, we will obtain a dispersion integral describing the multi-Regge limit of the $2\to5$ amplitude that is well-defined at any logarithmic accuracy, based on the eikonal approach of~\cite{Caron-Huot:2013fea}. We review the basic ingredients of this approach for the $2\to4$ amplitude in subsection \ref{sec:6ptBFKL}, before extending it the $2\to 5$ MHV amplitude in subsection \ref{sec:7ptBFKL}. The reader interested in the final result may jump directly to section \ref{sec:Rhhh}, where we present the generalization of the integral to any helicity configuration, along with certain redefinitions of its building blocks, which will prove more convenient in what follows.

\subsection{6-points}\label{sec:6ptBFKL}
For $2\to4$ scattering in the multi-Regge limit, the six-point remainder function ${R}_6$ in the region where we analytically continue the energy components of all produced particles is given by the all-order dispersion relation\footnote{Note that we have $(z_i/\bar z_i)^{n/2}=(-1)^n(w_i/\bar w_i)^{n/2}$ when converting any of the dispersion relations considered in this paper between $z_i\leftrightarrow-w_i$, because any choice of branch on the square roots should also respect complex conjugation. For example, if we choose $\sqrt{z}=i\sqrt{w}$, then we must also have $\sqrt{\bar z}=-i\sqrt{\bar w}$.\label{footnote1}}
\beq
\label{eq:R6start}
e^{{R}_6(z) + i \delta_6(z)} =2\pi i f_{++}\,,
\eeq
\be\label{eq:f_omega2}
f_{++}= \frac{a}{2}  \sum_{n=-\infty}^{\infty}
\left(\frac{z}{\bar z}\right)^{\frac{n}{2}}
\int_{-\infty}^{\infty}\frac{d\nu}{2\pi}\,\tilde{\Phi}(\nu,n)|z|^{2i\nu}e^{-L \omega(\nu,n)},
\ee
where
\be\label{eq:R6LargeLog}
L=\log(\sqrt{u_{21} u_{31}})+i\pi=\log(\tau)+i\pi\,,
\ee
contains the logarithms that become large in the limit, whereas $\omega(\nu,n)$ is the BFKL eigenvalue and $\tilde{\Phi}(\nu,n)$ is the product of the impact factors respectively, to all orders in the coupling
\be\label{eq:acoupling}
a=2g^2=\frac{\lambda}{8\pi^2}\,.
\ee
The phase $\delta_6$ appearing on the left-hand-side is the BDS contribution defined in ref.~\cite{Lipatov:2010qf}, which is given to all orders in perturbation theory as
\beq\label{eq:delta6}
\delta_6(z) =\pi\,\Gamma \log\frac{u_2 u_3}{(1-u_1)^2}= \pi\,\Gamma
\log\frac{|z|^2}{|1-z|^4}\,,\quad \Gamma\equiv\frac{\gamma_K}{8}=\frac{\Gamma_\text{cusp}}{4}\,,
\eeq
where $\Gamma$ is proportional to the cusp anomalous dimension
\be\label{eq:gK}
\gamma_K(a) = 4 a - 4 \zeta_2 a^2  + 22 \zeta_4 a^3
- \Bigl( \frac{219}{2} \zeta_6 + 4 \zeta_3^2 \Bigr) a^4
+ \Bigl( \frac{1774}{3} \zeta_8 + 8 \zeta_2 \zeta_3^2 + 40 \zeta_3 \zeta_5
  \Bigr) a^5 +\cO(a^6)\,,
\ee
known to all loops from integrability (see \cite{Freyhult:2010kc} for a review).

{We note that \eqref{eq:f_omega2} differs from other results in the literature, in particular from \cite{Fadin:2011we}. This can be traced back to the choice of the integration contour to be used in \eqref{eq:f_omega2}, which we have not specified so far. The two formulations are in fact equivalent via a change of integration contour~\cite{Caron-Huot:2013fea}, as we explain below.}
We should also note that one could equally well use $\Gamma$ as an expansion parameter, rather than $a$. Based on this choice, and some other considerations we will review in what follows, there exist two different definitions of $\tilde{\Phi}(\nu,n)$ in the literature,
\be
\begin{aligned}\label{eq:Phi_redef}
\frac{a}{2} \tilde{\Phi}(\nu,n)&=\frac{a}{2}\frac{\Phi_{\text{reg}}(\nu,n)}{\nu^2+\frac{n^2}{4}}=\frac{\Gamma  \Phi(\nu,n)}{\nu^2+\frac{n^2}{4}-\pi^2\Gamma^2}\,,
\end{aligned}
\ee
where the first expression is due to Lipatov, Bartels and collaborators~\cite{Lipatov:2010ad,Fadin:2011we}, and the second one is due to Caron-Huot~\cite{Caron-Huot:2013fea}.

\begin{figure}[h]
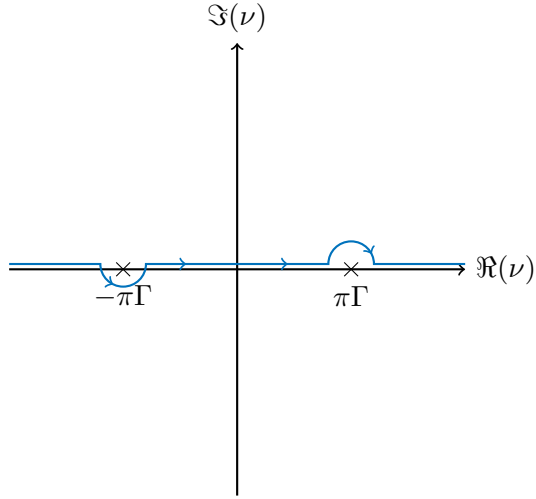

        \center
        \contourOne
        \caption{Integration contour for the six-gluon BFKL integral at finite coupling. {Here and in the following figures indicating different integration contours for the six-and seven-gluon BFKL integrals, we only depict the singularities on the integration contour, not the entire $\nu$-plane.}}
        \label{fig:contourOne}
\end{figure}

Strong constraints on the analytic structure of the integrand in \eqref{eq:f_omega2} at finite coupling can be derived by considering the soft limits $z\to0$ and $z\to\infty$.  {The correct soft limit behavior of the BDS ansatz implies that ${R}_6$ has to vanish there to all orders in perturbation theory}, and thus by virtue of~\eqref{eq:delta6} the left-hand-side of eq.~\eqref{eq:R6start} reduces to
\begin{align}
\lim_{z\to0}e^{{R}_6(z) + i \delta_6(z)} &= |z|^{2\pi i \Gamma}\label{eq:R6softlim1}\\
\lim_{z\to\infty} e^{{R}_6(z) + i \delta_6(z)} &= |z|^{-2\pi i \Gamma}.\label{eq:R6softlim2}
\end{align}
From this we can determine the behavior of the right-hand-side of eq.~\eqref{eq:R6start} for $n=0$ (terms coming from $n\neq0$ will be suppressed in the soft limit). It is evident that the integrand should have simple poles at $\nu = \pm \pi \Gamma$, with residues
\be
\mathrm{Res}_{\nu=\pm\pi \Gamma}\left(\tilde{\Phi}(\nu,0)e^{-L \omega(\nu,0)}\right)=\pm\frac{1}{\pi a}\,,
\ee
in order to capture the all order soft behavior of the left-hand-side.
In fact, a more detailed analysis of the soft limit reveals that it separately restricts $\omega(\nu,n)$ and $\tilde{\Phi}(\nu,n)$ to obey the \textit{exact bootstrap conditions}~\cite{Caron-Huot:2013fea},
\beq\label{eq:omega_phi_bootstrap}
\omega(\pm \pi \Gamma,0) = 0, \quad \textrm{and}\,\quad \mathrm{Res}_{\nu=\pm\pi \Gamma}\tilde{\Phi}(\nu,0)=\pm\frac{1}{\pi a},
\eeq
where by virtue of~\eqref{eq:Phi_redef}, the second relation may also be written as
\be\label{eq:phi_bootstrap}
\Phi(\pm\pi \Gamma,0) = 1\,.
\ee
Finally, since the integral~\eqref{eq:f_omega2} diverges if the poles are located right on the real axis, the soft limits~\eqref{eq:R6softlim1}-\eqref{eq:R6softlim2} also dictate
how the contour should be deformed in order to avoid these poles on the real axis:
Given that we need to close the contour on the lower (upper) half-plane in $\nu$ for $z$ small (large), it is also clear that the integration contour should run above the pole at $\pi \Gamma$ and below the pole at $-\pi \Gamma$, in other words
\be\label{Phi_contour}
\frac{\Phi(\nu,0)}{\nu^2-\pi^2\Gamma^2}\rightarrow \frac{\Phi(\nu,0)}{\nu^2-\pi^2\Gamma^2+i0}=\frac{\Phi(\nu,0)}{(\nu-\pi \Gamma+i0)(\nu+\pi \Gamma-i0)}\,,
\ee
as shown in fig.~\ref{fig:contourOne}.

{So far the discussion was restricted to finite coupling.}
As we will now review, the knowledge of the residues and integration contour that the soft limits provide at finite coupling, is also crucial for appropriately regularizing the divergences that appear in the weak coupling expansion of eq.~\eqref{eq:f_omega2}. In particular, given that at leading order $\Phi(\nu,n)\to1$, it is evident from~\eqref{Phi_contour} that the integrand becomes ill defined for $n=0$, since evaluating the residue in either the upper or lower half-plane leads to a divergence. In other words, while at finite coupling the contour runs between the poles at $\nu=-\pi \Gamma$ and $\nu=+\pi \Gamma$, in the weak coupling expansion, {where}  $\Gamma=\cO(a)$ {with} $a\to0$, these two poles will move towards $\nu=0$ and pinch the contour, leading to a divergence.

Consequently, we need to deform the contour at finite coupling before we are allowed to expand in the coupling.
There are two immediate choices for deforming the contour so that it does not pass between the two poles on the real axis anymore. We can either take plus the residue at $\nu=-\pi \Gamma$ in order to move the contour above the real line or we can take minus the residue at $\nu=+\pi \Gamma$ to move the contour below the real line.
It is possible to preserve the symmetry of the integral by averaging between both choices, in order to find
{
\begin{align}
\nonumber e^{{R}_6+i\delta_6} &= -\frac{\pi a}{2}\substack{\displaystyle\textrm{Res}\\\nu=-\pi \Gamma}\left(\tilde\Phi(\nu,0)|z|^{2i\nu}e^{-L\omega(\nu,0)}\right)+\frac{ia}{4}\sum_{n=-\infty}^{\infty}
        \left(\frac{z}{\bar z}\right)^{\frac{n}{2}}\int_{\uparrow}d\nu\,\tilde\Phi(\nu,n)|z|^{2i\nu}e^{-L\omega(\nu,n)}\\
\nonumber       &\quad+\frac{\pi a}{2}\substack{\displaystyle\textrm{Res}\\\nu=\pi \Gamma}\left(\tilde\Phi(\nu,0)|z|^{2i\nu}e^{-L\omega(\nu,0)}\right)+\frac{ia}{4}\sum_{n=-\infty}^{\infty}
        \left(\frac{z}{\bar z}\right)^{\frac{n}{2}}\int_{\downarrow}d\nu\,\tilde\Phi(\nu,n)|z|^{2i\nu}e^{-L\omega(\nu,n)}
\end{align}
where $\uparrow$ ($\downarrow$) denotes the contour running above (below) the real line, and the contours may be closed in either half-plane.}

{After we evaluate the residues with the help of the bootstrap conditions \eqref{eq:omega_phi_bootstrap}-\eqref{eq:phi_bootstrap}, combine the two contour integrals by introducing the Cauchy principal value $\cP$
\be\label{eq:CPV}
\mathcal{P}\left(\frac{1}{x}\right)=\frac{1}{2}\left(\frac{1}{x+i0}+\frac{1}{x-i0}\right)\,,
\ee
and reexpress the integrand with the help of ~\eqref{eq:Phi_redef}, we finally
obtain the separation into the conformal Regge pole and Regge cut contribution,\footnote{In fact, we could have equally well arrived at the expression~\eqref{eq:R6BFKL} from \eqref{eq:R6start} and \eqref{Phi_contour} by virtue of the Sokhotski–Plemelj theorem on the real line, \be \frac{1}{x\pm i0}=\mp \pi\delta(x)+\mathcal{P}\left(\frac{1}{x}\right)
\ee
for $x=\nu^2-\pi^2\Gamma^2$, together with $\delta(x^2-\alpha^2)=(\delta(x-\alpha)+\delta(x+\alpha))/(2|\alpha|)$.}
\beq\label{eq:R6BFKL}
e^{{R}_6+i \delta_6} = \cos(\log(|z|^2)\pi \Gamma) + i \frac{a}{2} \sum_{n=-\infty}^{+\infty}\left(\frac{z}{\bar z}\right)^{\frac{n}{2}}\mathcal{P}\int_{-\infty}^{\infty}d\nu\,\frac{\Phi_\text{reg}(\nu,n)}{\nu^2+\frac{n^2}{4}}|z|^{2i\nu}e^{-L\omega(\nu,n)}.
\eeq
This reproduces the well known expression from ref.~\cite{Fadin:2011we}, see also~\cite{Bartels:2008sc, Lipatov:2010qf, Lipatov:2010ad,Bartels:2010tx}. We see that the
regularization of the integral at $\nu=n=0$, which in the prescription \eqref{eq:CPV} amounts to the taking half the corresponding residue into account,   is intimately connected to the Regge pole contribution.}
It is worth emphasizing however that not just this symmetric choice, but any contour deformation
that avoids a pinching is equally valid for performing the weak coupling expansion. For example,
when evaluating $R_6$ for $z$ small, it is advantageous to pick the contour running below the real
axis, so that after closing it from below, the integral will no longer receive any contributions
from the poles near the real axis,\footnote{It is worth keeping in mind that there will still be
    contributions from poles in the interior of the contour for $n=0$.}
\beq\label{eq:R6_small_w}
e^{{R}_6+i \delta_6} = |z|^{2\pi i \Gamma} + i \frac{a}{2} \sum_{n=-\infty}^{+\infty}
\left(\frac{z}{\bar z}\right)^{\frac{n}{2}}\int_{\downarrow}d\nu\,\tilde \Phi(\nu,n)|z|^{2i\nu}e^{-L\omega(\nu,n)}.
\eeq

The generalization to the NMHV case is straightforward. Focusing on the helicity configuration most commonly found in the literature, see e.g. \cite{Dixon:2014iba}, the analogue of \eqref{eq:R6start}-\eqref{eq:f_omega2} is
\beq
\label{eq:R6NMHV}
\cR_{+-}e^{ i \delta_6(z)} =2\pi i f_{+-}\,,
\eeq
\be\label{eq:f_omega2NMHV}
f_{+-}= \frac{a}{2}  \sum_{n=-\infty}^{\infty}
\left(\frac{z}{\bar z}\right)^{\frac{n}{2}}
\int_{-\infty}^{\infty}\frac{d\nu}{2\pi}\,\tilde{\Phi}(\nu,n)\overline{H}(\nu,n)|z|^{2i\nu}e^{-L \omega(\nu,n)},
\ee
where the helicity flip kernel $H$ will be defined below in eq.~\eqref{eq:HelicityFlipH}.
At this point, it is sufficient to note that $H(\nu,0)=1$, which implies
that the MHV and NMHV integrands become identical for $n=0$,
and so our analysis of the poles on the integration contour as well as the prescription to avoid them
generalize straightforwardly to NMHV.
Finally, the $\cR_{-+}$ helicity configuration may be obtained from \eqref{eq:R6NMHV}-\eqref{eq:f_omega2NMHV} by a parity transformation \eqref{eq:ParitycR}.

\subsection{7-points MHV}\label{sec:7ptBFKL}
Armed with intuition from the six-point case, we now move on to propose an all-loop dispersion-type formula for the $2\to5$ amplitude in MRK, again in the region where we analytically continue the energy components of all produced particles. Our strategy will be as follows:\footnote{A similar strategy for obtaining dispersion integrals of higher-point amplitudes, also in different regions, has also been independently adopted by Basso,Caron-Huot and Sever, see \cite{BassoTalk}.}
\begin{enumerate}
\item We start with a formula that expresses the remainder plus conformal BDS contribution in the
    Multi-Regge limit at finite coupling, as a Regge cut (integral) only, i.e. without any explicit Regge poles.
\item We then examine the soft limits of the formula at finite coupling, which reveal to us the positions of the poles of the integral on the real axis, their residues, as well as the prescription for integrating around them\footnote{We assume that no other poles are present on the real axis but the ones dictated by the soft limits.\label{footn:minimalpoles}}.
\item In the weak-coupling expansion of the integrand, these poles will pinch the contour, leading to divergences. We may deform the original contour, prior to expanding in the coupling, to any contour that is not pinched in the weak-coupling limit, picking up the residues that are crossed in the process of this deformation at finite coupling, and then expand at weak coupling.
\end{enumerate}
In this manner, for any deformation described in the last step, we will obtain an integral that is well-defined at weak coupling, plus finite-coupling residue contributions, whose values we know from the soft limits.
\begin{figure}[h]
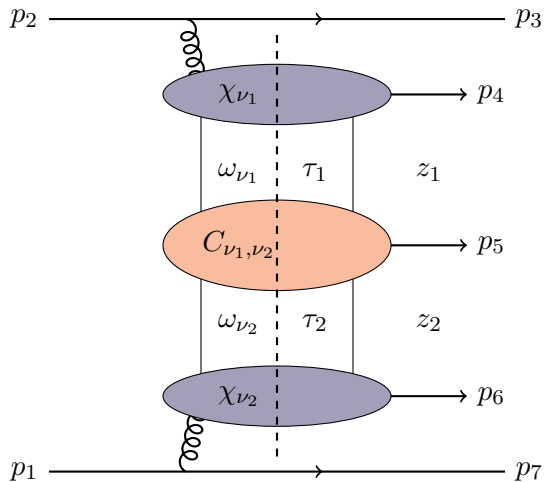

        \center
        \bfklViz
        \caption{Structure of the seven point amplitude in MRK.}
        \label{fig:bfklViz}
    \end{figure}

So let us start with the seven-point analogues of \eqref{eq:R6start}-\eqref{eq:f_omega2} shown
in Fig.~\ref{fig:bfklViz}\footnote{To make contact with other notations used in the literature, $f_{+++}$ is denoted as $f_{\omega_2\omega_3}$ in \cite{Bartels:2013jna,Bartels:2014jya}, and similarly the six-gluon analogue $f_{++}$ of the previous section is denoted as $f_{\omega_2}$.},
\beq
\label{eq:R7start}
e^{{R}_7(z_1,z_2) + i \delta_7(z_1,z_2)} =2\pi i f_{+++}\,,
\eeq
\be
\begin{aligned}\label{eq:f_omega2omega3}
f_{+++}= &\,\frac{a}{2} \sum_{n_1=-\infty}^\infty\sum_{n_2=-\infty}^\infty 
\left(\frac{z_1}{\bar z_1}\right)^{\frac{n_1}{2}} \left(\frac{z_2}{\bar z_2}\right)^{\frac{n_2}{2}}
\int \frac{d\nu_1 d\nu_2}{(2\pi)^2}
   |z_1|^{2i\nu_1}|z_2|^{2i\nu_2}
\\
&\quad \times e^{-L_1 \omega(\nu_1,n_1)}
e^{-L_2\omega(\nu_2,n_2)}\chi^+(\nu_1,n_1)C^+(\nu_1,n_1,\nu_2,n_2)
 \chi^-(\nu_2,n_2)\,,
\end{aligned}
\ee
where
\be\label{LtoTau}
L_i=\log\sqrt{u_{2i} u_{3i}}+i\pi=\log\tau_i+i\pi\,,
\ee
and the conformally invariant part of the 1-loop Regge cut coming from the BDS ansatz is\footnote{See for example  \cite{Bartels:2013jna}, where $\delta_7\to\delta_{14}$ is expressed in terms of the transverse momenta of the produced gluons, and the momentum transfer between them. It can be recast in terms of (transverse) cross ratios by virtue of the kinematic analysis of \cite{Bartels:2011ge}, as adapted to our conventions in \cite{DelDuca:2016lad}.}
\be
\begin{aligned}\label{eq:delta7}
\delta_7(z_1,z_2) &=2\pi\Gamma \log\frac{\sqrt{u_{21} u_{31} u_{22} u_{32}}}{1-\tilde u}=\pi\Gamma\log\frac{|z_1z_2|^2}{|1-z_2+z_1z_2|^4}\,.\\
\end{aligned}
\ee
In addition, apart from the BFKL eigenvalue encountered in the previous section, $\chi^{\pm}(\nu,n)$ are the two BFKL impact factors at the end of the ladder \cite{Lipatov:2010ad}, whose product with equal arguments also appeared in the hexagon case\footnote{The fact that the integrand contains impact factors of opposite helicity can be understood by thinking about the momentum flow along the ladder: If we assume that its actual direction is from the $\chi^+$ towards the $\chi^-$ impact factor, then in all-outgoing momentum conventions for these impact factors, the helicity assignments of the gluons associated with $\chi^+$ and $\chi^-$ will be $(-++)$ and $(--+)$, respectively. Hence the two must be related by parity.}
\be\label{eq:PhiTilde}
\chi^+(\nu,n)\chi^-(\nu,n)=\tilde\Phi(\nu,n)\,.
\ee
Finally, $C^+(\nu_1,\nu_2,n_1,n_2)$ is the central emission block, a new ingredient in the BFKL approach to the heptagon compared to the hexagon, first computed to leading order in~\cite{Bartels:2011ge}.

Next, we consider the three soft limits where the momentum of one of the produced particles goes to zero, and $R_7$ reduces to $R_6$. With the help of~\eqref{eq:delta7} we find that in the soft limits the left-hand-side of eq.~\eqref{eq:R7start} becomes,
\begin{align}
        \lim_{z_1\to 0}e^{{R}_7 + i \delta_7} &= e^{R_6(z_2)+i\delta_6(z_2)}|z_1|^{2\pi i \Gamma},\label{eq:R7_softlim1}\\
        \lim_{z_2\to \infty}e^{{R}_7 + i \delta_7} &= e^{R_6(z_1)+i\delta_6(z_1)}|z_2|^{-2\pi i \Gamma},\label{eq:R7_softlim2}\\
        \lim_{z_2\to 0,\, z_1z_2\, \textrm{fixed}}e^{{R}_7 + i \delta_7} &= e^{R_6(z_1 z_2) + i\delta_6(z_1 z_2)}\label{eq:R7_softlim3}.
\end{align}
Note that the last soft limit is also equivalent to $z_1\to\infty$ with $z_1 z_2$ fixed. From the behavior in the soft limits we can determine that the r.h.s of eq.~\eqref{eq:R7start}{, more precisely the integrand in eq.~\eqref{eq:f_omega2omega3},} needs to have a pole $\nu_1 =\pi \Gamma-i0$ for the $n_1=0$ term in the sum, a pole at $\nu_2 = -\pi \Gamma + i0$ for $n_2=0$, as well as at $\nu_1 = \nu_2 + i0$ (or equivalently at $\nu_2 = \nu_1 - i0$) for $n_1=n_2$. Furthermore, it is easy to check that the above relations hold if the residues on those poles are equal to
\begin{align}
\underset{{\nu_1=\pi \Gamma}}{\mathrm{Res}}\left(\chi^+(\nu_1,0)C^+(\nu_1,0,\nu_2,n_2)
 \chi^-(\nu_2,n_2)\right)&=i \tilde\Phi(\nu_2,n_2)\,,\label{eq:R7_bootstrapcond1}\\
\underset{{\nu_2=-\pi \Gamma}}{\mathrm{Res}}\left(\chi^+(\nu_1,n_1)C^+(\nu_1,n_1,\nu_2,0)
 \chi^-(\nu_2,0)\right)&=-i \tilde\Phi(\nu_1,n_1)\,,\label{eq:R7_bootstrapcond2}\\
\underset{{\nu_1=\nu_2}}{\mathrm{Res}}\left(\chi^+(\nu_1,n_2)C^+(\nu_1,n_2,\nu_2,n_2)
 \chi^-(\nu_2,n_2)\right)&=-i (-1)^{n_2}e^{i\pi\omega(\nu_2,n_2)}\tilde \Phi(\nu_2,n_2)\,.\label{eq:R7_bootstrapcond3}
\end{align}
where we already took the condition~\eqref{eq:omega_phi_bootstrap}{, as well as eq.~\eqref{eq:PhiTilde},} into account.
The previous equations can be seen as an integrand formulation in Fourier-Mellin space of the soft limits of the amplitude.
In particular, if the residues take the values in \eqref{eq:R7_bootstrapcond1}, \eqref{eq:R7_bootstrapcond2} and~\eqref{eq:R7_bootstrapcond3}, then the integrand with respect to the second integration variable becomes identical to the hexagon integrand\footnote{The exponential factor in the last relation is present because for $\nu_1=\nu_2$ the term multiplying $-\omega$ in~\eqref{eq:f_omega2omega3} becomes $L_1+L_2$, whereas the corresponding large logarithm in~\eqref{eq:R7_softlim3} should be \[L'=\log(\tau_1\tau_2)+i\pi=L_1+L_2-i\pi\,.\]}.
At this point we have to make a comment about these relations. Since the soft limits are valid for the integrated amplitude after Fourier-Mellin transformation, any integrand formulation is in principle valid only up to terms that vanish when computing the Fourier-Mellin transform. For example, we could add to~\eqref{eq:R7_bootstrapcond1} any function of $(\nu_1,n_1)$ which is mapped to zero by the Fourier-Mellin transform, without changing the soft limit of the amplitude. Since the Fourier-Mellin transform is invertible, any such function which maps to zero is necessarily trivial, and so the bootstrap conditions~\eqref{eq:R7_bootstrapcond1}, \eqref{eq:R7_bootstrapcond2} and~\eqref{eq:R7_bootstrapcond3} follow indeed from the soft limits of the full amplitude.

\begin{figure}[h]
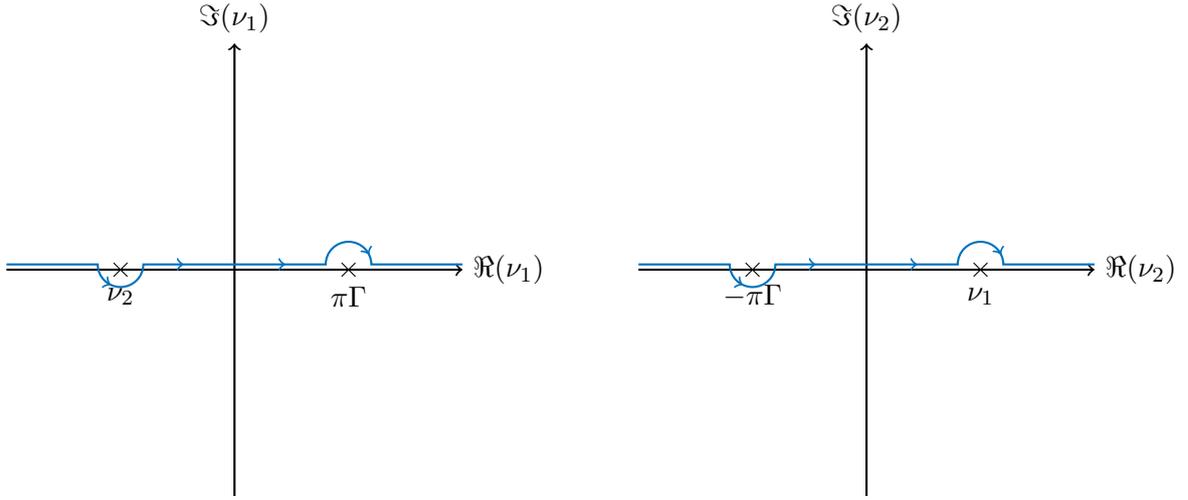

        \centering
        \begin{minipage}{0.45\textwidth}
        \centering
                \contourTwo
        \end{minipage}\hfill
        \begin{minipage}{0.45\textwidth}
                \centering
        \contourThree
\end{minipage}\caption{Integration contour for the seven-gluon BFKL integral.}
        \label{fig:contourRSevenInitial}
\end{figure}

\begin{figure}[h]
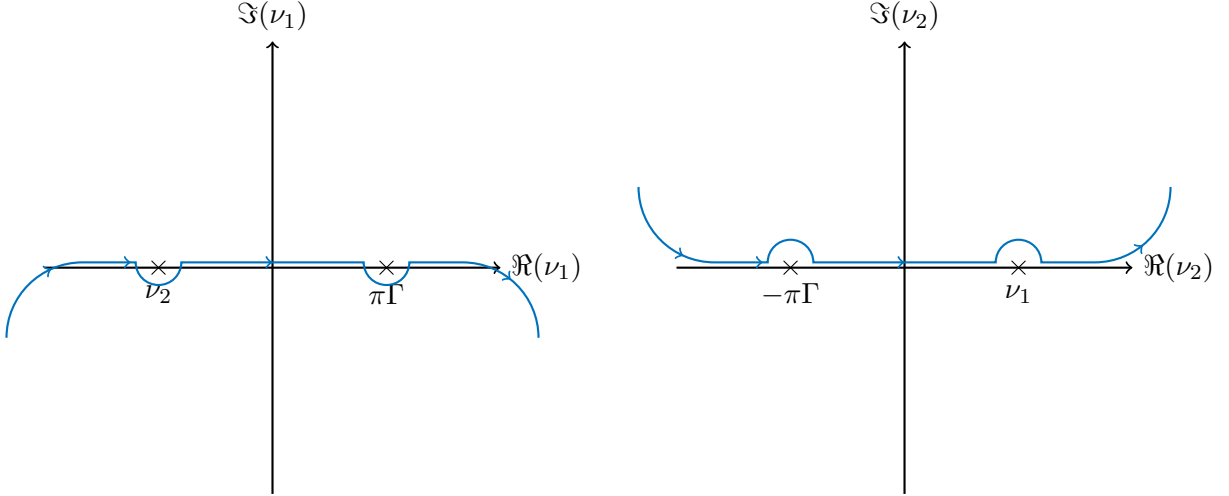

        \centering
        \begin{minipage}{0.45\textwidth}
        \centering
                \contourRSevenNuOne
        \end{minipage}\hfill
        \begin{minipage}{0.45\textwidth}
                \centering
        \contourRSevenNuTwo
\end{minipage}
        \caption{The deformed integration contour for the seven-gluon BFKL integral, which is convenient for the weak coupling expansion in the region $z_1\ll1,\,z_2\gg1$.}
        \label{fig:contourRSeven}
\end{figure}

Finally, we come to address the necessity for a contour deformation before performing the weak coupling expansion. {Given our assumption, stated in footnote \ref{footn:minimalpoles},  that no other singularities should be present on the real axis but the ones dictated by soft limits, eqs.~\eqref{eq:R7_bootstrapcond1}-\eqref{eq:R7_bootstrapcond3}, we observe that  when at least one of the $n_i$ is different from zero, we can have at most one pole on the real axis for either integration variable.  In this case one can safely expand at weak coupling, because no pinching can occur. Therefore we only need to consider a contour deformation for the case where $n_1=n_2=0$ simultaneously, depicted in figure \ref{fig:contourRSevenInitial}}. There, we see that if after we close the contour at infinity, we receive a contribution from a residue on the real line in any of the integration variables, the integral left to do in the other integration variable will have the same pole structure as the hexagon integral of figure \ref{fig:contourOne}, whereby the poles pinch the contour at weak coupling.
We will therefore have to deform the contour before expanding, and the simplest choice will be to
pick a new contour such that it does not contain any poles on the real axis. For example, in the
region $z_1\ll1, z_2\gg1$, where we close $\nu_1$ from below and $\nu_2$ from above, we can deform
the contour as follows. Let us first schematically rewrite the integrand in a way that exposes the
pole structure of the $n_1=n_2=0$ integrand in \eqref{eq:f_omega2omega3},
\be
\begin{aligned}
I&\equiv\frac{a}{2(2\pi)^2}|z_1|^{2i\nu_1}|z_2|^{2i\nu_2} e^{-L_1 \omega_1-L_2 \omega_2}\chi^+(\nu_1,0)C^+(\nu_1,0,\nu_2,0)\chi^-(\nu_2,0) \\
&=\frac{f(\nu_1,\nu_2)}{(\nu_1-\pi\Gamma)(\nu_1-\nu_2)(\nu_2+\pi\Gamma)}\,,
\end{aligned}
\ee
{where $f(\nu_1,\nu_2)$ is now regular for any real $\nu_i$.}
Then, we can rewrite our original contour as
{\begin{align}
\label{eq:R7ContourDeformation}
\int\tfrac{d\nu_1 d\nu_2 f(\nu_1,\nu_2)}{(\nu_1-\pi\Gamma+i0)(\nu_1-\nu_2-i0)(\nu_2+\pi\Gamma-i0)}&=\int\tfrac{d\nu_1 d\nu_2 f(\nu_1,\nu_2)}{(\nu_1-\pi\Gamma-i0)(\nu_1-\nu_2-i0)(\nu_2+\pi\Gamma-i0)}\\
\nonumber&+2\pi i\int \tfrac{d\nu_2 f(\pi \Gamma,\nu_2)}{(\nu_2-\pi\Gamma +i0)(\nu_2+\pi\Gamma-i0)}\\
\nonumber&=\int\tfrac{d\nu_1 d\nu_2 f(\nu_1,\nu_2)}{(\nu_1-\pi\Gamma-i0)(\nu_1-\nu_2-i0)(\nu_2+\pi\Gamma+i0)}\\
\nonumber&+2\pi i\int \tfrac{d\nu_1 f(\nu_1,-\pi \Gamma)}{(\nu_1-\pi\Gamma -i0)(\nu_1+\pi\Gamma-i0)}\\
\nonumber&+2\pi i\int \tfrac{d\nu_2 f(\pi \Gamma,\nu_2)}{(\nu_2-\pi\Gamma +i0)(\nu_2+\pi\Gamma-i0)}\\
\nonumber&=\int\tfrac{d\nu_1 d\nu_2 f(\nu_1,\nu_2)}{(\nu_1-\pi\Gamma-i0)(\nu_1-\nu_2-i0)(\nu_2+\pi\Gamma+i0)}\\
\nonumber&+2\pi i\int \tfrac{d\nu_1 f(\nu_1,-\pi \Gamma)}{(\nu_1-\pi\Gamma +i0)(\nu_1+\pi\Gamma-i0)}\\
\nonumber&+2\pi i\int \tfrac{d\nu_2 f(\pi \Gamma,\nu_2)}{(\nu_2-\pi\Gamma +i0)(\nu_2+\pi\Gamma-i0)}+(2\pi i)^2\frac{f(\pi\Gamma,-\pi\Gamma)}{2\pi\Gamma}\,,\nonumber
\end{align}}
where already in the second equality the double integral is free of poles on the real axis, and in going from the second to the third equality, we changed the contour of the single $\nu_1$ integral to make it identical to that of figure \ref{fig:contourOne}. The reason is that due to the bootstrap conditions \eqref{eq:omega_phi_bootstrap} and \eqref{eq:R7_bootstrapcond1}-\eqref{eq:R7_bootstrapcond3}, also the integrand of these simple integrals becomes identical to the hexagon integral in \eqref{eq:f_omega2}, up to factors independent of the integration variable,
{\be\label{eq:nuResidues}
\begin{aligned}
2\pi i\int \tfrac{d\nu_1 f(\nu_1,-\pi \Gamma)}{(\nu_1-\pi\Gamma +i0)(\nu_1+\pi\Gamma-i0)}=&|z_2|^{-2\pi i\Gamma} \,\frac{a}{2}\int \frac{d\nu_1}{2\pi} \tilde \Phi(\nu_1,0)|z_1|^{2i\nu_1}e^{-L_1 \omega(\nu_1,0)}\,,\\
2\pi i\int \tfrac{d\nu_2 f(\pi \Gamma,\nu_2)}{(\nu_2-\pi\Gamma +i0)(\nu_2+\pi\Gamma-i0)}=&|z_1|^{2\pi i\Gamma} \,\frac{a}{2}\int \frac{d\nu_2}{2\pi}  \tilde \Phi(\nu_2,0)|z_2|^{2i\nu_2}e^{-L_2 \omega(\nu_2,0)}\,.
\end{aligned}
\ee}
Similarly, for the double residue in the last line of \eqref{eq:R7ContourDeformation} we obtain
\be\label{eq:DoubleRes}
(2\pi i)^2\mathrm{Res}_{\nu_1=\pi \Gamma}\mathrm{Res}_{\nu_2=-\pi \Gamma}I= (2\pi i)^2\frac{f(\pi\Gamma,-\pi\Gamma)}{2\pi\Gamma}=-\frac{1}{2\pi i}\frac{|z_1|^{2\pi i \Gamma}}{|z_2|^{2\pi i \Gamma}}\,.
\ee
Choosing to deform the contour in the same fashion for the case where one of the $n_i$ is zero and the other nonzero, so that they contain no poles on the real axis, we observe that the summands combine nicely to yield
\be\label{eq:R7_w1_small_w2_large}
e^{{R}_7 + i \delta_7} =|z_1|^{2\pi i \Gamma}e^{R_6(z_2)+i\delta_6(z_2)}+|z_2|^{-2\pi i \Gamma}e^{R_6(z_1)+i\delta_6(z_1)}-\frac{|z_1|^{2\pi i \Gamma}}{|z_2|^{2\pi i \Gamma}}+2\pi i \tilde f_{+++}\,,
\ee
where $\tilde f_{+++}$ is defined precisely as in \eqref{eq:f_omega2omega3}, but now with the integration contour of {the last equality in} eq.~\eqref{eq:R7ContourDeformation}, which is also illustrated in figure~\ref{fig:contourRSeven}. Notice that the presence of the second-to-last term from the right, coming from eq.\eqref{eq:DoubleRes}, is necessary for reproducing the soft limits~\eqref{eq:R7_softlim1}-\eqref{eq:R7_softlim2}.
We stress that although the above formula, which is the 7-point analogue of \eqref{eq:R6_small_w}, holds independently of how we choose to close the integration contours, it is only valid in the region $z_1\ll1,\,z_2\gg1$ which is convenient for expanding at weak coupling. This is because the $\nu_1=\nu_2$ residue is a simple integral that diverges at weak coupling, and so it cannot be contained in our closed contour.
By deforming the contour in \eqref{eq:R7ContourDeformation} around the $\nu_1=\nu_2$ pole, we similarly find
\be\label{eq:R7_w1_large_w2_small}
e^{{R}_7 + i \delta_7} =e^{R_6(z_1 z_2) + i\delta_6(z_1 z_2)}+2\pi i \check f_{+++}\,,
\ee
where again $\check f_{+++}$ is defined as in \eqref{eq:f_omega2omega3}, but this time the contour is the one that results after we exchange $\nu_1\leftrightarrow\nu_2$ in figure~\ref{fig:contourRSeven}. The last formula is particularly suited for the weak coupling expansion in the $z_1\gg1,\,z_2\ll1$ region. Finally, one may tempted to take the average of \eqref{eq:R7_w1_small_w2_large}-\eqref{eq:R7_w1_large_w2_small} as the analogue of \eqref{eq:R6BFKL}, however at least currently it seems that it is not convenient for the weak coupling expansion, since it will lead to pinching in both contours.

\subsection{Summary and extension to any helicity}\label{sec:Rhhh}

In the previous section, for simplicity we focused on the MHV $2\to5$ amplitude. Here we will present the generalization of the all-loop dispersion integral~\eqref{eq:R7start}-\eqref{eq:delta7}, as well as the exact bootstrap conditions \eqref{eq:R7_bootstrapcond1}-\eqref{eq:R7_bootstrapcond3} that are obeyed by the building blocks, for arbitrary helicity configurations.

Using definitions of subsection~\ref{sec:kinematics}, the multi-Regge limit of the BDS-normalized $N=7$ particle amplitude~\eqref{eq:Rhhh_definition} will be given by
\beq
\label{eq:Rhhhstart}
 \cR_{h_1h_2h_{3}}e^{ i \delta_7(z_1,z_2)} =2\pi i f_{h_1h_2h_3}\,,
\eeq
\begin{align}
f_{h_1h_2h_3}= &\frac{a}{2} \sum_{n_1,n_2=-\infty}^\infty 
\left(\frac{z_1}{\bar z_1}\right)^{\frac{n_1}{2}} \left(\frac{z_2}{\bar z_2}\right)^{\frac{n_2}{2}}
\int \frac{d\nu_1 d\nu_2}{(2\pi)^2}
   |z_1|^{2i\nu_1}|z_2|^{2i\nu_2}\tilde \Phi(\nu_1,n_1)\tilde \Phi(\nu_2,n_2)
\nonumber\\
&\times e^{-L_1 \omega(\nu_1,n_1)-L_2\omega(\nu_2,n_2)}I^{h_1}(\nu_1,n_1)\tilde C^{h_2}(\nu_1,n_1,\nu_2,n_2)
 \bar I^{h_3}(\nu_2,n_2)\,, \label{eq:f_hhh}
\end{align}
 with
\be
L_i=\log\tau_i+i\pi\,, \quad \delta_7(z_1,z_2) =\pi\Gamma\log\frac{|z_1z_2|^2}{|1-z_2+z_1z_2|^4}\,,
\ee
as well as
\be\label{eq:PhiTildeAgain}
\tilde\Phi(\nu,n)=\chi^+(\nu,n)\chi^-(\nu,n)\,,
\ee
as defined previously, and the contour of integration as depicted in figure \ref{fig:contourRSevenInitial}. In addition, we have expressed the integrand in terms of the rescaled quantities\footnote{In more detail, the generalization of \eqref{eq:f_omega2omega3} to arbitrary helicity follows from $\chi^+ C^+\chi^-\to \chi^{h_1} C^{h_2}\chi^{-h3}$, which can then be recast in the form \eqref{eq:f_hhh} after we plug in the solution of \eqref{eq:Ih} and \eqref{eq:Ctilde} for $\chi^h$ and $C^h$ respectively, and finally use \eqref{eq:PhiTildeAgain}.}
\be\label{eq:Ctilde}
 \tilde C^h(\nu_1,n_1,\nu_2,n_2) \equiv\frac{C^h(\nu_1,n_1,\nu_2,n_2)}{\chi^-(\nu_1,n_1)\chi^+(\nu_2,n_2)}
\ee
and
\be\label{eq:Ih}
 I^h(\nu,n)\equiv\frac{\chi^h(\nu,n)}{\chi^+(\nu,n)}=
\begin{cases}
1,&h=+\\
H(\nu,n)&h=-
\end{cases}
 \\
\ee
with $\bar I^h$ denoting the complex conjugate of $I^h$. In the last equation,
\be\label{eq:HelicityFlipH}
H(\nu,n)=\frac{x\left(u(\nu)-\frac{in}{2}\right)}{x\left(u(\nu)+\frac{in}{2}\right)}\,,
\ee
is the helicity flip kernel, or NMHV form factor, known to all loops from integrability \cite{Basso:2014pla}, as is also the case for the hexagon impact factor $\tilde \Phi$. The precise definition of the Zhukowski variables $x$ and the rapidities $u$ will not be important for our purposes, and we will be explicitly  providing the weak-coupling expansion of $H$ in section \ref{sec:CNLO}. A crucial property that however follows immediately from the above representation, is that
\be\label{eq:H0I0}
H(\nu,0)=1 \,\,\Rightarrow \,\,I^h(\nu,0)=1\,.
\ee

The most significant advantage of defining a rescaled central emission block as in \eqref{eq:Ctilde}, is that it allows us to formulate separate exact bootstrap conditions for the latter: Along with the  conditions
\beq\label{eq:OmegaPhiBootStrapConditions}
\omega(\pm \pi \Gamma,0) = 0, \quad \textrm{and}\,\quad \mathrm{Res}_{\nu=\pm\pi \Gamma}\left(\tilde{\Phi}(\nu,0)\right)=\pm\frac{1}{\pi a},
\eeq
which as we reviewed in section \ref{sec:6ptBFKL} follow from the analysis of the six-gluon
amplitude, the consistency of soft limits of the seven-point amplitude requires
\begin{align}
\tilde C^h(\pi \Gamma,0,\nu_2,n_2)
 &=i \pi a \,I^h(\nu_2,n_2)\,,\label{eq:R7hhh_bootstrapcond1}\\
\tilde C^h(\nu_1,n_1,-\pi \Gamma,0)
&=-i \pi a \,\bar I^h(\nu_1,n_1)\,,\label{eq:R7hhh_bootstrapcond2}\\
\underset{{\nu_1=\nu_2}}{\mathrm{Res}}\tilde C^h(\nu_1,n_2,\nu_2,n_2)&=\frac{-i (-1)^{n_2}e^{i\pi\omega(\nu_2,n_2)}}{\tilde \Phi(\nu_2,n_2)}\,,\label{eq:R7hhh_bootstrapcond3}
\end{align}
as well as
\be\label{eq:R7hhh_bootstrapcond4}
C^h(-\pi \Gamma,0,\nu_2,n_2)=C^h(\nu_1,n_1,\pi \Gamma,0)=0\,.
\ee
In particular, eqs.~\eqref{eq:R7hhh_bootstrapcond1}-\eqref{eq:R7hhh_bootstrapcond3} for $h=+$ follow from \eqref{eq:R7_bootstrapcond1}-\eqref{eq:R7_bootstrapcond3} and \eqref{eq:OmegaPhiBootStrapConditions}, after also taking into account that $\tilde C^h$ must be regular at $n_1=0,\nu_1=\pi \Gamma$ and $n_2=0,\nu_2=-\pi \Gamma$ for the soft limits \eqref{eq:R7_softlim1}-\eqref{eq:R7_softlim2} to hold (e.g. a pole would lead to additional $\log z_i$ dependence that is incompatible with these limits). In a similar spirit, the regularity of the entire integrand at $n_1=0,\nu_1=-\pi \Gamma$ and $n_2=0,\nu_1=\pi \Gamma$ implies \eqref{eq:R7hhh_bootstrapcond4}, so as to cancel the poles of either of the $\tilde \Phi(\nu_i,n_i)$ there.

Then, the extension of these conditions to $\tilde C^-$ can be done by recalling that the MHV and $\overline{\text{MHV}}$ amplitudes must be equal to each other, $\cR_{+++}=\cR_{---}$, as a consequence of their equivalence to the same bosonic Wilson loop under the Wilson loop/amplitude duality. Imposing this on \eqref{eq:Rhhhstart}-\eqref{eq:f_hhh} implies
\be
\tilde C^-(\nu_1,n_1,\nu_2,n_2)=\overline{H}(\nu_1,n_1) C^+(\nu_1,n_1,\nu_2,n_2)H(\nu_2,n_2)\,,
\ee
allowing us to obtain \eqref{eq:R7hhh_bootstrapcond1}-\eqref{eq:R7hhh_bootstrapcond3} for $h=-$ from $h=+$, also with the help of \eqref{eq:H0I0}. Note that the last formula implies that we only need consider $\omega, \tilde \Phi, H$ and $\tilde C^+$ as the independent building blocks of the integrand, and then $\tilde C^-$ may be expressed in terms of the last two.

Finally, by deforming the contour, it is possible to equivalently write \eqref{eq:Rhhhstart}-\eqref{eq:f_hhh} as
\be\label{eq:R7hhh_w1_small_w2_large}
 \cR_{h_1h_2h_{3}}e^{ i \delta_7} =|z_1|^{2\pi i \Gamma}\cR_{h_2h_{3}}(z_2)e^{i\delta_6(z_2)}+|z_2|^{-2\pi i \Gamma}\cR_{h_1h_{2}}(z_1)e^{i\delta_6(z_1)}-\frac{|z_1|^{2\pi i \Gamma}}{|z_2|^{2\pi i \Gamma}}+2\pi i \tilde f_{h_1h_2h_3}\,,
\ee
where $\tilde f_{h_1h_2h_3}$ is defined as in \eqref{eq:Ctilde}, but with the integration contour
illustrated in figure \ref{fig:contourRSeven}. In what follows, we will almost exclusively be using
this form of the dispersion integral, which is particularly convenient for its weak-coupling
expansion in the region \mbox{$z_1\ll1,\,z_2\gg1$} we will consider.

\section{From symbols to functions in MRK}\label{sec:SymbolsToFunctions}
In the previous section, we succeeded in obtaining an all-order dispersion integral describing the multi-Regge limit of the $2\to5$ amplitude of any helicity configuration, that is well defined at arbitrary logarithmic accuracy. In order to complete the description we also need to determine the building blocks of the integrand, and while the ones associated to the six-particle amplitude are known to all loops, the (rescaled) central emission block \eqref{eq:Ctilde} is only known to leading order \cite{Bartels:2011ge}.

The aim of the next two sections will thus be to extract the central emission block at higher order from the known perturbative data for the amplitude, exploiting the fact that if we know the left-hand side of~\eqref{eq:Rhhhstart} at $\ell$ loops, we can determine all building blocks in \eqref{eq:f_hhh} at N$^{\ell-1}$LO, since they start at $\cO(a^0)$. In particular, in this section we will first promote the known 2-loop symbol of the MHV seven-particle amplitude~\cite{Bargheer:2015djt} to a function. From this, we will in fact obtain all 2-loop MHV amplitudes in the multi-Regge limit, by similarly promoting their interesting factorization property, i.e. their decomposition into building blocks associated with the six- and seven-particle amplitude, that was also discovered in ref.~\cite{Bargheer:2015djt}.

In subsection \ref{OPEDiscontinuity}, we first establish a necessary result for our subsequent
analysis, which is however expected to have other applications as well, since it holds in general
kinematics: Based on the framework of the Operator Product Expansion (OPE) for null polygonal Wilson
loops
\cite{Alday:2010ku,Gaiotto:2010fk,Gaiotto:2011dt,Sever:2011da,Basso:2013vsa,Basso:2013aha,Basso:2014koa,Basso:2014jfa,Basso:2014nra,Basso:2014hfa,Basso:2015rta,Basso:2015uxa},
we derive the maximal degree of logarithmic divergences of MHV amplitudes in the Euclidean region
for any number of particles $N$, extending the earlier analysis in ref.~\cite{Dixon:2011pw} of the $N=6$
case. In subsection \ref{sec:R72MRK}, we then use this property, together with information from soft
limits, in order to uniquely fix all beyond-the-symbol terms of an ansatz for the 2-loop MHV
seven-point amplitude, or more precisely remainder function $R_7^{(2)}$. Finally, in section \ref{sec:RN2MRK} we obtain all 2-loop MHV amplitudes in the multi-Regge limit, by proving that the NLLA factorization of the symbol observed in \cite{Bargheer:2015djt} must necessarily also hold at function level.

\subsection{Maximal degree of logarithmic divergence from the OPE}\label{OPEDiscontinuity}
Let us start by stating the main result of this subsection: We will prove that the $N$-point $L$-loop remainder function $R_N^{(L)}$ in general kinematics may always be written as
\be\label{eq:RNOPE}
R_N^{(L)}=\sum_{0\le j_1+\ldots +j_{N-5}\le L-1}\log^{j_1} U_1\ldots \log^{j_{N-5}} U_{N-5} f_{j_1,\ldots, j_{N-5}}\,.
\ee
{The arguments $U_i$, $1\le i,\le N=5$, in the logarithms in~\eqref{eq:RNOPE} are a specific subset of the cross ratios $U_{ij}$ defined in~\eqref{eq:Ucrossratios},
\be\label{eq:OPEcrossratios}
U_i=U_{N-\lceil \frac{i}{2}\rceil,\lfloor \frac{i}{2}\rfloor+2}\,,\quad i=1,\ldots, N-5\,.
\ee
In the last relation,
$\lfloor x \rfloor$ and $\lceil x\rceil$ are the floor and ceiling functions respectively. For example, for $N=6$ we have $U_1=U_{52}$, for $N=7$ there is $U_1=U_{62}$ and $U_2=U_{63}$ and for
$N=8$ we have $U_1=U_{72}$, $U_2=U_{73}$ and $U_3=U_{63}$ and so on for higher points.
The representation of the remainder function in~\eqref{eq:RNOPE} makes explicit some of the branch cuts in $R_N^{(L)}$. Indeed,
the functions $f_{j_1,\ldots, j_{N-5}}$ are analytic (i.e., free of branch cuts) for any of the cross ratios $U_i\to 0$. Note that representations of the remainder function similar to~\eqref{eq:RNOPE} exist also for other subsets of cross ratios, in particular any other set obtained
by acting on \eqref{eq:OPEcrossratios} with dihedral transformations. 
}

The main content of \eqref{eq:RNOPE} is twofold: First, the leading discontinuity of $R_N^{(L)}$, or equivalently its maximal degree of logarithmic divergence is $L-1$. And second, that one can unshuffle all logarithms in the $U_i$ simultaneously even in general kinematics, which is nontrivial because it cannot be done for general polylogarithmic functions.

We now proceed with the proof, and at the end of this section also mention the generalization of our result beyond the MHV case. We will be relying on the Wilson loop OPE \cite{Alday:2010ku} appoach and its subsequent refinements \cite{Basso:2013vsa,Basso:2013aha,Basso:2014koa,Basso:2014nra}, where the observable of interest is a ratio of bosonic Wilson loops $\cW_N$, which has a weak coupling expansion of the form
\be\label{eq:cW_expansion}
\cW_N=1+\sum_{L=1}^\infty a^L \cW_N^{(L)}\,,
\ee
and is related to $R_N$ by
\be\label{eq:RNtoWN}
R_N=\log \cW_N-\frac{\gamma_K}{4}\cW_N^{(1)} \,,
\ee
where $\gamma_K$ has already been defined in \eqref{eq:gK}. The last equation encodes the fact that although $\cW_N$ receives tree-level and one-loop contributions, $R_N$ only starts at two loops.

Within the Wilson loop OPE approach, the $N$-gon Wilson loop dual to the MHV amplitude is tessellated into $N-5$ consecutive squares, where the two segments of each square that are part of the original Wilson loop can be thought of as a quark-antiquark pair sourcing a color-electric flux tube. We can then decompose the Wilson loop into excitations of this flux tube $\psi_i$, with energy $E_i$, momentum $p_i$ and helicity $m$, corresponding to the three isometries of the square,
\be\label{eq:WNOPE}
\cW_N=\sum_{\psi_1,\ldots, \psi_{N-5}}e^{\sum_{j}^{N-5}(-\tau_j E_j+i p_j \sigma_j+i m_j\phi_j)}\cP(0|\psi_1)\cP(\psi_1|\psi_2)\ldots\cP(\psi_{N-6}|\psi_{N-5})\cP(\psi_{N-5}|0)\,.
\ee
In the above formula, the exponential factors describe the propagation of the excitations, whereas the \emph{pentagon transitions}  $\cP(\psi_i|\psi_{i+1})$ describe the transition amplitude when an excitation of the $i$-th square crosses to the neighboring square $i+1$, and the name derives from the fact that the union of two squares forms a pentagon.

The $3N-15$ algebraically independent variables $\tau_i,\sigma_i$ and $\phi_i$ is a {natural, from the point of view of the OPE, parametrization of the conformally invariant kinematics \cite{Basso:2013vsa}. When expressing the cross-ratios in terms of these variables, it can be shown} that the $N-5$ cross ratios of \eqref{eq:OPEcrossratios} take the form
\be
U_i=\frac{1}{1+e^{2\tau_i}}\,,
\ee
This formula can be obtained from the following equivalent parametrization of the exponential
factors in terms of so-called 4-brackets $\langle ijkl\rangle$ of momentum twistors\footnote{See in
    particular appendix A of \cite{Basso:2013aha} for more details, which also contains definitions of momentum
    twistors. Using the momentum twistor parametrizations of the latter reference, it is possible to relate the OPE variables $\tau_i,\sigma_i$ and $\phi_i$ to any other set of independent variables, for example those in eq.~\eqref{eq:UcrossratiosMRK}.}
\begin{eqnarray}
e^{2\tau_{2j+1}}&\equiv&{\< {-j-1}, {j+1}, {j+2}, {j+3}\>\< {-j-2}, {-j-1}, {-j}, {j+2}\>\over\< {-j-2}, {-j-1}, {j+2}, {j+3}\>\< {-j-1}, {-j}, j+1, {j+2}\>}\, ,\quad j=0,\ldots \lfloor \frac{N-6}{2}\rfloor\,,\nn\\
e^{2\tau_{2j}}&\equiv&{\< {-j}, j+1,  {j+2}, {j+3}\>\< {-j-1}, {-j}, {-j+1}, {j+2}\>\over\< {-j-1}, {-j}, {j+2}, {j+3}\>\< {-j}, {-j+1}, j+1, {j+2}\>}\, ,\quad j=1,\ldots \lfloor \frac{N-5}{2}\rfloor\,,
\end{eqnarray}
For our purposes, all the information we will need is the property
\be
x_{ij}^2\propto \langle i-1, i, j-1, j\rangle\,,
\ee
where the proportionality factors cancel when considering the conformal cross ratios \eqref{eq:Ucrossratios}, together with the following identity between six momentum twistors,
\be
\<cdef\>\<abef\>-\<bdef\>\<acef\>=-\<adef\>\<bcef\>\,.
\ee
The latter is a consequence of the Schouten identity\,,
\be
\<ab\>\<cd\>-\<ac\>\<bd\>=-\<bc\>\<ad\>\,,
\ee
since six points in $\mathbb{CP}^3$ (the twistors) are equivalent to six points in $\mathbb{CP}^1$ (the spinors), as can be seen by replacing the spinor bracket of two points with a 4-bracket of its complement, $\<ab\>\to \<cdef\>$ etc. With the help of the last two formulas, we can show that indeed
\be\label{eq:OPEcroddeven}
\begin{aligned}
\frac{1}{1+e^{2\tau_{2j+1}}}&=\frac{\< {-j-2}, {-j-1}, {j+2}, {j+3}\>\< {-j-1}, {-j}, j+1, {j+2}\>}{\< {-j-2}, {-j-1}, {j+1}, {j+2}\>\< {-j-1}, {-j}, j+2, {j+3}\>}=U_{-j-1,j+2}\,,\\
\frac{1}{1+e^{2\tau_{2j}}}&=\frac{\< {-j-1}, {-j}, {j+2}, {j+3}\>\< {-j}, {-j+1}, j+1, {j+2}\>}{\< {-j-1}, {-j}, {j+1}, {j+2}\>\< {-j}, {-j+1}, j+2, {j+3}\>}=U_{-j,j+2}\,,\\
\end{aligned}
\ee
and \eqref{eq:OPEcrossratios} neatly combines the separate odd and even cases of the last equation.

At weak coupling, the tree-level term in \eqref{eq:cW_expansion} comes from the vacuum state, whereas excitations on top of it have energies
\be
E_i=M_i+\gamma_i a+\cO(a^2)\,,
\ee
where $M_i$ is the excitation number. {The value of the one-loop correction to the energy $\gamma_i$  depends on the type of the excitation, but will not be relevant for our discussion.}
Thus expanding \eqref{eq:WNOPE} at weak coupling, and given that
\be
\tau_i=-\frac{1}{2}\log{U_i}+\frac{1}{2}\log(1-U_i)\,,
\ee
we see that at $L$ loops the terms that maximize the sum of powers of $\log U_i$ will be
\begin{multline}\label{eq:OPEMaxLog}
\sum_{M_i} \prod_{i=1}^{N-5} \frac{(-a \gamma_i \tau_i)^{j_i}}{j_i!} e^{\sum_i(-\tau_i M_i+i p_i \sigma_i+i m_i\phi_i)}a[\cP(0|\psi_1)\ldots\cP(\psi_{N-5}|0)]^{(1)}\,,\quad \sum_i^{N-5}j_i=L-1\,,\\
=a^L\sum_{l_i=0}^{j_i}\prod_{i=1}^{N-5}\log^{l_i} U_i \times (\text{terms analytic as $U_i=0$})\,.\qquad\qquad\qquad\qquad\quad\qquad
\end{multline}
All other terms in the OPE will also have the same general structure as \eqref{eq:OPEMaxLog} but with fewer powers of $\log{U_i}$ (this also includes so-called small fermions which have $\gamma_i=0$ and in fact start at $\cO(a^3)$, see for example \cite{Basso:2014koa}), which proves that $\cW_N^{(L)}$ has the structure of the right-hand side of eq.~\eqref{eq:RNOPE}. Then by virtue of \eqref{eq:cW_expansion} and \eqref{eq:RNtoWN}, the same will be true for $R_N^{(L)}$, which thus completes the proof.

A very similar statement also holds true beyond the MHV case, where it is more convenient to consider the entire superamplitude, rather than its gluonic component alone. This is then dual to a super-Wilson loop, whose OPE is equal to the same expansion for MHV,  times non-MHV form factors \cite{Basso:2013aha,Basso:2014hfa,Basso:2015rta,Basso:2015uxa}. These form factors may increase or decrease the order at which OPE excitations begin to contribute, and indeed, some components of the superamplitude, or more precisely ratio function, are nontrivial at tree level. By repeating the arguments presented above, if a component of the dual super-Wilson loop receives its first nontrivial OPE contribution at $k$ loops, then the total degree of its logarithmic divergence at $L$ loops should be $L-k$.

\subsection{The function-level 7-particle 2-loop MHV amplitude in MRK}\label{sec:R72MRK}

In this subsection, we will promote the known 2-loop symbol of the heptagon remainder function in the multi-Regge limit of ref.~\cite{Bargheer:2015djt}, as defined in~\eqref{eq:RN} for $N=7$, and in the region corresponding to the analytic continuation~\eqref{eq:Ucontinuation}, to a function $R_7^{(2)}$.

Let us start by reviewing the relevant information from the aforementioned paper, where it was shown that
\begin{equation}\label{eq:R72}
\frac{R_7^{(2)}(z_1,z_2)}{2\pi i}=\sum_{i=1}^2 \left(2f(\rho_{i})\log\tau_i  +
\tilde f(\rho_i)\right)+g(\rho_1,\rho_2)\,,
\end{equation}
where
\be
\rho_1=-\frac{z_1 z_2}{1-z_2}\,,\quad \rho_2=(1-z_1)z_2\,,
\ee
are the coordinates defined in \eqref{eq:def_rho}-\eqref{ztorho} specialized to $N=7$, and the functions $f, \tilde f$ are the LLA and NLLA parts of the hexagon remainder function,
\begin{equation}\label{ffunctiondef}
\frac{R_6^{(2)}(z_1)}{2\pi i}= 2 f(z_1)\log\tau_1+\tilde f(z_1)\,,
\end{equation}
this time in the region corresponding to the analytic continuation \eqref{eq:Ucontinuation}, but for $N=6$ (two-particle cut). Explicitly we have~\cite{Lipatov:2010ad}, in the coupling normalization of~\eqref{eq:acoupling},
\begin{align}\label{eq:fftilde}
4f(z)&=4f(1/z)=\frac{1}{2} \log |1-z|^2 \log \frac{|1-z|^2}{|z|^2}={-\frac{1}{2}\cG_{0}(z)\cG_{1}(z)+\frac{1}{2}\cG^2_{1}(z)}\,,\nn\\
4\tilde f(z)&=4\tilde f(1/z)=-4 \text{Li}_3(z)-4 \text{Li}_3(\bar z)+2 \log |z|^2 (\text{Li}_2(z)+\text{Li}_2(\bar z))\\
&\mspace{21mu}+\frac{1}{3} \log ^2|1-z|^2 \log \frac{|z|^6}{|1-z|^4}-\frac{1}{2} \log |1-z|^2 \log \frac{|1-z|^2}{|z|^2} \log \frac{|z|^2}{|1-z|^4}\nn\\
&={-2 \cG_0(z) \cG_{0,1}(z)+4 \cG_{0,0,1}(z)+\frac{1}{3} \cG_1(z){}^3-\frac{1}{2} \cG_0(z) \cG_1(z){}^2+\frac{1}{2} \cG_0(z){}^2 \cG_1(z)}\,.\nn
\end{align}
Finally, the symbol, as well as a 25-parameter functional representative for the genuinely heptagonal NLLA function $g$, were found in \cite{Bargheer:2015djt}. Here we fix all remaining ambiguity, and show that
\be\label{eq:gfunction}
\begin{aligned}
4g(\rho_1,\rho_2)=&2 \cG_{0,1,1/\rho _1}\left(1/\rho _2\right)-2 \cG_{1,1,1/\rho _1}\left(1/\rho _2\right)-\cG_{1/\rho _1}\left(1/\rho _2\right) \cG_{0,1}\left(1/\rho _2\right)\\
&-\cG_{0}\left(\rho _1\right) \cG_{0,1/\rho _1}\left(1/\rho _2\right)+\cG_{1}\left(\rho _1\right) \cG_{0,1/\rho _1}\left(1/\rho _2\right)\\
&+\cG_{0}\left(\rho _1\right) \cG_{1,1/\rho _1}\left(1/\rho _2\right)+\cG_{0}\left(\rho _2\right) \cG_{1,1/\rho _1}\left(1/\rho _2\right)\\
&-\cG_{1}\left(\rho _1\right) \cG_{1,1/\rho _1}\left(1/\rho _2\right)-\cG_{1/\rho _1}\left(1/\rho _2\right) \cG_{0,1}\left(\rho _1\right)\\
&-\cG_{1}\left(\rho _1\right) \cG_{0,1}\left(1/\rho _2\right)+\cG_{1,1/\rho _1}\left(1/\rho _2\right) \cG_{1}\left(1/\rho _2\right)\\
&+\frac{1}{2} \cG_{0}\left(\rho _1\right) \cG_{1}\left(\rho _1\right) \cG_{1/\rho _1}\left(1/\rho _2\right)-\frac{1}{2} \cG_{0}\left(\rho _1\right) \cG_{1}\left(\rho _1\right) \cG_{1}\left(1/\rho _2\right)\\
&-\frac{1}{2} \cG_{0}\left(\rho _2\right) \cG_{1}\left(\rho _1\right) \cG_{1}\left(1/\rho _2\right)-\frac{1}{2} \cG_{0}\left(\rho _2\right) \cG_{1/\rho _1}\left(1/\rho _2\right) \cG_{1}\left(1/\rho _2\right)\\
&+\cG_{0,1}\left(\rho _1\right) \cG_{1}\left(1/\rho _2\right)\,,
\end{aligned}
\ee
with the definitions \eqref{eq:ExampleSVMPL}, etc. for the single-valued $\cG$-functions appearing here, thus fully specifying $R_7^{(2)}$ at function level.

For the remainder of this subsection, we will describe how we have obtained eq.~\eqref{eq:gfunction}. Specializing the discussion of subsection \ref{sec:SVMPL} to the seven-particle amplitude, we infer that the relevant class of functions for describing it are single-valued $A_2$ polylogarithms. The construction of a basis of such functions at any weight is given in appendix \ref{sec:A2SVPolylogs}. Then, if we know the symbol of any function in this space (in this case, the symbol of $g$), we can find a representative function either by matching it against the symbol of an ansatz made of the basis functions of the same weight, as was done in \cite{Bargheer:2015djt}, or, even better, by directly integrating the symbol along a given contour, as also reviewed in appendix \ref{sec:A2SVPolylogs}.

The actual function may differ from the representative function by beyond-the-symbol terms, namely transcendental constants multiplying lower-weight functions of the same type. Assuming that the only transcendental constants appearing here are multiple zeta values (MZV), we thus form an ansatz for the actual function by augmenting the representative function with all products of MZVs with the bases of lower-weight functions, multiplied by yet-to-be-determined coefficients.

In more detail, we may form separate ans\"atze for the imaginary and real parts of $R_7^{(2)}$, which at two loops will have weight three and two, respectively. However, by virtue of the property~\eqref{eq:RNOPE}, $R_7^{(2)}$ has vanishing real part {in the region where all produced particles have a negative energy}. The functions $f$ and $\tilde f$ are determined from the six-point amplitude and are known to be real. Hence,
the function $g$ must also be real.
We thus only need form an ansatz for the {real} part of $g$, which after taking into account parity and projectile symmetry, also discussed in appendix \ref{sec:A2SVPolylogs}, will contain just four undetermined coefficients: a constant $\zeta_3$ term, plus $\zeta_2$ times the three parity and flip symmetric weight-one functions
\be\label{eq:weight1SVMPL}
\cG_{0}(\rho_1)+\cG_{0}(1/\rho_2)\,,\quad \cG_{1}(\rho_1)+\cG_{1}(1/\rho_2)\,,\quad \cG_{\rho_2}(\rho_1)\,.\\
\ee

The final piece of information from subsection \ref{sec:SymmetriesLimits} we will rely on in order to arrive at a unique answer will be the expected behavior of the amplitude in soft limits. {From~\eqref{eq:fftilde} we can easily see that
\be\label{eq:softf}
f(0)=\tilde f(0)=0\,,
\ee
in agreement with \eqref{eq:wsoftlimits}-\eqref{eq:vsoftlimits} and the fact that $R_5=0$.}
Similarly, the three soft limits of the heptagon building block $g$ require
\begin{align}
g(0,\rho_2)&=0\,,\label{eq:soft1}\\
g(\rho_1,\infty)&=0\,,\label{eq:soft2}\\
g(\rho_1,\rho_1)&=-\tilde f(\rho_1)\,.\label{eq:soft3}
\end{align}

As explained in the last paragraph of appendix \ref{sec:A2SVPolylogs}, taking the limits on the left-hand side is straightforward, after expressing it in the Lyndon basis \eqref{2dHPLbasis}. The first limit  sets the coefficients of the constant $\zeta_3$ terms and the first two logarithms in \eqref{eq:weight1SVMPL} to zero. The second limit is related to the first one by target-projectile symmetry, which leaves $g$ invariant, $g(\rho_1,\rho_2)=g(1/\rho_2,1/\rho_1)$. Therefore it will not provide any new information, since our ansatz already respects this symmetry. Finally, the third limit $\rho_2\to \rho_1$ also sets the coefficient of the third logarithm in \eqref{eq:weight1SVMPL} to zero, since there it is the only term that becomes divergent.

We thus arrive at the unique answer \eqref{eq:gfunction} for the function $g$, or equivalently the heptagon remainder function in MRK. Both of them can be found in the attached ancillary file \texttt{R2MRK.m}.

\subsection{All function-level 2-loop MHV amplitudes in MRK}\label{sec:RN2MRK}

Quite interestingly, from the result of the previous section, we can also obtain \emph{all} 2-loop MHV amplitudes, in any region in which the adjacent particles $k+3,k+4,\ldots,l+3$ have their energy signs flipped.  In particular, we will show that in the region in question, we have\begin{equation} \label{eq:Rnkl}
\frac{R_{N[k+3,l+3]}^{{(2)}}}{2\pi i}
=
\sum_{i=k}^{l-1}\left(2f(v_{i})\log\tau_{i} +
\tilde f(v_{i})\right) + \sum_{i=k}^{l-2}g(v_{i},v_{i+1})\,,
\end{equation}
where the hexagon $f, \tilde f$ and heptagon $g$ building blocks have already been provided in \eqref{eq:fftilde}-\eqref{eq:gfunction}, and the variables $v_i$ are slight generalizations of the $\rho_i$ variables defined in \eqref{eq:def_rho}-\eqref{ztorho}, corresponding to simplicial coordinates on the Riemann sphere with $l-k+3$ marked points.\footnote{That is, for the long cut with $k=1$, $l=N-4$ the two sets of variables coincide, $v_i=\rho_i$.} They are related to the usual transverse cross ratios $z_i$ by
\begin{equation}\label{wtov}
z_j=\frac{(v_{j-1}-v_j)(1-v_{j+1})}{(v_{j+1}-v_j)(1-v_{j-1})}\,,
\quad
j\in\brc{k,\dots,l-1}\,,
\end{equation}
with the boundary conditions $v_{k-1}=0$, $v_{l}=\infty$.

The above result has already been established at symbol level in \cite{Bargheer:2015djt},\footnote{Note however that we have modified the conventions slightly: $z_i=-w_{i+3}$ and  $v_i^\text{here}=-v_{i+3}^\text{there}$.} so more precisely here we will prove that if it holds at symbol level, then must necessarily also hold at function level.

It can be seen that by virtue of \eqref{eq:softf}-\eqref{eq:soft3}, the right-hand side of
\eqref{eq:Rnkl} has correct soft limit behavior, namely it reduces to the same function with one leg
less. Therefore if this factorization is to break down beyond symbol level, it can only be through terms that vanish in the soft limit.

From the results of subsection \eqref{OPEDiscontinuity} we know that the 2-loop remainder function has a vanishing real part for any $N$, so the only factorization-violating beyond-the-symbol terms are transcendental constants or weight-1 functions. Given that the latter are single-valued, they can never turn into transcendental constants in the soft limit, and therefore one cannot add transcendental constants to the right-hand side of \eqref{eq:Rnkl} without violating soft limits.

With the weight-1 SVMPLs remaining as the only allowed beyond-the symbol terms not captured in \eqref{eq:Rnkl}, we will now show that there exists no linear combination thereof that vanishes in all soft limits, and therefore they too should be absent. Forgetting target-projectile symmetry momentarily, the weight-1 SVMPLs that can appear in $N$-point scattering in MRK will be
\be\label{eq:w1SVMPL}
\log |v_i|\,,\quad \log |1-v_i|\,,\quad \log |v_i-v_j|\,,\quad i<j=k,\ldots,l-1\,.
\ee
This is a consequence of the fact that in the regions we are considering, the multi-Regge limit is described by a configuration of $l-k+3$ points ${\bf x}_i$ (out of a total of $N-2$) in $\mathbb{CP}^1$, which can only have singularities when two of the points coincide, i.e. of the form $\log ({\bf x}_i-{\bf x}_j)$, plus complex conjugates. Equation~\eqref{eq:w1SVMPL} then follows from single-valuedness and the fact that the $v_i$ coordinates are a particular set of simplicial coordinates parametrizing this space, in absolute analogy to \eqref{eq:def_rho} for the $k=1$, $l=N-4$ case.

Soft limits prohibit logarithms of adjacent $v_i$, $\log |v_{i-1}-v_i|$, to appear, since for each $i$ they will be the only ones that diverge in the $v_{i-1}=v_i$ limit, with all remaining terms being finite. Then $\log |v_{i-2}-v_i|$ are also prohibited, since in the same limit they will be the only ones that reduce to $\log |v_{i-1}-v_i|$, an independent function whose coefficient should vanish. By extending the argument to differences of $v_i$ with larger and larger separation, we thus sequentially exclude all $\log |v_{i}-v_j|$ from appearing as extra beyond-the-symbol terms not captured by \eqref{eq:Rnkl}.

We are thus left with the logarithms of the first two types appearing in eq.~\eqref{eq:w1SVMPL}. All of them but $\log|1-v_1|$ can be similarly eliminated by the $v_1=0$ limit, namely they will either diverge, or remain independent functions. And going back to any soft limit not involving $v_1$ will also force $\log|1-v_1|$ to be absent, since it will remain nonzero.

Concluding our analysis, we have shown that there exist no beyond-the-symbol terms respecting soft limits that can be added to the right hand side of eq.~\eqref{eq:Rnkl} for $N\ge 7$. Therefore the interesting NLLA factorization structure observed at symbol level also holds at function level, and the latter equation accurately describes the MRK of 2-loop MHV amplitudes with any number of points $N$, in terms of the functions $f,\tilde f$ and $g$.

\section{Extracting the NLO central emission block}\label{sec:CNLO}
Let us now combine the knowledge of seven-gluon amplitudes in MRK we have gathered so far, namely the function-level 2-loop MHV case of section \ref{sec:SymbolsToFunctions}, and the dispersion integral governing any helicity configuration to all loops \eqref{eq:Rhhhstart}-\eqref{eq:f_hhh}. By matching the perturbative two-loop result to the weak coupling expansion of the dispersion integral we determine a main result of this paper: the (rescaled) central emission block \eqref{eq:Ctilde} to next-to-leading order. This is a result which cannot be obtained from the well-studied six-gluon amplitudes.

We start by presenting the result, and describe the details of our calculation in the following subsections. If we denote the perturbative expansion of the rescaled central emission block as
\be\label{eq:CtildeExpansion}
\tilde C^+(\nu_1,n_1,\nu_2,n_2)=\tilde C^{(0)}(\nu_1,n_1,\nu_2,n_2)+a \tilde C^{(1)}(\nu_1,n_1,\nu_2,n_2) +\cO(a^2)\,,
\ee
we find the  result for the $\mathcal{O}(a)$ correction to the central emission block:
\begin{align}
\frac{\tilde C^{(1)}(\nu_1,n_1,\nu_2,n_2) }{\tilde C^{(0)}(\nu_1,n_1,\nu_2,n_2) } =&\frac{1}{2}\left[ DE_1-DE_2   + E_1 E_2  +\tfrac{1}{4} (N_1+ N_2)^2 + V_1 V_2 \right.\label{eq:CtildeNLO}\\
&\left.+ (V_1-V_2)\bigl(M-E_1-E_2)+2\zeta_2+i\pi (V_2-V_1-E_1-E_2) \right]\,.\notag
\end{align}
Here we normalized to the known leading order result, translated to our
conventions~\cite{Bartels:2011ge},
\be\label{eq:CtildeLO}
\tilde C^{(0)}(\nu_1,n_1,\nu_2,n_2) = \frac{\Gamma \bigl(1-i \nu_1 - \tfrac{n_1}{2}\bigr)\Gamma \bigl(1+i \nu_2 + \tfrac{n_2}{2}\bigr) \Gamma \bigl(i \nu_1 - i \nu_2 - \tfrac{n_1}{2} + \tfrac{n_2}{2}\bigr)}{\Gamma \bigl(i \nu_1 - \tfrac{n_1}{2}\bigr) \Gamma \bigl(- i \nu_2 + \tfrac{n_2}{2}\bigr) \Gamma \bigl(1 - i \nu_1 + i \nu_2 - \tfrac{n_1}{2} + \tfrac{n_2}{2}\bigr)}\,.
\ee
In the above, we have expressed the answer in terms of the hexagon BFKL building blocks~\cite{Dixon:2012yy}
\be\begin{aligned}\label{eq:buildingblocks}
 E(\nu,n)&=-\frac12\frac{|n|}{\nu^2+\frac{n^2}{4}}+\psi\left(1+i\nu+\frac{|n|}{2}\right) +\psi\left(1-i\nu+\frac{|n|}{2}\right)-2\psi(1)\,,
\\
V(\nu,n)&\equiv \frac{i\nu}{\nu^2+\frac{n^2}{4}}\,,\qquad N(\nu,n) = \frac{n}{\nu^2+\frac{n^2}{4}}\,,
\qquad D_\nu=-i\partial/\partial \nu\,,
\end{aligned}\ee
 with the shorthand $E_1 = E(\nu_1,n_1)$ etc, as well a new quantity involving mixed polygamma functions,
\be\label{eq:BuidingBlockM}
M(\nu_1,n_1,\nu_2,n_2)=\psi(i(\nu_1-\nu_2)-\tfrac{n_1-n_2}{2}) + \psi(1-i(\nu_1-\nu_2)-\tfrac{n_1-n_2}{2})-2\psi(1)\,.
\ee
Note however that when one changes the integration variables from angular momenta $\nu_i$ to integrability-wise more natural rapidities $u_i$ \cite{Basso:2014pla} (taking into account that our $\nu_i$ differ with the ones used in the latter reference by a factor of $1/2$)
\be\label{NuToU}
\nu_i = u_i + a \frac{i V_i}{2}	 + O(a^2)\,,
\ee
then any dependence on mixed polygamma functions drops out. In other words they only appear when we expand the arguments of the gamma functions in the inverse transformation from the $u_i$ to the $\nu_i$.

We may readily check that our expression \eqref{eq:CtildeExpansion}-\eqref{eq:BuidingBlockM}  indeed obeys the $\mathcal{O}(a)$ expansion of the exact bootstrap conditions \eqref{eq:R7hhh_bootstrapcond1}-\eqref{eq:R7hhh_bootstrapcond4}. For completeness, let us also mention the weak coupling expansion of the BFKL eigenvalue \cite{Bartels:2008sc,Fadin:2011we}, hexagon measure \cite{Lipatov:2010qg,Dixon:2012yy,Dixon:2014voa} and NMHV helicity flip kernel \cite{Lipatov:2012gk,Dixon:2014iba,Basso:2014pla}  (see also the last paper for all-loop expressions of these quantities),
\begin{align}
-\omega(\nu,n)  =&\, \,  a E - \frac{a^2}{4}\left(D^2E-2VDE+4\zeta_2 E+12\zeta_3\right) + \mathcal{O}(a^{3})\,,\label{eq:omegaofnu}\\
\tilde \Phi(\nu,n)=&\frac{\Phi_{\text{reg}}(\nu,n)}{\nu^2+\frac{n^2}{4}}=\frac{1}{\nu^2+\frac{n^2}{4}}\left[1-\frac{a}{2}\left(E^2+\frac{3}{4}N^2+\frac{\pi^2}{3}\right)+\cO(a^2)\right]\,,\label{eq:Phitildeofnu}\\
H(\nu,n)=&\frac{\nu-\frac{in}{2}}{\nu+\frac{in}{2}}\left[1-\frac{a}{2}NV+\cO(a^2)\right]\,.\label{eq:Hofnu}
\end{align}
Plugging these formulas\ back into \eqref{eq:Rhhhstart}-\eqref{eq:f_hhh} or \eqref{eq:R7hhh_w1_small_w2_large}, we may obtain predictions for the heptagon to NLLA at higher loop orders. We detail how to evaluate the relevant integrals to obtain explicit expressions in momentum space in section~\ref{sec:HigherLoopPredictions}. Finally, we may invert~\eqref{eq:Ctilde}-\eqref{eq:Ih} in order to obtain equivalent perturbative expansions for the $\chi^{\pm}$ and $C^+$ building blocks of the BFKL approach,{\footnote{{Note that here we have redefined $\chi^{\pm}$ and $C^+$, compared to e.g. \cite{DelDuca:2016lad}, as follows: $\chi^{\pm}_{\text{here}}=i [\chi^{\pm}_{\text{there}}+\cO(a)]$, and $C^+_{\text{here}}=-[C^+_{\text{there}}+\cO(a)]$.}}
\begin{align}
\chi^+(\nu,n)=\sqrt{\frac{\tilde \Phi(\nu,n)}{H(\nu,n)}}&=\frac{1}{\nu-\frac{i n}{2}}\left[1-\frac{a}{4}\left(E^2+\frac{3}{4}N^2-NV+\frac{\pi^2}{3}\right)+\cO(a^2)\right]\,,\nonumber\\
\chi^-(\nu,n)=\chi^+(\nu,n) H(\nu,n)&=\frac{1}{\nu+\frac{i n}{2}}\left[1-\frac{a}{4}\left(E^2+\frac{3}{4}N^2+NV+\frac{\pi^2}{3}\right)+\cO(a^2)\right]\,,\label{eq:chiNLO}
\end{align}
and
\begin{multline}
C^+(\nu_1,n_1,\nu_2,n_2) =-\frac{\Gamma \bigl(1-i \nu_1 - \tfrac{n_1}{2}\bigr)\Gamma \bigl(i \nu_2 + \tfrac{n_2}{2}\bigr) \Gamma \bigl(i \nu_1 - i \nu_2 - \tfrac{n_1}{2} + \tfrac{n_2}{2}\bigr)}{\Gamma \bigl(1+i \nu_1 - \tfrac{n_1}{2}\bigr) \Gamma \bigl(- i \nu_2 + \tfrac{n_2}{2}\bigr) \Gamma \bigl(1- i \nu_1 + i \nu_2 - \tfrac{n_1}{2} + \tfrac{n_2}{2}\bigr)}\times\\
\times\left[1+a\left(\frac{\tilde C^{(1)}}{\tilde C^{(0)}}-\tfrac{1}{4}(E_1^2+E_2^2+N_1 V_1-N_2 V_2)-\tfrac{3}{16}(N_1^2+N_2^2)-\zeta_2\right)+\cO(a^2)\right]\,,
\end{multline}
where in the first line of \eqref{eq:chiNLO} we picked the branch $\sqrt{\left(\nu-\frac{i n}{2}\right)^2}=\nu-\frac{i n}{2}$.}

\subsection{Building the Fourier-Mellin representation}
\label{sec:leading singularities}
Here we would like to describe a procedure that can take us from the amplitude in multi-Regge kinematics to its corresponding Fourier-Mellin (FM) representation. As we have recalled in subsections \ref{sec:kinematics} and \ref{sec:SVMPL},  in multi-Regge kinematics the amplitude exhibits divergent logarithms which take the form of powers of the $\log\tau_i$, whose coefficients are SVMPLs in the variables $z_i$. When investigating the heptagon amplitudes we find it useful to define
\be
\hat{z}_2 = 1/z_2
\ee
so that the target-projectile symmetry becomes simply $z_1 \leftrightarrow \hat{z}_2$.

The first step in our analysis is to focus solely on the holomorphic part of the heptagon amplitude in MRK, as defined in \eqref{eq:HolomorphicPart}, since we can reconstruct the full kinematic dependence of the latter with the help of the single-valued map, cf.~\eqref{eq:SingleValuedMap}. One is then left with a five-letter $A_2$ or $\mathcal{M}_{0,5}$ alphabet for the holomorphic part of the form
\be
\label{heptholpartalphabet}
\{z_1 ,\hat{z}_2, 1-z_1 , 1-\hat{z}_2 , 1- z_1 - \hat{z}_2\}\,.
\ee
In addition to restricting to the holomorphic part we also focus on the terms in the Taylor expansion of amplitude with strictly positive powers of $z_1$ and $\hat{z}_2$. In other words we decompose terms in the perturbative expansion of the amplitude in a similar manner to that shown in eq. (\ref{eq:R7hhh_w1_small_w2_large}) and keep only the final term corresponding to $(2 \pi i) \tilde{f}_{h_1 h_2 h_3}$. Concretely, this amounts to taking the symbol expression for $A =e^{R_7 + i\delta_7}\Big|_{L \text{ loops}}$ and forming the combination
\be
\mathcal{A}^{(L)}(z_1,\hat{z}_2) = A^h(z_1,\hat{z}_2) - A^{h}(z_1,0) - A^{h}(0,\hat{z}_2) + A^{h}(0,0)\,,
\ee
in order to remove contributions that reduce to lower-point objects.
We remind the reader that the superscript $h$ refers to taking the holomorphic part as in eq. (\ref{eq:HolomorphicPart}).
The quantity $\mathcal{A}^{(L)}$ contains all the information necessary to construct the Fourier-Mellin representation we require.


The holomorphic part $\mathcal{A}^{(L)}(z_1,\hat{z}_2)$ will have logarithmic branch cuts around $z_1=0$ and $\hat{z}_2=0$, in addition to exhibiting the large logarithms $\log\tau_i$ associated with taking the multi-Regge limit.
We may render such branch cuts explicit by employing the shuffle relations to obtain each function as a polynomial in $\log z_1$ and $\log \hat{z}_2$ with coefficients which are analytic around $z_1 = 0 $ and $\hat{z}_2 = 0$ respectively. Thus we obtain an expression for the holomorphic part of the amplitude in MRK of the form
\be
\mathcal{A}^{(L)}(z_1,\hat{z}_2) = \sum_{p,q,r,s} \log^p \tau_1 \log^q \tau_2 \log^r z_1 \log^s \hat{z}_2 f_{pqrs}(z_1,\hat{z}_2)\,,
\ee
where $f_{pqrs}(z_1,\hat{z}_2)$ are linear combinations of polylogarithms which are analytic at $z_1=\hat{z}_2=0$.


For each analytic function $f_{p,q,r,s}(z_1,\hat{z}_2)$ we now Taylor expand around the origin $z_1= \hat{z}_2=0$.
We may do this simply by employing the following general formula for the Taylor expansion about $z=0$ of a $G$-function from e.g. \cite{Brown:2009qja}
\begin{align} \label{texp}
G_{0^{n_r}a_{i_r}\ldots0^{n_{1}}a_{i_{1}}}(z) =\sum \frac{(-1)^r}{m_1^{n_1+1} \ldots m_r^{n_r+1}}\Bigl[\frac{z}{a_{i_1}}\Bigr]^{m_1}\!\Bigl[\frac{z}{a_{i_2}}\Bigr]^{m_2-m_1}\ldots \Bigl[\frac{z}{a_{i_r}}\Bigr]^{m_r-m_{r-1}}\,,
\end{align}
where the nested summation is performed over the region $1\leq m_1 < \ldots <m_r$ and the $a_{i_r}\neq0$.
Here $0^{n}$ denotes a length $n$ sequence of $0$.
While the formula (\ref{texp}) provides the explicit Taylor expansions, we empirically find that $f_{pqrs}$ is always decomposable into sums of the following much simpler type involving only simple harmonic sums of depth one,
\begin{align} \label{gform}
    H_{k_1,k_2,\{r_i\},\{s_i\},\{t_i\}}(z_1,z_2)=\sum\limits_{n_1,n_2>0} \Biggl[\frac{z_1^{n_1} \hat{z}_2^{n_2}}{n_1^{k_1}n_2^{k_2}}\frac{\Gamma(n_1+n_2)}{\Gamma(1+n_1)\Gamma(1+n_2)} \notag \\
 \biggl(\prod_i Z_{r_i}(n_1-1)\biggr) \biggl(\prod_i Z_{s_i}(n_2-1)\biggr) \biggl( \prod_i Z_{t_i}(n_1+n_2-1)\biggr)\Biggr]\,.
\end{align}
Here we have adopted the following notation for harmonic sums of depth one,
\be \label{ancon}
Z_{r}(n)=\sum_{i=1}^{n}\frac{1}{i^r}=\sum_{i=1}^{\infty}\frac{1}{i^r}-\frac{1}{(i+n)^r}=\frac{(-1)^{r-1}}{(r-1)!}\left(\psi^{(r-1)}(n+1)-\psi^{(r-1)}(1)\right)\,.
\ee
Note that the polygamma function $\psi^{(r)}(z)$ serves as an analytic continuation of the harmonic sums of depth one. The total weight of the representation (\ref{gform}) is \begin{equation}k_1+k_2+\sum_i r_i + \sum_i s_i +\sum_i t_i\,.\end{equation}

We stress that the fact that the functions $f_{pqrs}$ are always expressible as linear combinations of terms of the form (\ref{gform}) is not at all trivial. Certainly not all polylogarithms of the form (\ref{Gform}) are expressible in the form (\ref{gform}), even restricting to those which are analytic about $z_1=\hat{z}_2=0$.
For low weights it is possible to use various binomial and harmonic sum identities to go from the general Taylor expansion to the reduced form (\ref{gform}). Unfortunately this requires cancellations among sums that do not have a simple closed form, and it becomes increasingly intractable at higher weights. However it is simple to explicitly evaluate the Taylor expansions (\ref{texp}) up to a finite order and thus generate enough terms to reduce the $f_{pqrs}$ to the form (\ref{gform}) by means of an ansatz and linear algebra. This requires building the vector space spanned by the sums of the form (\ref{gform}) for each weight required.

The reason why we are interested in solving the linear problem to arrive at the form (\ref{gform})
is two-fold. Firstly, it gives us a double infinite sum that is reminiscent of the BFKL LLA form.
Note that this structure is not automatic just from using the form (\ref{texp}). Secondly, the
reduced sum~(\ref{gform}) is particularly well suited to expressing its single-valued completion through
a Fourier-Mellin integral.

To find a Fourier-Mellin representation for the single-valued completion of each of the $f_{pqrs}$ we begin by specifying the following prescription to be applied to the summations in  terms of the form (\ref{gform}),
\begin{align}
\label{presint}
\sum_{n_1,n_2>0} \frac{z_1^{n_1} \hat{z}_2^{n_2}}{n_1^{k_1} n_2^{k_2}} &\to\sum_{-\infty <n_1,n_2<\infty} \int_{-\infty}^{\infty}\frac{d\nu_1}{2\pi}\int_{-\infty}^{\infty }\frac{d\nu_2}{2\pi} \frac{z_1^{i\nu_1+\frac{n_1}{2}} \bar{z}_1^{i\nu_1-\frac{n_1}{2}} \hat{z}_2^{i\nu_2+\frac{n_2}{2}}  \hat{\bar{z}}_2^{i\nu_2-\frac{n_2}{2}}}{(i\nu_1+\frac{n_1}{2})^{k_1} (i\nu_2+\frac{n_2}{2})^{k_2}}\,.
\end{align}
Here the contours of integration should be taken to be slightly below the real axes in $\nu_1$ and $\nu_2$. Next we specify how to continue the harmonic sums,
\begin{align}
\label{pres}
&Z_{r}(n_j-1)\to \notag \\
&\quad \frac{(-1)^{r-1}}{(r-1)!}\Bigl[\psi^{(r-1)}\bigl(\tfrac{n_j}{2}+i\nu_j\bigr)+(-1)^{r-1}\psi^{(r-1)}\bigl(1+\tfrac{n_j}{2}-i\nu_j\bigr)  -2\delta_{r,{\rm odd}}\psi^{(r-1)}(1)\Bigr]\,.
\end{align}
Finally we provide a prescription for the binomial coefficients,
\begin{align}
\label{presbinomial}
\frac{\Gamma(n_1+n_2)}{\Gamma(1+n_1)\Gamma(1+n_2)}& \to \frac{\Gamma(\frac{n_1}{2}-i\nu_1)\Gamma(\frac{n_2}{2}-i\nu_2)\Gamma(i\nu_1+i\nu_2+\frac{n_1}{2}+\frac{n_2}{2})}{\Gamma(1+i\nu_1+\frac{n_1}{2})\Gamma(1+i\nu_2+\frac{n_2}{2})\Gamma(1-i\nu_1-i\nu_2+\frac{n_1}{2}+\frac{n_2}{2})} \notag \\
&= \frac{\Gamma(-\frac{n_1}{2}-i\nu_1)\Gamma(-\frac{n_2}{2}-i\nu_2)\Gamma(i\nu_1+i\nu_2-\frac{n_1}{2}-\frac{n_2}{2})}{\Gamma(1+i\nu_1-\frac{n_1}{2})\Gamma(1+i\nu_2-\frac{n_2}{2})\Gamma(1-i\nu_1-i\nu_2-\frac{n_1}{2}-\frac{n_2}{2})} \notag \\
&= \frac{\tilde{C}^{(0)}(\nu_1,n_1,-\nu_2,-n_2)}{\bigl(i\nu_1 + \tfrac{n_1}{2}\bigr)\bigl(i \nu_1- \tfrac{n_1}{2}\bigr)\bigl(i\nu_2+ \tfrac{n_2}{2}\bigr)\bigl(i\nu_2-\tfrac{n_2}{2}\bigr)} \,.
\end{align}
The equality of the first and second lines above holds for integer $n_1$ and $n_2$ due to the following identity obeyed by the Gamma function,
\be\label{eq:GammaRatioIdentity}
\frac{\Gamma \left(x-\frac{k}{2}\right)}{\Gamma \left(1-x-\frac{k}{2}\right)}=(-1)^k\frac{ \Gamma \left(x+\frac{k}{2}\right)}{\Gamma \left(1-x+\frac{k}{2}\right)}\,.
\ee

The combination of gamma functions provides poles at $\nu_j = -i\left(\frac{n_j}{2}+h\right)$ that recover the holomorphic part for $h=0$ by closing the contours in the lower half-planes. It is easy to see that for the holomorphic pole the right hand side of (\ref{pres}) reduces to their initial quantities once we make use of (\ref{ancon}) and similarly for (\ref{presbinomial}).

The prescription (\ref{presint}), (\ref{pres}), (\ref{presbinomial}) is a conjectural method for producing the single-valued completion of a given holomorphic function. Once the prescription is applied one may then evaluate the non-holomorphic residues and verify that they correctly reproduce the corresponding terms in the expansion of the initial single-valued function. This procedure has been applied and verified for the two-loop heptagon amplitude in MRK as well as many other single-valued polylogarithms.

The final step is to promote the power of $\log z_1$ and $\log \hat{z}_2$ to their single-valued versions $\log |z_1|^2$ and $\log |\hat{z}_2|^2$ and then absorb the log terms into the integrand by writing them as derivatives as follows,
\be
\log^{n}|z|^2 \int_{-\infty}^{\infty}\frac{d\nu}{2\pi} z^{i\nu+\frac{n}{2}}\bar{z}^{i\nu-\frac{n}{2}}F(\nu,n)=\int_{-\infty}^{\infty}\frac{d\nu}{2\pi} (-i)^n\frac{\partial^n}{\partial \nu^n}\left(z^{i\nu+\frac{n}{2}}\bar{z}^{i\nu-\frac{n}{2}}\right)F(\nu,n)\,.
\ee
Then we may use integration by parts, ignoring surface terms, to shift the derivatives onto the rest of the integrand $F(\nu,n)$. By construction this operation does not spoil the single valuedness of the integrand. However this can also be seen from the fact that the structure on the RHS of (\ref{pres}) is closed with respect to derivatives. This is immediately evident except for the Gamma functions, for which we have
\begin{align}
&(-i)\frac{\partial}{\partial \nu}\frac{\Gamma(\frac{n}{2}-i\nu)}{\Gamma(1+i\nu+\frac{n}{2})} = -\frac{\Gamma(\frac{n}{2}-i\nu)}{\Gamma(1+i\nu+\frac{n}{2})}\left(\psi(\tfrac{n}{2}-i\nu)+\psi(1+i\nu+\tfrac{n}{2})\right)\notag\\
&= -\frac{\Gamma(\frac{n}{2}-i\nu)}{\Gamma(1+i\nu+\frac{n}{2})}\left(\psi(1+\tfrac{n}{2}-i\nu)+\psi(i\nu+\tfrac{n}{2})+ \frac{1}{\tfrac{n}{2}+i\nu}-\frac{1}{\tfrac{n}{2}-i\nu}\right)\,.
\end{align}
The derivative increases the order of the holomorphic pole and as is consistent with recovering $\log |z|^2$ from the contour integration. Note that manipulating the arguments of the polygamma functions introduces only rational terms in the integrand and thus does not spoil single-valuedness. Furthermore the derivative of the gamma functions with mixed arguments is already of the single-valued form and requires no rational terms. Altogether this makes it easy to express the integrand in terms of the $D,N,V,E,M$ basis.

It remains to note that to return to the variable $z_2$ instead of $\hat{z}_2=1/z_2$ we simply replace $(\nu_2,n_2) \rightarrow (-\nu_2,-n_2)$ in the integrand. In expressions written in the $D,N,V,E,M$ basis this amounts to simply replacing $D_2$, $N_2$ and $V_2$ with $(-D_2)$, $(-N_2)$ and $(-V_2)$.

Applying the above procedure to the finite part of the two-loop heptagon amplitude in MRK yields the correction to the integrand $\tilde \Phi \tilde C^+ \tilde \Phi$, and by dividing with the known expansion \eqref{eq:Phitildeofnu}, we arrive at the expression \eqref{eq:CtildeNLO}-\eqref{eq:CtildeLO}. Similarly applying the procedure to the symbol of the amplitude in MRK, obtained from the results of \cite{Drummond:2014ffa}, yields the NNLO central emission vertex (up to beyond-the-symbol terms). We will analyse the NNLO results in future work. Since the intermediate expressions in these calculations can be slightly cumbersome, in the next section we give a worked example of all the steps we have outlined here on a simpler example of a weight-three polylogarithm.

\subsection{A worked example}
We provide here a demonstration of the entire algorithm on a simple weight-three SVMPL whose holomorphic part admits a representation of the form (\ref{gform}). We begin with a function with only positive powers of $z_1$ and $\hat{z}_2$ in its Taylor expansion,
\begin{align}
\label{exG}
G_{1-z_1,1-z_1,0}(\hat{z}_2) - G_{1,1,0}(\hat{z}_2) &= \log (\hat{z}_2) [G_{1-z_1,1-z_1}(\hat{z}_2) - G_{1,1}(\hat{z}_2)] - G_{1-z_1,0,1-z_1}(\hat{z}_2)  \notag\\
&\quad +G_{1,0,1}(\hat{z}_2) - G_{0,1-z_1,1-z_1}(\hat{z}_2)  +G_{0,1,1}(\hat{z}_2)\,.
\end{align}
On the RHS of (\ref{exG}) we have made the logarithmic branch cut at $\hat{z}_2=0$ explicit by shuffling out the trailing zeros.

By comparing the explicit sum representation for (\ref{exG}) against terms of the form (\ref{gform}) of weight three, we find we can write
\begin{align}
\label{exgform}
&G_{1-{z}_1,1-{z}_1,0}(\hat{z}_2)-G_{1,1,0}(\hat{z}_2)=\notag \\
& \sum_{n_1,n_2>0}  z_1^{n_1} \hat{z}_2^{n_2} \frac{\Gamma(n_1+n_2)}{\Gamma(1+n_1)\Gamma(1+n_2)}
\left[Z_{1}(n_2-1)\Bigl(\log \hat{z}_2-\frac{1}{n_2}\Bigr)-Z_{2}(n_2-1)\right]\,.
\end{align}
Note that the example chosen can be expressed in terms of harmonic polylogarithms and thus the form on the RHS of (\ref{exgform}) is easy to derive. We emphasise again that it is not always simple to derive such a form and in general we have to resort to comparing against an ansatz of terms of the form (\ref{gform}).

We now pass to the FM representation for the single-valued completion. Following the prescription in (\ref{pres}) we have
\begin{align}
\label{svexample}
&\mathcal{G}_{1-{z}_1,1-{z}_1,0}(\hat{z}_2) - \mathcal{G}_{1,1,0}(\hat{z}_2) \notag \\
&\qquad =\sum_{-\infty<n_1,n_2<\infty} \int_{-\infty}^{\infty}\frac{d\nu_1}{2\pi}\int_{-\infty}^{\infty}\frac{d\nu_2}{2\pi}z_1^{i\nu_1+\frac{n_1}{2}} \bar{z}_1^{i\nu_1-\frac{n_1}{2}} \hat{z}_2^{i\nu_2+\frac{n_2}{2}}  \hat{\bar{z}}_2^{i\nu_2-\frac{n_2}{2}} \notag \\
&\qquad \qquad \qquad \qquad \qquad \qquad \qquad \times \Big[\log|\hat{z}_2|^2I_1(\nu_1,\nu_2,n_1,n_2)-I_2(\nu_1,\nu_2,n_1,n_2)\Big]\notag\\
&\qquad =\sum_{-\infty<n_1,n_2<\infty} \int_{-\infty}^{\infty}\frac{d\nu_1}{2\pi}\int_{-\infty}^{\infty}\frac{d\nu_2}{2\pi}z_1^{i\nu_1+\frac{n_1}{2}} \bar{z}_1^{i\nu_1-\frac{n_1}{2}} \hat{z}_2^{i\nu_2+\frac{n_2}{2}}  \hat{\bar{z}}_2^{i\nu_2-\frac{n_2}{2}} \notag \\
& \qquad \qquad \qquad \qquad \qquad \qquad \qquad \times \bigl[i \partial_{\nu_2} I_1(\nu_1,\nu_2,n_1,n_2)-I_2(\nu_1,\nu_2,n_1,n_2)\bigr]\,.
\end{align}
For convenience we have split the integrand into two pieces,
\begin{align}
I_1(\nu_1,\nu_2,n_1,n_2)=&\, \hat{C}^{(0)}(\nu_1,-\nu_2,n_1,-n_2)\Bigl[\psi\left(\tfrac{n_2}{2}+i\nu_2\right)+\psi\left(1+\tfrac{n_2}{2}-i\nu_2\right)-2\psi(1)\Bigr]\,,\notag\\
I_2(\nu_1,\nu_2,n_1,n_2)=&\, \hat{C}^{(0)}(\nu_1,-\nu_2,n_1,-n_2) \biggl[-\psi^{(1)}\left(\tfrac{n_2}{2}+i\nu_2\right)+\psi^{(1)}\left(1+\tfrac{n_2}{2}-i\nu_2\right) \notag \\
&\qquad  + \frac{1}{\tfrac{n_2}{2}+i\nu_2}\bigl(\psi\left(\tfrac{n_2}{2}+i\nu_2\right)+\psi\left(1+\tfrac{n_2}{2}-i\nu_2\right)-2\psi(1)\bigr)\biggr]\,,
\end{align}
where $\hat{C}^{(0)}$ is the quantity which arises from the prescription (\ref{presbinomial}),
\be
\hat{C}^{(0)}(\nu_1,-\nu_2,n_1,-n_2) = \frac{\tilde{C}^{(0)}(\nu_1,-\nu_2,n_1,-n_2)}{\bigl(i\nu_1 + \tfrac{n_1}{2}\bigr)\bigl(i \nu_1- \tfrac{n_1}{2}\bigr)\bigl(i\nu_2+ \tfrac{n_2}{2}\bigr)\bigl(i\nu_2-\tfrac{n_2}{2}\bigr)} \,.
\ee
After performing the differentiation we obtain the desired Fourier-Mellin integrand for the single-valued polylogarithm (\ref{svexample}) integrand, here expressed in terms of the  $N,V,E,M$ basis.
\begin{align}
\label{exampleresult}
\hat{C}^{(0)}(\nu_1,-\nu_2,n_1,-n_2) \times \left(E_2+V_2\right)\left(E_2 -M-\frac{1}{2}N_2\right)\,.
\end{align}
Finally, returning to the $z_2$ variable instead of $\hat{z}_2$ means flipping the sign of $\nu_2$ and $n_2$. In the above expression this means that $\tilde{C}^{(0)}$ and $M$ acquire arguments with canonical sign and the sign of $N_2$ and $V_2$ get flipped.

\section{Higher-loop NLLA predictions}\label{sec:HigherLoopPredictions}
In the previous section, we used the 2-loop MHV heptagon amplitude in the multi-Regge limit, that we
promoted from symbol to function in section \ref{sec:SymbolsToFunctions}, in order to extract the
NLO central emission block \eqref{eq:CtildeNLO}-\eqref{eq:CtildeLO}. Here, we will use this result,  together with the analogous weak coupling expansion of the BFKL eigenvalue \eqref{eq:omegaofnu}, hexagon impact factor \eqref{eq:Phitildeofnu}, and helicity flip kernel \eqref{eq:Hofnu},
to compute explicit analytic expressions for the heptagon at higher loops in NNLA, from the
dispersion integral \eqref{eq:Rhhhstart}-\eqref{eq:f_hhh}.

Let us start by introducing some useful notation to express the perturbative expansion of the amplitude. At weak coupling, it is evident that the dispersion integral naturally organizes itself into a double expansion in the coupling and in the large logarithms $\log \tau_k$. Separating the coefficients of this expansion into real and imaginary parts, we may define them as
\begin{align}\label{eq:perturbativeExpansion}
\cR_{h_1,h_2,h_3}\left(\tau_{1},{z_{1}},\tau_{2},{z_{2}}\right)e^{i\delta_7(z_1,z_2)} &=1\,+ 2\pi i \sum_{\ell=1}^{\infty}\sum_{i_1,i_{2}=0}^{\ell-1}a^\ell \,\left(\prod_{k=1}^{2}\frac{1}{i_k!}\log^{i_k}\tau_k\right) \\
&\times \left( \,{\tilde g}_{h_1,h_2,h_{3}}^{(\ell;i_1,i_2)}(z_1,z_{2}) + 2 \pi i \, {\tilde h}_{h_1,h_2,h_{3}}^{(\ell;i_1,i_{2})}(z_1,z_{2}) \right)\,. \nonumber
\end{align}
Note in particular that we have defined the perturbative coefficients not of $\cR_{h_1,h_2,h_3}$ alone, but with its combination with a phase, that is equal to the dispersion integral.

The LLA contribution amounts to the coefficients with $i_1+i_2=\ell-1$, for which it is easy to show that
\begin{equation}
{\tilde h}_{h_1,h_2,h_3}^{(\ell;i_1,\ell-1-i_1)} = 0\,,\quad i_1=0,\ldots,\ell-1 \,.
\end{equation}
In this section, we will be obtaining new results for the coefficients ${\tilde g}_{h_1,h_2,h_{3}}^{(\ell;i_1,i_{2})}$ and ${\tilde h}_{h_1,h_2,h_{3}}^{(\ell;i_1,i_{2})}$ with $i_1+i_2=\ell - 2$, or in other words the NLLA contribution.

We will work in the region $z_1\ll1,\,z_2\gg1$, for which we saw in section \ref{sec:7ptBFKL} that it is advantageous to deform the contour of the dispersion integral before the weak coupling expansion, so that the latter becomes equal to~\eqref{eq:R7hhh_w1_small_w2_large}, with
\begin{align}
\tilde f_{h_1h_2h_3}= &\frac{a}{2} \sum_{n_1,n_2=-\infty}^\infty (-1)^{n_1+n_2}
\left(\frac{z_1}{\bar z_1}\right)^{\frac{n_1}{2}} \left(\frac{z_2}{\bar z_2}\right)^{\frac{n_2}{2}}
\int \frac{d\nu_1 d\nu_2}{(2\pi)^2}
   |z_1|^{2i\nu_1}|z_2|^{2i\nu_2}\tilde \Phi(\nu_1,n_1)\tilde \Phi(\nu_2,n_2)
\nonumber\\
&\times e^{-L_1 \omega(\nu_1,n_1)-L_2\omega(\nu_2,n_2)}I^{h_1}(\nu_1,n_1)\tilde C^{h_2}(\nu_1,n_1,\nu_2,n_2)
 \bar I^{h_3}(\nu_2,n_2)\,, \label{eq:ftilde_hhh}
\end{align}
and the integration contour goes below (above) the poles on the real axis for $\nu_1$ ($\nu_2$), as shown in figure \ref{fig:contourRSeven}. The perturbative coefficients \eqref{eq:perturbativeExpansion} will be a linear combination of the respective coefficients of all the terms in the right-hand side of~ \eqref{eq:R7hhh_w1_small_w2_large}. {However for the hexagon amplitudes $\cR_{h_1h_2}$ they have already been obtained up to at least 8 loops to NLLA ~\cite{Dixon:2012yy,Caron-Huot:2016owq}, and more generally the holomorphic part may evaluated in terms of harmonic polylogarithms with the method of~\cite{Papathanasiou:2013uoa}, see also~\cite{Drummond:2015jea,Broedel:2015nfp}. So we only need to focus on the last term in \eqref{eq:perturbativeExpansion}, that contains the genuine heptagon contributions.}

As we will detail in the next sections, we will compute~\eqref{eq:ftilde_hhh} with the help of two complementary methods. First, we will use nested sum evaluation algorithms, which are easier to apply for the heptagon to high loop order. Then, we will also rely on Fourier-Mellin convolutions, which are particularly suited for applying to higher-point amplitudes, a topic we plan to address in more detail in subsequent work.

Before we proceed with the description of our methods, let us briefly summarize the checks we have performed on our results. First of all, we have confirmed that the two methods yield the same expressions for the $2\to5$ amplitude to NLLA, {through 3 loops in the $\cR_{-++}$ NMHV helicity configuration, and through 4 loops in the MHV case. Up to the same loop orders, we have also checked that under soft limits, the amplitude in any helicity configuration reduces to the known $2\to 4$ amplitude \cite{Dixon:2012yy}}. Finally, at 3 and 4 loops we have compared the symbol of our expressions for the MHV amplitude with the MRK limit \cite{HeptLimits} of the known symbol in general kinematics~\cite{Drummond:2014ffa,Dixon:2016nkn}, finding perfect agreement.

\subsection{A nested sum evaluation algorithm}

After we expand the integrand in~\eqref{eq:ftilde_hhh} at weak coupling, we close the integration contour below (above) the real axis for $\nu_1$ ($\nu_2$), and use Cauchy's theorem to express it as a sum over the enclosed residues, with the infinite semicircles giving a vanishing contribution due to $|z_1|^{2|\text{Im}(\nu_1)|}, |z_2|^{-2|\text{Im}(\nu_2)|}\to 0$ in the region $z_1\ll1,\,z_2\gg1$ we are considering.

In reality, the fact that amplitudes in the multi-Regge limit, and thus also $\tilde f_{h_1h_2h_3}$, are expressible in terms of SVMPLs, allows us to compute the latter by only considering the subset of poles $\nu_1=-in_1/2$ and $\nu_2=-in_2/2$, with $n_1>0$ and $n_2<0$, which is equal to its holomorphic part, in the sense of~\eqref{eq:HolomorphicPart}, with respect to the variables $z_1,1/z_2$~\cite{Drummond:2015jea}\footnote{Note that we need to multiply the integrand with $-(2\pi i)^2$ due to the orientation of our contours.}
\be
\begin{aligned}\label{eq:f_omega2omega3antihol}
    \tilde f^{ h}_{h_1h_2h_3}= &\frac{a}{2} \sum_{n_1=1}^\infty\sum_{n_2=-1}^{-\infty}\,
\underset{\nu_i=\frac{-in_i}{2}}{\text{Res}}
   \Big( z_1^{i\nu_1+\frac{n_1}{2}}  z_2^{i\nu_2+\frac{n_2}{2}}\tilde \Phi(\nu_1,n_1)\tilde \Phi(\nu_2,n_2) \times
\\
&\quad \times e^{-L_1 \omega(\nu_1,n_1)}
e^{-L_2\omega(\nu_2,n_2)}I^{h_1}(\nu_1,n_1)\tilde C^{h_2}(\nu_1,n_1,\nu_2,n_2)
 \bar I^{h_3}(\nu_2,n_2)\Big)\,.
\end{aligned}
\ee
That is, in what follows we will focus on computing $\tilde f^{h}_{h_1h_2h_3}$, and then recover $\tilde f_{h_1h_2h_3}$
with the help of the single-valued map~\eqref{eq:SingleValuedMap} at the very end\footnote{To be precise, this equality holds if $\tilde f_{h_1h_2h_3}$ is a pure function. Subtleties when this is not the case are discussed at the end of this section.}.

After we substitute \eqref{eq:CtildeNLO}-\eqref{eq:CtildeLO}, \eqref{eq:omegaofnu}-\eqref{eq:Hofnu} in \eqref{eq:f_omega2omega3antihol}, extracting the residues becomes in practice very easy after we also use the symmetry of  \eqref{eq:CtildeNLO}-\eqref{eq:CtildeLO} in order to replace $n_i\to-n_i$ there. In particular,  it is manifest that only the rational denominators and $\Gamma(-i\nu_1+\frac{n_1}{2})$ and $\Gamma(i\nu_2-\frac{n_2}{2})$ will have poles, whereas all polygamma functions will have positive arguments. In this manner, and after we set $k=|n_1|, l=|n_2|$, \eqref{eq:f_omega2omega3antihol} becomes a sum of terms of the general form
\be\label{eq:HeptSum}
\sum_{k,l=1}^\infty  \frac{z_1^k}{k^{r_1}}\frac{z_2^{-l}}{l^{r_2}}\frac{\Gamma(k+l)}{\Gamma(1+k)\Gamma(1+l)}\prod_{m_i,m'_i,m''_i}\psi^{(m_i)}(k+1)\psi^{(m'_i)}(l+1)\psi^{(m''_i)}(k+l)\,,
\ee
for different choices of integers $r_1,r_2,m_i,m'_i,m''_i\ge 0$, not necessarily different from each
other, times factors that do not depend on the summation variables.

Next, we express the polygamma functions in terms of generalizations of harmonic numbers known as $S$- or $Z$-sums \cite{Moch:2001zr}, via
\be\label{psi_to_S}
\begin{aligned}
\psi(k+1)&\equiv\psi^{(0)}(k+1)=-\gamma_E+S(k;1;1)\\
\psi^{(m-1)}(k+1)&=(-1)^{m} (m-1)! [\zeta_{m}-S(k;m;1)]\,,
\end{aligned}
\ee
where $\zeta_{m}$ the Riemann zeta function, $ \gamma_E=-\psi(1)\simeq 0.577$ the Euler-Mascheroni constant, and
\begin{align}
S(n;m_1,\ldots,m_j;x_1,\ldots,x_j)&=\sum_{n\geq i_1\geq i_2\geq\ldots\geq i_j\geq1}\frac{x_1^{i_1}}{i_1^{m_1}}\ldots\frac{x_j^{i_j}}{i_j^{m_j}}\,,\label{Ssum}\\
Z(n;m_1,\ldots,m_j;x_1,\ldots,x_j)&=\sum_{n\ge i_1>i_2>\ldots>i_j>0}\frac{x_1^{i_1}}{i_1^{m_1}}\ldots\frac{x_j^{i_j}}{i_j^{m_j}}\,,\label{Zsum}
\end{align}
with the generalized harmonic numbers corresponding to the special case $S(k;m;1)=Z(k;m;1)$. This substitution allows us to use the quasi-shuffle algebra relations of $S$- or $Z$-sums, in order to express their products with the same outer summation index, in terms of linear combinations thereof.

As we will see very shortly, it proves advantageous to replace $\psi^{(m'_i)}(l+1)$, $\psi^{(m_i)}(k)$ by $S$- and $\psi^{(m''_i)}(k+l)$ by $Z$-sums respectively. After soaking up the gamma function dependence of \eqref{eq:HeptSum} into a rational factor times a binomial coefficient,
\be
{k+l \choose k}=\frac{\Gamma(k+l+1)}{\Gamma(k+1)\Gamma(l+1)}\,,
\ee
shifting the summation variable $l\to j=k+l$,  and partial fractioning with respect to $k$, the latter formula splits into terms that look like
\begin{multline}\label{eq:residuestoSsums}
\sum_{j=1}^\infty \frac{(z_2)^{-j}}{j^{n_1}}Z(j-1;n_2,\ldots;1,\ldots,1)\times\\
\times \sum_{k=1}^{j-1}\binom{j}{k}\frac{(z_1  z_2)^k}{k^{n'_1}}S(k;n'_2,\ldots;1,\ldots,1)\frac{1}{(j-k)^{n''_1}}S(j-k;n''_2,\ldots;1,\ldots,1)\,,
\end{multline}
where we also extended the summation range to include $j=1$, since $Z(j-1;\ldots)$ vanishes there.

Very crucially, the sum on the second line of \eqref{eq:residuestoSsums} can be evaluated in terms of $Z$-sums with the help of algorithm D of \cite{Moch:2001zr}, which has already been implemented in \texttt{GiNaC} \cite{Bauer20021} and \texttt{FORM}  \cite{Vermaseren:2000nd} symbolic computation  frameworks, as part of the \texttt{nestedsums} library \cite{Weinzierl:2002hv} or \texttt{XSummer} package \cite{Moch2006759} respectively. We make use of the former by directly interfacing it to \texttt{Mathematica}, in particular by sequentially calling the \texttt{transcendental\_sum\_type\_D} and \texttt{Ssum\_to\_Zsum} commands for each sum of this type.

The $Z$-sums we obtain in this manner may have outer summation index $j-a$ for $a\ge 0$, which from
the definition \eqref{Zsum} is equivalent to the statement that in reality the outermost summation
range should be $j\ge\max(1,a)$. They may also come with $(z_1  z_2)^j$ or $(1-z_1  z_2)^j$
prefactors, products/powers of $(j-b)$ denominators with $a-1\ge b\ge 0$, as well as factors that do
not depend on $j$. After shifting the summation variable $j\to i=j-\max(1,a)+1$ for each different $a$, and partial fractioning in $i$, we reduce all terms \eqref{eq:residuestoSsums} in our expression for $\tilde f^{h}_{h_1h_2h_3}$ into simple sums of the form
\be\label{eq:DoubleSumsToSimpleSums}
\sum_{i=1}^\infty \frac{x^{i}}{(i+c)^{n_1}}Z(i+o-1;n_2,\ldots;1,\ldots,1)Z(i-1;n'_2,\ldots;x_2,\ldots)\,.
\ee

We then synchronize the $Z$-sums, namely remove the offset $o$ of the first of the two, by recursive definition of the identity\footnote{Since in this case the $Z$-sums with offset have their origin in the terms $\psi^{(m''_i)}(k+l)$ in  \eqref{eq:HeptSum}, we could have alternatively left them in this form, shift their arguments with the identity
\be\label{psi_recurrence}
\psi ^{(n)}(z+1)=\psi ^{(n)}(z)+(-1)^n n! z^{-n-1}
\ee
at this point, and only then use \eqref{psi_to_S}-\eqref{Zsum} to express them as $Z$-sums.}
\begin{multline}
Z(i+o-1;m_1,...;x_1,...) \\
=  Z(i-1;m_1,...;x_1,...)
      + \sum\limits_{j=0}^{o-1} x_1^j \frac{x_1^i}{(i+j)^{m_1}} Z(i-1+j;m_2,...;x_2,...)\,,
\end{multline}
and once again eliminate any products with quasi-shuffle algebra relations. Similarly, we remove the offset from the denominators with the help of
\be
     \sum\limits_{i=1}^\infty \frac{x^i}{(i+c)^m} Z(i-1,...)
=
       \sum\limits_{i=1}^\infty \frac{x^{i-1}}{(i+c-1)^m} Z(i-1,...)
 - \sum\limits_{i=1}^{\infty} \frac{x^i}{(i+c)^m} \frac{x_1^i}{i^{m_1}} Z(i-1,m_2,...)
       \,,
\ee
or
\be
    \sum\limits_{i=1}^\infty \frac{x^i}{(i+c)^m}
     =
       \frac{1}{x} \sum\limits_{i=1}^\infty \frac{x^i}{(i+c-1)^m}
       \mbox{} - \frac{1}{c^m} .
\ee
if no $Z$-sums are present. After these steps, the expression \eqref{eq:f_omega2omega3antihol} for $\tilde f^{h}_{h_1h_2h_3}$ may be readily evaluated in terms of multiple polylogarithms, thanks to the definition
\be\label{LitoZ}
\text{Li}_{m_1,\ldots,m_j}(x_1,\ldots,x_j)=\sum_{i=1}^\infty \frac{x_1^{i}}{i^{m_1}}Z(i-1;m_2,\ldots,m_j;x_2,\ldots,x_j)\,.
\ee
The procedure we have described for evaluating the Fourier-Mellin integrand is the same for both the MHV and NMHV case, the only difference being that in the former the powers of the denominators $r_1,r_2$ are strictly positive, whereas in the latter they can also be zero. This difference is closely related to the fact that the MHV amplitude is a pure function, whereas the NMHV ones also contain some rational factors in the $z_i$ variables.

It is only with respect to these rational factors, that some additional care is needed when considering the projection to the holomorphic part \eqref{eq:f_omega2omega3antihol}, since this will also set the antiholomorphic rational factors to constants, possibly zero. Particularly for the $\cR_{-++}$  NMHV amplitude, it was shown in \cite{DelDuca:2016lad} that the rational factors contain no $\bar z_i$ dependence, so similarly to the MHV case, the full $\tilde f_{-++}$ may also be obtained with the help of~\eqref{eq:SingleValuedMap}, when the rational factors are considered as constants with respect to the single-valued map.

Using the techniques we have described in this section, we have obtained the MHV $\cR_{+++}$ and NMHV $\cR_{-++}$ heptagons to NLLA though 5 and 4 loops respectively. The treatment of  $\cR_{+-+}$ will follow in the next section, with the method of convolutions.


\subsection{Evaluation by Fourier-Mellin convolutions}
In this section, we will give a brief overview of the main aspects of a convolution-based method to compute amplitudes in MRK, introduced in \cite{DelDuca:2016lad}, and how it can be adapted for computations beyond LLA. For proofs and a more detailed review, we refer the reader to \cite{DelDuca:2016lad}.

The dispersion integral \eqref{eq:ftilde_hhh}, describing the nontrivial part of $\cR_{h_1h_2 h_3}$, corresponds to the two-fold application of an inverse Fourier-Mellin transform,

\begin{equation}
\cF[F(\nu,n)] = \sum_{n=-\infty}^{\infty}\left(\frac{z}{\zb}\right)^{n/2}\,\int_{-\infty}^{+\infty}\frac{d\nu}{2\pi}\,|z|^{2i\nu}\,F(\nu,n) \,.
\end{equation}
The Fourier-Mellin transform maps products into convolutions, so that for $\cF[F]=f$ and $\cF[G]=g$ we have
\begin{align}\label{eq:conv_thm}
\cF[F\cdot G] = \cF[F]\ast\cF[G] = f\ast g\,,
\end{align}
where the convolution is defined as
\begin{align}\label{eq:conv_def}
(f\ast g)(z) = \frac{1}{\pi}\int \frac{d^2w}{|w|^2}\,f(w)\,\,g\left(\frac{z}{w}\right)\,.
\end{align}

So as to render the computation by convolutions more transparent, we will define the first few orders of the following building blocks separately,
\begin{equation}\label{eq:intro_All_Order_Building_Blocks}
\begin{split}
\omega(\nu,n) &= -a(E_{\nu,n} + a E^{(1)}_{\nu,n} +\mathcal{O}(a^2)) \,, \\
\chi^\pm(\nu,n) &= \chi_0^\pm(\nu,n) (1 + a \kappa^\pm_1(\nu, n)  + \mathcal{O}(a^2)) \,, \\
C^\pm(\nu, n,\mu, m) &= C_0^\pm(\nu, n,\mu, m)(1 + a c^\pm_1(\nu, n,\mu, m) + \mathcal{O}(a^2)) \,,
\end{split}
\end{equation}
and we define
\begin{align}
E_i &\equiv E_{\nu_i n_i}, &
\chi_{0,i}^\pm &\equiv \chi_0^\pm  (\nu_i,n_i), & \kappa_{0,i}^\pm &\equiv \kappa_0^\pm  (\nu_i,n_i) \\
C_{0, ij}^\pm &\equiv C_{0}^\pm (\nu_i,n_i,\nu_j,n_j), & c_{1, ij}^\pm &\equiv c_{1}^\pm (\nu_i,n_i,\nu_j,n_j), \nonumber
\end{align}
We also define a shorthand for the product of leading order impact factors and central emission blocks at seven points,
\begin{equation}
\varpi_7 \equiv \varpi_7^{h_1 h_2 h_3} = \chi^{h_1}_{0,1} C_{0,12}^{h_2} \chi^{-h_3}_{0,2} \,,
\end{equation}
where we drop explicit dependence on the helicities.

Then at LLA, (i.e. for $i_1+i_2 = \ell - 1$) we find
\begin{equation}\label{eq:g_LLA_def}
\begin{split}
{\tilde g}_{h_1h_2 h_3}^{(\ell;i_1,i_2)}(z_1,z_2) &= \frac{1}{2} \,\cF_2 \left[\varpi_7 E^{i_1}_1E^{i_2}_2\right],
\end{split}
\end{equation}
where
\begin{equation}
\cF_2 \left[ F \right] = \sum_{n_1,n_2 =-\infty}^{\infty}\left(\frac{z_1}{\zb_1}\right)^{\frac{n_1}{2}} \left(\frac{z_2}{\zb_2}\right)^{\frac{n_2}{2}}\,
\int_{-\infty}^{+\infty}\frac{d\nu_1}{2\pi} \frac{d\nu_2}{2\pi}\,\,|z_1|^{2i\nu_1} |z_2|^{2i\nu_2} F
\end{equation}
denotes the two-fold Fourier-Mellin transform. At NLLA, (i.e. for $i_1 + i_2 = \ell - 2$),{ we write the perturbative coefficients as
\begin{align}
\label{eq:inrto_hatted_PertCoef}
{\tilde g}_{h_1 h_2 h_{3}}^{(\ell;i_1,i_{2})}(z_1,z_{2}) &= \sum_{j=1}^{2} i_j {\tilde{g}}_{h_1 h_2 h_{3}}^{j;(\ell;i_1, i_{2})}(z_1,z_{2}) + \sum_{j=1}^{3} {\tilde{g}}_{j;h_1h_2 h_{3}}^{(\ell;i_1, i_{2})}(z_1,z_{2}) + P_{h_1 h_2 h_{3}}^{(\ell;i_1,i_{2})}(z_1,z_{2}) \,, \nonumber \\
{\tilde h}_{h_1 h_2 h_{3}}^{(\ell;i_1,i_{2})}(z_1,z_{2}) &= \sum_{j=1}^{2} {\tilde{h}}_{h_1 h_2 h_{3}}^{j;(\ell;i_1, i_{2})}(z_1,z_{2}) + \sum_{j=1}^{3} {\tilde{h}}_{j;h_1 h_2 h_{3}}^{(\ell;i_1, i_{2})}(z_1,z_{2}) + Q_{h_1 h_2 h_{3}}^{(\ell;i_1,i_{2})}(z_1,z_{2})\,,
\end{align}
where $P$ and $Q$ are due to the contributions from the first three terms of eq.~\eqref{eq:R7hhh_w1_small_w2_large}, explicitly given by
\begin{align}
P_{h_1 h_2 h_{3}}^{(\ell;i_1,i_{2})}(z_1,z_{2}) =& \,\delta_{i_2,0} \, {\tilde{g}}_{h_1h_2}^{(\ell;i_1)}(z_1) +  \delta_{i_1,0}\,{\tilde{g}}_{h_2h_3}^{(\ell;i_{2})}(z_2) \\
Q_{h_1 h_2 h_{3}}^{(\ell;i_1,i_{2})}(z_1,z_{2}) =& \, \delta_{i_2,0} \left({\tilde{h}}_{h_1h_2}^{(\ell;i_1)}(z_1) - \frac{1}{4}\cG_{0}(z_2){\tilde{g}}_{h_1h_2}^{(\ell-1;i_1)}(z_1)\right)\\
&+\delta_{i_1,0}\left( {\tilde{h}}_{h_2h_3}^{(\ell;i_{2})}(z_2) + \frac{1}{4}\cG_{0}(z_1){\tilde{g}}_{h_2h_3}^{(\ell-1;i_{2})}(z_2)\right)\nonumber\\
&+ \delta_{\ell,2}\frac{1}{16}\cG_{0}(z_1)\cG_{0}(z_2)\nonumber
\end{align}
and  we have also introduced \emph{corrected perturbative coefficients} describing different contributions to the expansion of the purely heptagonal $\tilde f_{h_1h_2h_3}$ term of the latter equation.} Perturbative coefficients with an additional upper index correspond to insertions of the NLO corrections to the BFKL eigenvalue and perturbative coefficients with an additional lower index correspond to insertions of NLO corrections to the impact factors or central emission blocks. Then these corrected perturbative coefficients are given by
  \begin{equation}
  \begin{split}\label{eq:g_NLLA_def}
  {\tilde g}_{h_1h_2 h_3}^{j;(\ell;i_1,i_2)}(z_1,z_2) =& \,
  \frac{1}{2} \,\cF_2 \left[\varpi_7 \, E^{i_1-\delta_{1j}}_1E^{i_2-\delta_{2j}}_2 E^{(1)}_j \right],\\
  {\tilde g}_{1;h_1h_2 h_3}^{(\ell;i_1,i_2)}(z_1,z_2) =& \,
  \frac{1}{2} \,\cF_2 \left[\varpi_7 \, E^{i_1}_1E^{i_2}_2 \kappa^{h_1}_{1,1}\right],\\
  {\tilde g}_{3;h_1h_2 h_3}^{(\ell;i_1,i_2)}(z_1,z_2) =& \,
  \frac{1}{2} \,\cF_2 \left[\varpi_7 \, E^{i_1}_1E^{i_2}_2 \kappa^{-h_3}_{1,2}\right],\\
  {\tilde g}_{2;h_1h_2 h_3}^{(\ell;i_1,i_2)}(z_1,z_2) =& \,
  	\frac{1}{2} \,\cF_2 \left[\varpi_7 \, E^{i_1}_1E^{i_2}_2 \Re \left( c^{h_2}_{1,12}\right)\right].
  \end{split}
  \end{equation}
and
\begin{equation}
\begin{split}\label{eq:h_NLLA_def}
{\tilde h}_{h_1h_2 h_3}^{j;(\ell;i_1,i_2)}(z_1,z_2) =& \, -\frac{1}{4} \,\cF_2 \left[\varpi_7 \, E^{i_1}_1E^{i_2}_2 E_j\right],\\
{\tilde h}_{j;h_1h_2 h_3}^{(\ell;i_1,i_2)}(z_1,z_2) =& \; 0, \quad j \in \{1,3\},\\
{\tilde h}_{2;h_1h_2 h_3}^{(\ell;i_1,i_2)}(z_1,z_2) =& \,
	\frac{1}{4 \pi} \,\cF_2 \left[\varpi_7 \,  E^{i_1}_1E^{i_2}_2 \Im \left( c^{h_2}_{1,12}\right)\right].
\end{split}
\end{equation}
Let us now focus on MHV amplitudes to explain how to use convolutions to compute the perturbative coefficients and extend it later to the case of non-MHV amplitudes. Using the convolution theorem for Fourier-Mellin transforms \eqref{eq:conv_thm}, we can identify simple relations between the perturbative coefficients at different loop orders. At six points and at LLA, for example, we have \cite{DelDuca:2016lad}
\begin{align}\label{eq:convolution_recursion}
g_{++}^{(\ell;\ell-1)} = -\frac{1}{2}\cF \left[ \chi_{0,1}^+ E_1^{\ell-1} \chi_{0,1}^- \right] = g_{++}^{(\ell-1;\ell-2)} * \cF[E] \,.
\end{align}
By repeatedly convoluting with leading-order eigenvalues, we can build up the desired perturbative coefficients recursively from lower-loop expressions.
Exploiting further that the perturbative coefficients are single-valued functions \cite{DelDuca:2016lad}, we can simplify the evaluation of the convolution integrals a residue computation \cite{Schnetz:2013hqa}. Consider a function $f(z)$, consisting of single-valued hyperlogarithms and rational functions with possible singularities at $z=a_i$ and $z = \infty$. Close to these singularities, $f$ can be written as
\begin{align}
f(z) &\,= \sum_{k,m,n}\,c^{a_i}_{k,m,n}\,\log^k\left|1-\frac{z}{a_i}\right|^2\,(z-a_i)^m\,(\zb-\bar{a}_i)^n\,, \quad z\to a_i\,,\\
f(z) &\,= \sum_{k,m,n}\,c^{\infty}_{k,m,n}\,\log^k\frac{1}{|z|^2}\,\frac{1}{z^m}\,\frac{1}{\zb^n}\,, \quad z\to \infty\,.
\end{align}
The \emph{holomorphic residue} of $f$ at $z = a_i$ and $z = \infty$ are defined as the coefficient of the simple holomorphic pole without logarithmic singularities,
\begin{align}
\textrm{Res}_{z=a_i}f(z) &\equiv c^{a_i}_{0,-1,0} & \textrm{Res}_{z=\infty}f(z) &\equiv c^{\infty}_{0,-1,0}\,.
\end{align}
It was also shown in \cite{Schnetz:2013hqa} that the integral of $f$ over the whole complex plane, if it exists, can be computed by the holomorphic residues of its (single-valued) antiholomorphic primitive $F$, i.e. if $\bar{\partial} F = f$, then
\begin{equation}
\int \frac{d^2z}{\pi}\,f(z) = \textrm{Res}_{z=\infty}F(z) - \sum_i\text{Res}_{z=a_i}F(z)\,.
\end{equation}
Starting form the two-loop amplitude, the only thing missing to compute higher order MHV amplitudes is the Fourier-Mellin transform of the leading-order BFKL Eigenvalue $\cF(E)$, which is given by
\begin{equation}
\cE(z) \equiv \cF[E] = -\frac{z+\zb}{2\,|1-z|^2}\,.
\end{equation}
The convolution formalism can also be extended to the computation of non-MHV amplitudes using the helicity-flip kernel (\ref{eq:Hofnu}). Its Fourier-Mellin transform reads

\begin{equation}
\begin{split}
\cH(z) &= \cF[H(\nu,n)] \\
&= \, -\frac{z}{(1-z)^2} + \frac{a}{4} \left(\cG_1(z)+\frac{z }{(1-z)}\cG_0(z)+\frac{z }{(1-z)^2} \cG_{0,0}(z)\right) + \cO(a^2)\\
&= \, \Biggl(-\frac{z}{(1-z)^2}\Biggr) * \Biggl(\delta^{(2)}(1-z) +  \frac{a}{4} \frac{z - \bar{z} }{(1-z)(1-\bar{z})}\cG_0(z) + \cO(a^2)\Biggr) \,,
\end{split}
\end{equation}
where $H(\nu,n)$ is the helicity flip kernel in Fourier-Mellin space, defined in~\eqref{eq:Hofnu}.

Given the form of the recursion relation \eqref{eq:convolution_recursion}, all we need in addition to the integration kernels $\cE$ and $\cH$ is a starting point for the recursion. Starting from the two-loop NLLA amplitude~\eqref{eq:R72} we computed the perturbative coefficients $\tilde{g}$ and $\tilde{h}$ through four loops in the MHV case and through three loops in all other helicity configurations.

We will conclude this section by commenting on some details of this computation.
At NLLA, we had to introduce the terms $|z_1|^{2 \pi i \Gamma} e^{R_6(z_2)+i \delta_6(z_2)}$ (the same discussion applies to the corresponding terms with $z_1 \leftrightarrow z_2$) and $|z_1|^{2\pi i \Gamma}/ |z_2|^{2\pi i \Gamma}$ in~\eqref{eq:R7_w1_small_w2_large} in order to avoid a pinching of our integration contour. Since only the term $\tilde f_{+++}$ corresponds to the two-fold Fourier-Mellin integral, when relating different perturbative coefficients via the recursion~\eqref{eq:convolution_recursion}, these additional terms should be subtracted before performing convolutions and added back afterwards. At NLLA, for example, the term $|z_1|^{2\pi i \Gamma}/|z_2|^{2\pi i \Gamma}$ only contributes at two loops, as it is independent of the large logarithms $\log \tau_i$. Naively, convoluting over this term will introduce additional terms at three-loop order that should not be there. It turns out, however, that these terms only interfere with our computations when raising the loop order of the real part from two- to three loops. This is easy to see when analyzing how the individual parts of these terms behave under convolutions with the $\cE$ and $\cH$ kernels. Expanding the extra terms in $a$ only yields powers of logarithms $\cG_0(z_i)^k$, with $0 \leq k \leq \ell$ and six-point perturbative coefficients $\tilde{g}$ and $\tilde{h}$ at any given order $\ell$, and though NLLA, we can limit our analysis to $k \leq 2$. When convoluted with $\cE$ and $\cH$, these logarithms yield the following results.
\begin{equation}\label{eq:EKer_Convolution}\bsp
1 \ast \cE(z) &= 0 \,, \\
 \cG_0(z) \ast \cE(z) &= 0 \,, \\
\cG_0(z)^2 \ast \cE(z) &= - 4 \zeta_3 \,,
\esp\end{equation}

\begin{equation}\label{eq:HKer0_Convolution}\bsp
1 \ast \cH^{(0)}(z) &= 1 \,, \\
\cG_0(z) \ast \cH^{(0)}(z) &= \cG_0(z) \,, \\
\cG_0(z)^2 \ast \cH^{(0)}(z) &= \cG_0(z)^2 \,,
\esp\end{equation}

\begin{equation}\label{eq:HKer1_Convolution}
1 \ast \cH^{(1)}(z) = \cG_0(z) \ast \cH^{(1)}(z) =\cG_0(z)^2 \ast \cH^{(1)}(z) = 0 \,.
\end{equation}

Let us now have a look at the two different kinds of extra terms individually. We will start with the fraction term $|z_1|^{2\pi i \Gamma}/ |z_2|^{2\pi i \Gamma}$, which has no dependence on the large logarithms $\log \tau_i$ and therefore, at NLLA, should only affect the 2-loop amplitude. As convolutions with $\cE(z_i)$ and $\cH^{(1)}(z_i)$ both appear with a factor of $a$ in the Fourier-Mellin integrand, they both result in a higher-order contribution and should evaluate to zero when convoluted with the extra term. Furthermore, it does not depend on the helicity configuration of the amplitude, which suggests that it should be invariant under convolution with leading order helicity flip kernels $\cH^{(0)}(z_i)$. Expanding the term in $a$, we find
\begin{equation}
\begin{split}
\frac{|z_1|^{2\pi i \Gamma}}{|z_2|^{2\pi i \Gamma}} &= 1 +  \frac{a}{2} i \pi  (\cG_0(z_1)-\cG_0(z_2)) - \frac{a^2}{12} i \pi ^3 (\cG_0(z_1)-\cG_0(z_2)) \\
&+ \frac{a^2}{8} \pi ^2 \left( \cG_0(z_1)^2 - 2 \cG_0(z_1) \cG_0(z_2) + \cG_0(z_2)^2\right) + \cO(a^3)\,.
\end{split}
\end{equation}
Considering \eqref{eq:EKer_Convolution} - \eqref{eq:HKer1_Convolution}, we see immediately that the aforementioned criteria are only violated by the terms $\cG_0(z_i)^2$ appearing in the 2-loop real part. We will therefore have to subtract these terms before performing convolutions with $\cE(z_i)$. \\

We will now focus on the extra terms $|z_1|^{2\pi i \Gamma}e^{R_6(z_2)+i \delta_6(z_2)}$ (and the one with $z_1 \leftrightarrow z_2$) containing the 6-point amplitude. At LLA, it contributes only through the exponential ${e^{R_6(z_2)+i \delta_6(z_2)}}\bigr|_{LLA} $, which by definition transforms correctly under convolutions with $\cE(z_2)$ and $\cH(z_2)$. Since the term only comes with large logarithms $\log \tau_2$, it should vanish when convoluting with $\cE(z_1)$ so that its presence will not spoil the terms proportional to both large logarithms. Furthermore, it should be invariant under leading order helicity flips $\cH^{(0)}(z_1)$ and should vanish under first oder helicity flips $\cH^{(1)}(z_1)$. This follows again from (\ref{eq:EKer_Convolution}) - \eqref{eq:HKer1_Convolution}. At NLLA the situation is similar. Here we will again encounter terms that only arise from ${e^{R_6(z_2)+i \delta_6(z_2)}}\bigr|_{NLLA} $\footnote{Note that $e^{R_6(z_2)+i \delta_6(z_2)}$ also contains an extra term that is not part of the Fourier-Mellin transform. This term is the 6-point equivalent of the fractional term we discussed before and all observations apply here, too.}, i.e. the NLLA hexagon contributions. In addition, we will find terms arising from $\log |z_1|^2 {e^{R_6(z_2)+i \delta_6(z_2)}}\bigr|_{LLA} $. Since Fourier-Mellin convolutions are also suited for the computation of hexagon NLLA amplitudes, convolutions in $z_2$ will behave in the desired fashion. Once again, we have to ensure that convolutions in $z_1$ do not spoil our results, which means that convolutions with $\cE(z_1)$ and $\cH^{(1)}(z_1)$ should vanish and that leading order helicity flips $\cH^{(0)}(z_1)$ should have no impact. This is again given by \eqref{eq:EKer_Convolution} - \eqref{eq:HKer1_Convolution}.
We see therefore, that these potentially dangerous terms appearing in the amplitude are sufficiently well behaved and we can therefore completely ignore the presence of these terms and perform our convolutions without taking further precautions.

Let us now briefly summarize the previous observations. We have seen that at NLLA, we need to introduce extra terms to our amplitudes that are not part of the two-fold Fourier-Mellin integral~\eqref{eq:R7_w1_small_w2_large} due to the presence of Regge poles. Even though these extra terms contribute to the perturbative coefficients at all orders, we only have to subtract these terms when convoluting $\tilde{h}_{h_1h_2}^{(2;0,0)}$ with $\mathcal{E}$ in order to raise its loop order.
In all other cases, and in particular for all helicity flips, convolutions can be applied directly to the full perturbative coefficients.

\section{Conclusions}\label{sec:Conclusions}
In this paper, we took a decisive step towards the description of $\cN=4$ super-Yang Mills amplitudes with more than six legs in the multi-Regge limit, to arbitrary logarithmic accuracy. Focusing on the $2\to5$ amplitude, we first succeeded in describing it in this limit in terms of the all-loop dispersion integral \eqref{eq:Rhhhstart}-\eqref{eq:f_hhh}, or equivalently \eqref{eq:R7hhh_w1_small_w2_large}, which has a well-defined weak-coupling expansion, thus overcoming regularization issues that arose in previous attempts beyond LLA.

The dispersion integral factorizes into certain building blocks, the determination of which is necessary in order to fully specify the amplitude. While most of them are known to all loops \cite{Basso:2014pla}, this is not true for the BFKL central emission vertex, which has only been worked out to leading order \cite{Bartels:2011ge}.
The determination of the central emission vertex in eq.~\eqref{eq:CtildeNLO} to NLO, is the second important result of this paper.

We extracted the NLO central emission block from the 2-loop $2\to5$ MHV amplitude in MRK, which we first had to promote from symbol~\cite{Bargheer:2015djt} to function level, see eq.~\eqref{eq:R72}-\eqref{eq:gfunction}. The latter is a single-valued function in momentum space, and by exploiting this property we were able to transform it to Fourier-Mellin space,
allowing us to extract the NLO correction to the BFKL central emission block.
It is also worth noting that in intermediate steps of our calculation, we obtained two results for the general $N$-particle MHV amplitude that are interesting in their own right:
its explicit expression at two loops, eq.~\eqref{eq:Rnkl}-\eqref{wtov} at function level, and also its maximal degree of logarithmic divergence, or leading OPE discontinuity, eq.~\eqref{eq:RNOPE}-\eqref{eq:OPEcrossratios}, at any loop order.

Finally, the knowledge of the NLO central emission vertex, and the weak-coupling expansion of the dispersion integral, allows for the determination of the $2\to5$ amplitude to NLLA, in principle at any loop order. Indeed, we put our results into good use, and evaluated the dispersion integral with a combination of nested sum algorithms and convolutions, in order to end up with explicit predictions for the perturbative coefficients of the amplitude, eq.~\eqref{eq:perturbativeExpansion}, in momentum space. In particular, we computed the NLLA coefficients of the MHV case $\cR_{+++}$  through 5 loops, and of the NMHV $\cR_{+-+}$ and $\cR_{-++}$ helicity configurations through 3 and 4 loops, respectively.
Considering the length of these expressions, we give them explicitly through two loops in appendix~\ref{sec:ExplicitResults}, and provide the rest as ancillary files accompanying the submission of this article on the \texttt{arXiv}. In particular the files \texttt{gTilde.m} and \texttt{hTilde.m} contain the imaginary and real NLLA perturbative coefficients of eq.\eqref{eq:perturbativeExpansion}, respectively, \texttt{RFactors.m} contains the leading singularities appearing in the NMHV amplitudes, as defined in appendix B, and finally \texttt{R2MRK} contains the 2-loop heptagon remainder function $R_7^{(2)}$ of eq.~\eqref{eq:R72}.

As we mentioned in the main text, a straightforward extension of the work presented here is the determination of the central emission vertex at higher loops. Using the method described in \ref{sec:leading singularities}, we have already extracted the NNLO correction to the central emission vertex, up to transcendental constants, from the 3-loop heptagon symbol \cite{Drummond:2014ffa}, and although beyond the scope of this paper, it is a straightforward extension to do the same at N$^3$LO, from the corresponding 4-loop symbol~\cite{Dixon:2016nkn}.
Beyond NLO however, we cannot only rely on soft limits in order to complete these results with all beyond-the-symbol terms:
Starting at weight 2, there exists a kernel of single-valued functions that vanish in all soft limits. Nevertheless, it may be possible to fix all remaining ambiguities by exploiting the perturbative overlap between the collinear and multi-Regge limit, as was done e.g.~in~\cite{Bargheer:2016eyp},  relying on earlier observations on this overlap in the six-gluon case~\cite{Hatsuda:2014oza}.

Equally importantly, the procedure of section \ref{sec:BFKLFinite} can be generalized to provide a well-defined dispersion integral describing the multi-Regge limit of the general $N$-particle amplitude, at least in the region where the energies of all produced particles have been analytically continued. Assuming that the factorization of this integral into building blocks in Fourier-Mellin space persists beyond LLA, from the knowledge of the central emission vertex up at N$^k$LO, as described in the previous paragraph, we will thus be able to compute amplitudes with any number of gluons up to N$^k$LLA.

The aforementioned factorization in Fourier-Mellin space also leaves its imprint in momentum space: In \cite{DelDuca:2016lad}, we showed that amplitudes in the multi-Regge limit of $\cN = 4$ SYM decompose at LLA into momentum-space building blocks, naturally associated to amplitudes with fewer legs (see also \cite{Bargheer:2016eyp}).
Using the amplitudes computed in this paper, we observe a generalization of this factorization theorem to hold also at NLLA, allowing some 7-point perturbative coefficients at NLLA to be written in terms of 6-point objects. A full study of this factorization theorem involves going to higher numbers of particles, and a detailed discussion will be presented in future work. {While in the region \eqref{eq:Ucontinuation} the amplitude will always receive contributions from bound states of up to two reggeized gluons independent of the number of external particles $N$, generally there will also exist regions where up to  $\lfloor N/2-1\rfloor$ reggeized gluons contribute \cite{Lipatov:2009nt}. It will thus be very interesting to also study this new qualitative feature, starting with the three-Reggeon state for $N=8$, see also the recent work \cite{Chachamis:2018bys}.}

\section*{Acknowledgements}

The authors are grateful to Benjamin Basso, Simon Caron-Huot and Lance Dixon for useful discussions and for comments on the manuscript.
VDD, CD, FD, GP, RM and BV acknowledge the hospitality and support of the Munich Institute for Astro- and Particle Physics (MIAPP) of the DFG cluster of excellence ``Origin and Structure of the Universe. All authors thank the Higgs Centre for Theoretical Physics at the University of Edinburgh for hospitality during the workshop ``Iterated integrals and the Regge Limit".
CD, RM and BV acknowledge the hospitality of the ETH Zurich at various stages of this
project, and VDD, GP, RM, BV acknowledge the hospitality of the Theory Department at CERN. This
work is supported by the European Research Council (ERC) under the Horizon 2020 Research and
Innovation Programme through the grants 637019 (MathAm) and 648630 (IQFT), and by the U.S. Department of Energy (DOE) under contract  DE-AC02-76SF00515.


\appendix

\section{(Single-valued) $A_2$ polylogarithms}
\subsection{A convenient basis}\label{sec:A2SVPolylogs}
As we have reviewed in the main text, the seven-particle amplitude in MRK is expressible in terms of single-valued $A_2$ polylogarithms. A  generating set for their holomorphic parts may be chosen as
\begin{equation}\label{2dHPLbasis}
\mathcal{L}=\Big\{G_{\vec a}(\rho_1)|a_i\in \{0,1\}\Big\} \cup
\Big\{G_{\vec a}(1/\rho_2)|a_i\in \{0,1,1/\rho_1\}\Big\}\,,
\end{equation}
where the vectors $\vec a$ are Lyndon words. In what follows, it will be convenient to define
\be
\hat{\rho}_2=1/\rho_2\,,
\ee
so that in the variables $(\rho_1,\hat{\rho}_2)$ the symbol alphabet of the above $A_2$ polylogarithms becomes
\be
\label{alphabetrho}
\{ \rho_1 , 1-\rho_1, \hat{\rho}_2 , 1- \hat{\rho}_2 , 1- \rho_1 \hat{\rho}_2\}\,.
\ee
Let us briefly comment on the relation between this basis, and the one already considered in \cite{Bargheer:2015djt}. After taking into account that in the notations of the latter paper,
\be\label{eq:xytorho}
x=1/\hat \rho_2\,,\quad y=1/\rho_1\,,
\ee
it is evident that the two bases are related by the transformation
 $\rho_1\to 1/\rho_1$, $\hat \rho_2\to 1/\hat \rho_2$. Clearly, this transformation preserves the alphabet \eqref{alphabetrho}, so that each element in one of the bases, can be expressed in terms of the other basis. As we will see however, the basis generated by $\mathcal{L}$ is advantageous for taking soft limits.

Given an integrable symbol over the alphabet (\ref{alphabetrho}) we may straightforwardly produce a functional representative in the basis generated by $\mathcal{L}$. This can be done by iteratively integrating the symbol along a contour of our choosing as follows (see e.g. \cite{symbolsC,Brown:2009qja,Bogner:2012dn}). For each letter $\phi_i(\rho_1,\hat{\rho}_2)$ in the symbol we form the one-form $w_i = d \log \phi_i(t_1,t_2)$. Then term by term in the symbol we apply the map
\be
\label{intmap}
\phi_1\otimes \ldots \otimes \phi_n  \mapsto \int_\gamma w_n \circ \ldots \circ w_1\,,
\ee
where the integration is performed iteratively along the contour $\gamma$. We take $\gamma$ to run from the origin $t_1=t_2=0$ along the $t_1$ axis to the point $t_1={\rho}_1$ and then in the $t_2$ direction to the point $(t_1,t_2) = (\rho_1,\hat{\rho}_2)$. This choice of integration contour produces linear combinations of terms of the form
\be
\label{Gform}
G_{\vec{m}_1}({\rho}_1) G_{\vec{m}_2}(\hat{\rho}_2) \,,
\ee
where $\vec{m}_1$ is a word in the letters $\{0,1\}$ and $\vec{m}_2$ is a word in the letters $\{0,1,1/\rho_1\}$, which are the terms generated by $\mathcal{L}$. Although the representation obtained depends on the choice of the contour $\gamma$, the actual function obtained only depends on the choice of base point (in this case the origin in $(\rho_1,\hat{\rho}_2)$ coordinates. This is the statement of homotopy invariance and holds provided our initial symbol is integrable. Different contour choices may be used to generate different representations and hence identities among different polylogarithms.
Applying this procedure to the symbol of the holomorphic part of $g$ in (\ref{eq:R72}) produces the expression (\ref{eq:gfunction}), but with each $\mathcal{G}$ replaced by the holomorphic part $G$. Then (\ref{eq:gfunction}) is obtained simply replacing each holomorphic $G$ with the single-valued $\mathcal{G}$. Note that this is not yet a derivation of eq. (\ref{eq:gfunction}) because it remains to show that there are no beyond-the-symbol terms to be added to the expression (\ref{eq:gfunction}). This is determined through analysis of the behaviour of potential beyond-the-symbol terms in the various soft limits, as described in section \ref{sec:R72MRK}.

Since every multiple polylogarithm can be uniquely completed to its single-valued analogue with the help of the map $\bf s$, as we also reviewed in subsection \ref{sec:SVMPL}\footnote{For the 2-loop analysis of section \ref{sec:R72MRK}, the examples of single-valued completions \eqref{eq:ExampleSVMPL} are in fact sufficient.}, we obtain a generating set of SVMPLs $\cL_{SV}\equiv\text{\bf s}\left(\cL\right)$. To parametrize potential beyond-the-symbol terms we then form a complete basis of (single-valued) MPLs by adding to $\cL$ ($\cL_{SV}$) all distinct products of lower-weight functions from the same generating
set, and with the same total weight. It is easy to show that this basis has dimension $3^{n+1}-2^{n+1}$ at weight $n$, also illustrated in the first line of table \ref{tab:t1} through weight 5.
\renewcommand{\arraystretch}{1.25}
\begin{table}[!ht]
\begin{center}
\begin{tabular}{|l|>{\hfill}p{1.46cm}|>{\hfill}p{1.46cm}|>{\hfill}p{1.46cm}|>{\hfill}p{1.46cm}|>{\hfill}p{1.46cm}|>{\hfill}p{1.46cm}|}
\hline
\multicolumn{1}{|c|}{$~$ \hfill Weight $k=$}
&\multicolumn{1}{c|}{$1$}
&\multicolumn{1}{c|}{$2$}
&\multicolumn{1}{c|}{$3$}
&\multicolumn{1}{c|}{$4$}
&\multicolumn{1}{c|}{$5$}\\
\hline\hline
$A_2\times A_2$  SVMPL & 5& 19& 65& 211 & 665\\
\hline
Parity& 5 & 15 & 45 &130 & 326\\
\hline
Flip symmetry &3 & 9 & 25 & 70 & 170\\
\hline\hline
Including MZV &3 & 10 & 29 & 83 & 209\\
\hline
\end{tabular}
\end{center}
\caption{Dimensionality of basis of SVMPL relevant for seven-particle scattering in MRK, sequentially reduced by imposing the discrete symmetries of the MHV remainder function.} \label{tab:t1}
\end{table}

Next, we may impose the discrete symmetries of the seven-particle amplitude, in order to further reduce this
basis. As we reviewed in section \ref{sec:SymmetriesLimits}, the remainder function is parity
invariant, which in MRK translates to invariance under the exchange of barred and unbarred
variables, \eqref{eq:Parityzrho}, which is equivalent to demanding that the basis
functions can be written as real parts of single-valued polylogarithms. Imposing this on an ansatz of functions from our original basis, we arrive at the second line of table \ref{tab:t1}. As may be seen from \eqref{eq:ExampleSVMPL}, this also requires using shuffle identities in order to express the antiholomorphic $G$-functions in a Lyndon basis.

Finally, as we also reviewed in subsection \ref{sec:SymmetriesLimits}, the $\cO(\log^0 \tau_i)$ term of the remainder function will be invariant under target-projectile symmetry \eqref{eq:TargetProjectileUzrho}. Particularly at two loops, discussed in subsection \ref{sec:R72MRK}, the symmetries of $f, \tilde f$, \eqref{eq:fftilde}, as well as~\eqref{eq:R72}, implies that the function $g$ defined there should separately obey
\be
g(\rho_1,\rho_2)=g(1/\rho_2,1/\rho_1)\,.
\ee
Enforcing the target-projectile symmetry requires knowledge of how the elements of $\cL$
($\cL_{SV}$) get mapped to the same Lyndon basis, after we act with the transformation
$\rho_1\leftrightarrow \hat \rho_2$, that also preserves the alphabet~\eqref{alphabetrho}, on them. These identities
can be constructed using coproduct techniques~\cite{Duhr:2012fh,Anastasiou:2013srw}, as reviewed in appendix \ref{appx:A2Inversions}.

In this manner, we derive a basis of relevant SVMPL respecting both parity and target-projectile symmetry, whose dimensionality can be found in the third line of table \ref{tab:t1}. Finally, assuming that the only transcendental numbers appearing on the beyond-the-symbol terms are multiple zeta values (MZV), we may tensor a complete basis thereof with functions of lower weight in order to arrive at the fourth line of the table. From this complete basis, we form an ansatz for the beyond-the-symbol terms of the function $g$, or more generally its higher loop analogues. One could also arrive at (\ref{eq:gfunction}) via matching the symbol to an ansatz of the above functions at weight three.

Since one of the most important constraints for fixing the coefficients of our ansatz are the soft limits $\rho_1\to 0, \rho_2\to\infty$ and  $\rho_2\to \rho_1$, let us close this section by briefly mentioning how our basis elements behave under these limits. The first limit is very easy to take on our Lyndon basis \eqref{2dHPLbasis}, since all elements except $G_{0}(\rho_1)$ will vanish as $\rho_1\to 0$. The second limit is related to the first one by target-projectile symmetry, so we need not consider it, since our basis already respects this symmetry. Finally, in the third limit $\rho_1\to \rho_2=1/\hat \rho_2$ we may reduce our Lyndon basis \eqref{2dHPLbasis} to a basis of harmonic polylogarithms (HPL) $G_{\vec a}(\hat \rho_2)$, $\vec a\in\{0,1\}$. This can be done either through HPL inversion identities, or by relating this limit to the $\hat \rho_2\to 1$ limit with the help of the property
\begin{equation}\label{Grescale}
G_{a_1,\dots,a_k}(z)
=
G_{ x a_1,\dots,x a_k}( x z)\,.
\end{equation}
for $a_k\ne0$ and $\hat \rho_2\in\mathbb{C}^*$, as can be seen in the following example,
\be
G_{0,1,\hat \rho_2}(\hat \rho_2)=G_{0,1/\hat \rho_2,1}(1)\,.
\ee
We may then use the identities discussed in appendix \ref{appx:A2Inversions}, in order to reduce the left-hand side to HPLs.

\subsection{Exchange identities}
\label{appx:A2Inversions}
Here we discuss in more detail the $\rho_1\leftrightarrow \hat \rho_2$ exchange identities of the basis $A_2$ polylogarithms we constructed in appendix \ref{sec:A2SVPolylogs}.

Up to weight 3, the only nontrivial identity needed is
\begin{equation}
G_{1,1/\hat \rho_2}(\rho_1)=G_{1}(\hat \rho_2) G_{1}(\rho_1)+G_{0,1/\rho_1}(\hat \rho_2)-G_{1,1/\rho_1}(\hat \rho_2)\,,
\end{equation}
while the rest are consequences of the property \eqref{Grescale}.

At higher weights, we may derive exchange identities following the coproduct approach of~\cite{Duhr:2012fh,Anastasiou:2013srw}, see also the review \cite{Duhr:2014woa}, as well as \cite{Dixon:2013eka} for similar applications to $A_3$ polylogarithm identities. In other words, if we know all identities at weight $k-1$, we may equate a transformed weight $k$ function with an ansatz in terms of the complete basis at this weight, and take the differential, or ${k-1,1}$ coproduct on both sides of the equation. Expressing the weight $k-1$ functions in terms of products of irreducibles and using the weight $k-1$ identities for the latter, we may thus determine all free parameters of our ansatz, given that the coefficients of each independent element of our basis should be separately equal to each other on the left- and right-hand side.

This recursive procedure starts with the known weight-1 identities between logarithms, and in principle at each step we also need to determine a constant term, since we are equating the differentials of two quantities. By taking the $\hat \rho_2\to 0$ limit, it is easy to show that this constant should always be zero, as a consequence of the fact that with the exception of $G(0;\hat \rho_2)$, all functions of the original and transformed Lyndon basis vanish at that point. As an example, we list all nontrivial exchange identities at weight 4 below.
\begin{align}
G_{0,1,1/\hat \rho_2}(\rho_1)=& G_{1}(\hat \rho_2) G_{0,1}(\rho_1)-G_{1}(\hat \rho_2) G_{0,1/\rho_1}(\hat \rho_2)+G_{0,0,1/\rho_1}(\hat \rho_2)\nonumber\\
&+G_{0,1,1/\rho_1}(\hat \rho_2)+G_{0,1/\rho_1,1}(\hat \rho_2)\nonumber\\
G_{0,1/\hat \rho_2,1}(\rho_1)=& G_{1}(\rho_1) G_{0,1/\rho_1}(\hat \rho_2)+G_{0,1}(\hat \rho_2) (-G_{1}(\rho_1))-G_{1}(\hat \rho_2) G_{0,1}(\rho_1)\nonumber\\
&+G_{1}(\hat \rho_2) G_{0,1/\rho_1}(\hat \rho_2)-2 G_{0,0,1/\rho_1}(\hat \rho_2)-G_{0,1/\rho_1,1}(\hat \rho_2)\\
G_{1,1,1/\hat \rho_2}(\rho_1)=& -\frac{1}{2} G_{1}(\hat \rho_2)^2 G_{1}(\rho_1)+\frac{1}{2} G_{1}(\hat \rho_2) G_{1}(\rho_1)^2+G_{0,1}(\hat \rho_2) G_{1}(\rho_1)\nonumber\\
&-G_{1}(\hat \rho_2) G_{0,1/\rho_1}(\hat \rho_2)+G_{0,0,1/\rho_1}(\hat \rho_2)+G_{0,1/\rho_1,1}(\hat \rho_2)+G_{1,1,1/\rho_1}(\hat \rho_2)\nonumber\\
G_{1,1/\hat \rho_2,1/\hat \rho_2}(\rho_1)=& \frac{1}{2} G_{1}(\hat \rho_2)^2 G_{1}(\rho_1)+G_{1}(\hat \rho_2) G_{0,1/\rho_1}(\hat \rho_2)-G_{0,1,1/\rho_1}(\hat \rho_2)\nonumber\\
&-G_{0,1/\rho_1,1}(\hat \rho_2)+G_{0,1/\rho_1,1/\rho_1}(\hat \rho_2)-G_{1,1,1/\rho_1}(\hat \rho_2)-G_{1,1/\rho_1,1/\rho_1}(\hat \rho_2)\nonumber
\end{align}
The above identities can also be thought of as an expression of homotopy invariance. The equivalence between the LHS and RHS in the identities above can be shown by simply shifting the contour of integration from being first along the $\hat \rho_2$ axis then in the $\rho_1$ direction to being first along the $\rho_1$-axis then in the $\hat \rho_2$ direction when applying the map (\ref{intmap}) (this time using the $(\hat \rho_2,\rho_1)$ variables) to the corresponding symbol. The integrability of the symbol guarantees the equality of the two results.

\section{Explicit Results}\label{sec:ExplicitResults}

In this section, we are presenting explicitly the NLLA perturbative coefficients $\tilde{g}$ and $\tilde{h}$ at two loops both in the MHV case and the NMHV case for the helicity configurations $-++$ and $-+-$. Together with target-projectile symmetry and conjugation, these span the full set of two-loop perturbative coefficients at NLLA in all helicity configurations. In the following, for compactness of the results, we will use the notation
\begin{equation}
\cG_{\vec{a}}^i  \equiv \cG_{\vec{a}}(\rho_i) \,.
\end{equation}
The leading singularities $R_{bac}$ introduced in \cite{DelDuca:2016lad} and correspond to
\begin{align}
R_{234} &= \frac{\rho_1 (\rho_2-1)}{(\rho_1-1) \rho_2}
&R_{235} &= \frac{\rho_1}{\rho_1-1}
&R_{345} &= \frac{\rho_2-\rho_1}{\rho_2-1} \,.
\end{align}

\begin{align}
\tilde{g}_{+++}^{(0,0)} \left( \rho_1, \rho_2 \right) = &\frac{1}{4}\cG_{0,0,1}^1+\frac{1}{4}\cG_{0,0,1}^2-\frac{1}{4}\cG_{0,1,0}^1-\frac{1}{4}\cG_{0,1,0}^2-\frac{1}{4}\cG_{0,1,1}^1-\frac{1}{4}\cG_{0,1,1}^2-\frac{1}{8}\cG_{1,0}^2\cG_{\rho_2}^1\\
 \nonumber &+\frac{1}{8}\cG_{0,1,\rho_2}^1-\frac{1}{8}\cG_{0,\rho_2,1}^1+\frac{1}{4}\cG_{1,0,0}^1+\frac{1}{4}\cG_{1,0,0}^2-\frac{1}{4}\cG_{1,0,1}^1-\frac{1}{4}\cG_{1,0,1}^2+\frac{1}{8}\cG_1^1\cG_{1,0}^2\\
 \nonumber &-\frac{1}{8}\cG_{1,0,\rho_2}^1-\frac{1}{4}\cG_{1,1,0}^1-\frac{1}{4}\cG_{1,1,0}^2+\frac{1}{2}\cG_{1,1,1}^1+\frac{1}{2}\cG_{1,1,1}^2-\frac{1}{8}\cG_{1,\rho_2,0}^1+\frac{1}{8}\cG_1^2\cG_{1,0}^1\\
 \nonumber &+\frac{1}{4}\cG_{1,\rho_2,1}^1-\frac{1}{8}\cG_{\rho_2,0,1}^1+\frac{1}{8}\cG_{\rho_2,1,0}^1-\frac{1}{4}\zeta_2\cG_0^1+\frac{1}{4}\zeta_2\cG_1^1+\frac{1}{4}\zeta_2\cG_1^2-\frac{1}{8}\cG_0^2\cG_{1,0}^1\\
 \nonumber &+\frac{1}{8}\cG_0^2\cG_{0,1}^1-\frac{1}{8}\cG_1^2\cG_{0,1}^1-\frac{1}{8}\cG_1^1\cG_{0,1}^2+\frac{1}{8}\cG_{0,1}^2\cG_{\rho_2}^1-\frac{1}{4}\cG_1^2\cG_{1,\rho_2}^1+\frac{1}{4}\cG_1^2\cG_{0,\rho_2}^1
\end{align}
\begin{align}
\tilde{h}_{+++}^{(0,0)} \left( \rho_1, \rho_2 \right) = &\frac{1}{16}\cG_{0,0}^1-\frac{1}{16}\cG_{0,1}^1-\frac{1}{16}\cG_{1,0}^1+\frac{1}{16}\cG_{1,1}^1+\frac{1}{16}\cG_{1,1}^2-\frac{1}{16}\cG_0^1\cG_1^2+\frac{1}{16}\cG_1^1\cG_1^2
\end{align}

\beq\bsp
\tilde{g}_{-++}^{(0,0)} \left(\rho_1,\rho_2\right) = &\fa^{\text{(0,0)}}_{-++}\left(\rho_1,\rho_2\right)+R_{234}\fb_{1,-++}^{\text{(0,0)}}\left(\rho_1,\rho_2\right)+R_{235}\fb_{2,-++}^{\text{(0,0)}}\left(\rho_1,\rho_2\right)
\esp\eeq
\begin{align}
\fa^{\text{(0,0)}}_{-++}\left(\rho_1,\rho_2\right) = &\frac{1}{4}\cG_{0,0,1}^2+\frac{1}{8}\cG_{0,1,0}^1-\frac{1}{4}\cG_{0,1,0}^2-\frac{1}{4}\cG_{0,1,1}^2-\frac{1}{4}\cG_{1,0,0}^1+\frac{1}{4}\cG_{1,0,0}^2-\frac{1}{4}\cG_{1,0,1}^2\\
 \nonumber &+\frac{1}{8}\cG_{1,0,\rho_2}^1-\frac{1}{4}\cG_{1,1,0}^1-\frac{1}{4}\cG_{1,1,0}^2+\frac{1}{2}\cG_{1,1,1}^2-\frac{1}{4}\cG_{1,1,\rho_2}^1+\frac{1}{8}\cG_{1,\rho_2,0}^1-\frac{1}{4}\cG_{1,\rho_2,\rho_2}^1\\
 \nonumber &-\frac{1}{4}\cG_{\rho_2,0,1}^1-\frac{1}{8}\cG_{\rho_2,1,0}^1-\frac{1}{4}\cG_{\rho_2,1,1}^1+\frac{1}{4}\cG_{\rho_2,1,\rho_2}^1-\frac{1}{4}\zeta_2\cG_0^1-\frac{1}{4}\zeta_2\cG_1^1+\frac{1}{4}\zeta_2\cG_1^2\\
 \nonumber &-\frac{1}{4}\cG_1^1\cG_{0,0}^2+\frac{1}{4}\cG_1^1\cG_{0,1}^2+\frac{1}{8}\cG_0^2\cG_{1,0}^1-\frac{1}{8}\cG_1^2\cG_{1,0}^1+\frac{1}{2}\cG_1^1\cG_{1,0}^2-\frac{1}{4}\cG_{1,0}^2\cG_{\rho_2}^1\\
 \nonumber &-\frac{1}{4}\cG_0^2\cG_{1,1}^1+\frac{1}{4}\cG_1^2\cG_{1,1}^1-\frac{1}{2}\cG_1^1\cG_{1,1}^2-\frac{1}{4}\cG_0^2\cG_{1,\rho_2}^1+\frac{1}{4}\cG_1^2\cG_{1,\rho_2}^1+\frac{1}{4}\cG_1^2\cG_{\rho_2,0}^1\\
 \nonumber &+\frac{1}{4}\cG_0^2\cG_{\rho_2,1}^1
\end{align}
\begin{align}
\fb_{1,-++}^{\text{(0,0)}}\left(\rho_1,\rho_2\right) = &-\frac{1}{4}\cG_{0,0,1}^2+\frac{1}{8}\cG_{0,1,0}^2-\frac{1}{4}\cG_{0,1,1}^1+\frac{3}{8}\cG_{0,1,1}^2+\frac{1}{8}\cG_{0,1,\rho_2}^1-\frac{1}{8}\cG_{0,\rho_2,1}^1\\
 \nonumber &+\frac{1}{8}\cG_{1,0,1}^2-\frac{1}{8}\cG_{1,1,0}^2-\frac{1}{4}\cG_{1,1,1}^2+\frac{1}{8}\cG_{1,\rho_2,1}^1+\frac{1}{8}\cG_{\rho_2,0,1}^1+\frac{1}{4}\cG_{\rho_2,1,1}^1-\frac{1}{8}\cG_{\rho_2,1,\rho_2}^1\\
 \nonumber &+\frac{1}{8}\cG_0^2\cG_{0,1}^1+\frac{1}{8}\cG_1^2\cG_{0,1}^1+\frac{1}{8}\cG_0^1\cG_{0,1}^2-\frac{1}{8}\cG_1^1\cG_{0,1}^2+\frac{1}{8}\cG_1^2\cG_{0,\rho_2}^1+\frac{1}{8}\cG_1^2\cG_{1,0}^1\\
 \nonumber &-\frac{1}{8}\cG_0^1\cG_{1,0}^2+\frac{1}{8}\cG_{1,0}^2\cG_{\rho_2}^1-\frac{1}{4}\cG_0^1\cG_{1,1}^2+\frac{1}{8}\cG_1^1\cG_{1,1}^2+\frac{1}{8}\cG_{1,1}^2\cG_{\rho_2}^1-\frac{1}{8}\cG_1^2\cG_{1,\rho_2}^1\\
 \nonumber &-\frac{1}{8}\cG_1^2\cG_{\rho_2,0}^1-\frac{1}{8}\cG_0^2\cG_{\rho_2,1}^1-\frac{1}{8}\cG_1^2\cG_{\rho_2,1}^1-\frac{1}{8}\cG_{1,0,1}^1
\end{align}
\begin{align}
\fb_{2,-++}^{\text{(0,0)}}\left(\rho_1,\rho_2\right) = &\frac{3}{8}\cG_{0,0,0}^1-\frac{3}{8}\cG_{0,0,0}^2+\frac{1}{4}\cG_{0,0,1}^2-\frac{1}{8}\cG_{0,0,\rho_2}^1-\frac{1}{4}\cG_{0,1,0}^1+\frac{3}{8}\cG_{0,1,0}^2+\frac{1}{4}\cG_{0,1,1}^1\\
 \nonumber &-\frac{1}{4}\cG_{0,1,1}^2-\frac{1}{8}\cG_{0,\rho_2,0}^1+\frac{1}{4}\cG_{0,\rho_2,\rho_2}^1-\frac{1}{4}\cG_{1,0,0}^1+\frac{1}{4}\cG_{1,0,0}^2+\frac{1}{8}\cG_{1,0,1}^1-\frac{1}{4}\cG_{1,0,1}^2\\
 \nonumber &+\frac{1}{8}\cG_{1,0,\rho_2}^1+\frac{1}{2}\cG_{1,1,0}^1+\frac{1}{8}\cG_{\rho_2,0,0}^1+\frac{1}{8}\cG_{\rho_2,0,1}^1-\frac{1}{8}\cG_{\rho_2,0,\rho_2}^1+\frac{1}{2}\zeta_2\cG_0^1-\zeta_3\\
 \nonumber &-\frac{1}{8}\cG_0^2\cG_{0,0}^1+\frac{1}{8}\cG_1^2\cG_{0,0}^1+\frac{1}{4}\cG_0^1\cG_{0,0}^2-\frac{1}{8}\cG_1^1\cG_{0,0}^2+\frac{1}{8}\cG_{0,0}^2\cG_{\rho_2}^1-\frac{1}{4}\cG_1^2\cG_{0,1}^1\\
 \nonumber &-\frac{1}{4}\cG_0^1\cG_{0,1}^2+\frac{1}{8}\cG_1^1\cG_{0,1}^2-\frac{1}{8}\cG_{0,1}^2\cG_{\rho_2}^1+\frac{1}{4}\cG_0^2\cG_{0,\rho_2}^1-\frac{1}{4}\cG_1^2\cG_{0,\rho_2}^1+\frac{1}{8}\cG_0^2\cG_{1,0}^1\\
 \nonumber &-\frac{1}{4}\cG_1^2\cG_{1,0}^1-\frac{1}{4}\cG_0^1\cG_{1,0}^2+\frac{1}{2}\cG_0^1\cG_{1,1}^2-\frac{1}{8}\cG_0^2\cG_{\rho_2,0}^1
\end{align}
\beq\bsp
\tilde{h}_{-++}^{(0,0)} \left(\rho_1,\rho_2\right) = &\fd^{\text{(0,0)}}_{-++}\left(\rho_1,\rho_2\right)+R_{234}\fe_{1,-++}^{\text{(0,0)}}\left(\rho_1,\rho_2\right)+R_{235}\fe_{2,-++}^{\text{(0,0)}}\left(\rho_1,\rho_2\right)
\esp\eeq
\begin{align}
\fd^{\text{(0,0)}}_{-++}\left(\rho_1,\rho_2\right) = &\frac{1}{16}\cG_{0,0}^1+\frac{1}{16}\cG_{0,1}^1-\frac{1}{16}\cG_{1,0}^1+\frac{1}{16}\cG_{1,1}^1+\frac{1}{16}\cG_{1,1}^2-\frac{1}{16}\cG_0^1\cG_1^2-\frac{1}{16}\cG_1^1\cG_1^2
\end{align}
\begin{align}
\fe_{1,-++}^{\text{(0,0)}}\left(\rho_1,\rho_2\right) = &0
\end{align}
\begin{align}
\fe_{2,-++}^{\text{(0,0)}}\left(\rho_1,\rho_2\right) = &-\frac{1}{8}\cG_{0,0}^1-\frac{1}{8}\cG_{0,1}^1+\frac{1}{8}\cG_{1,0}^1+\frac{1}{8}\cG_0^1\cG_1^2
\end{align}
\beq\bsp
\tilde{g}_{-+-}^{(0,0)} \left(\rho_1,\rho_2\right) = &\fa^{\text{(0,0)}}_{-+-}\left(\rho_1,\rho_2\right)+R_{234}\fb_{1,-+-}^{\text{(0,0)}}\left(\rho_1,\rho_2\right)+\bar{R}_{345}\fb_{2,-+-}^{\text{(0,0)}}\left(\rho_1,\rho_2\right)\\
&+R_{234}\bar{R}_{345}\fc_{1,-+-}^{\text{(0,0)}}\left(\rho_1,\rho_2\right)
\esp\eeq
\begin{align}
\fa^{\text{(0,0)}}_{-+-}\left(\rho_1,\rho_2\right) = &-\frac{1}{4}\cG_{0,0,1}^2+\frac{1}{8}\cG_{0,1,0}^2-\frac{1}{8}\cG_{0,1,1}^1-\frac{1}{4}\cG_{0,1,1}^2+\frac{1}{8}\cG_{0,\rho_2,1}^1+\frac{1}{8}\cG_{1,0,1}^1\\
 \nonumber &+\frac{5}{8}\cG_{1,1,1}^1-\frac{3}{8}\cG_{1,\rho_2,1}^1-\frac{1}{8}\cG_{\rho_2,0,1}^1-\frac{1}{8}\cG_{\rho_2,1,0}^1-\frac{1}{2}\cG_{\rho_2,1,1}^1+\frac{1}{4}\cG_{\rho_2,\rho_2,1}^1-\frac{1}{4}\zeta_2\cG_0^1\\
 \nonumber &+\frac{1}{4}\zeta_2\cG_1^1-\frac{1}{4}\zeta_2\cG_1^2+\frac{1}{8}\cG_{1,1,0}^1+\frac{1}{2}\cG_1^2\cG_{\rho_2,1}^1-\frac{1}{4}\cG_1^2\cG_{\rho_2,\rho_2}^1+\frac{3}{8}\cG_1^2\cG_{1,\rho_2}^1+\frac{1}{8}\cG_1^2\cG_{\rho_2,0}^1\\
 \nonumber &+\frac{1}{8}\cG_1^2\cG_{0,1}^1+\frac{1}{8}\cG_1^1\cG_{0,1}^2-\frac{1}{8}\cG_{0,1}^2\cG_{\rho_2}^1-\frac{1}{8}\cG_1^2\cG_{0,\rho_2}^1-\frac{1}{8}\cG_1^2\cG_{1,0}^1-\frac{5}{8}\cG_1^2\cG_{1,1}^1\\
 \nonumber &+\frac{1}{4}\cG_1^1\cG_{1,1}^2-\frac{1}{4}\cG_{1,1}^2\cG_{\rho_2}^1
\end{align}
\begin{align}
\fb_{1,-+-}^{\text{(0,0)}}\left(\rho_1,\rho_2\right) = &\frac{1}{4}\cG_{0,0,1}^1+\frac{3}{4}\cG_{0,0,1}^2-\frac{1}{8}\cG_{0,1,0}^1-\frac{1}{8}\cG_{0,1,0}^2-\frac{1}{2}\cG_{0,1,1}^1-\frac{1}{2}\cG_{0,1,1}^2\\
 \nonumber &-\frac{3}{8}\cG_{1,0,1}^1-\frac{3}{8}\cG_{1,0,1}^2+\frac{5}{8}\cG_{1,1,1}^2+\frac{3}{8}\cG_{1,\rho_2,1}^1+\frac{1}{8}\cG_{\rho_2,0,1}^1+\frac{1}{8}\cG_{\rho_2,1,0}^1\\
 \nonumber &-\frac{1}{4}\cG_{\rho_2,\rho_2,1}^1+\frac{1}{2}\zeta_2\cG_1^2-\frac{1}{8}\cG_{0,\rho_2,1}^1-\frac{1}{2}\cG_1^2\cG_{\rho_2,1}^1+\frac{1}{4}\cG_1^2\cG_{\rho_2,\rho_2}^1+\frac{1}{2}\cG_{\rho_2,1,1}^1\\
 \nonumber &+\frac{1}{4}\cG_1^2\cG_{0,0}^1-\frac{3}{8}\cG_0^1\cG_{0,1}^2+\frac{1}{4}\cG_1^1\cG_{0,1}^2+\frac{1}{8}\cG_{0,1}^2\cG_{\rho_2}^1+\frac{1}{8}\cG_1^2\cG_{0,\rho_2}^1-\frac{1}{8}\cG_1^2\cG_{1,0}^1\\
 \nonumber &+\frac{1}{2}\cG_1^2\cG_{1,1}^1+\frac{1}{8}\cG_0^1\cG_{1,1}^2-\frac{3}{8}\cG_1^1\cG_{1,1}^2+\frac{1}{4}\cG_{1,1}^2\cG_{\rho_2}^1-\frac{3}{8}\cG_1^2\cG_{1,\rho_2}^1-\frac{1}{8}\cG_1^2\cG_{\rho_2,0}^1
\end{align}
\begin{align}
\fb_{2,-+-}^{\text{(0,0)}}\left(\rho_1,\rho_2\right) = &\frac{3}{8}\cG_{0,0,0}^2-\frac{1}{4}\cG_{0,0,1}^2+\frac{1}{8}\cG_{0,1,0}^1-\frac{1}{4}\cG_{0,1,0}^2+\frac{1}{8}\cG_{0,1,1}^1+\frac{1}{2}\cG_{0,1,1}^2\\
 \nonumber &-\frac{1}{8}\cG_{0,\rho_2,0}^1-\frac{1}{8}\cG_{0,\rho_2,1}^1+\frac{1}{8}\cG_{0,\rho_2,\rho_2}^1-\frac{1}{4}\cG_{1,0,0}^1-\frac{1}{8}\cG_{1,0,1}^1-\frac{3}{8}\cG_{1,1,0}^1-\frac{5}{8}\cG_{1,1,1}^1\\
 \nonumber &+\frac{1}{4}\cG_{1,1,\rho_2}^1+\frac{1}{4}\cG_{1,\rho_2,0}^1+\frac{1}{4}\cG_{1,\rho_2,1}^1-\frac{3}{8}\cG_{1,\rho_2,\rho_2}^1+\frac{1}{4}\cG_{\rho_2,0,0}^1+\frac{1}{8}\cG_{\rho_2,0,1}^1\\
 \nonumber &+\frac{1}{2}\cG_{\rho_2,1,1}^1-\frac{1}{8}\cG_{\rho_2,1,\rho_2}^1-\frac{1}{8}\cG_{\rho_2,\rho_2,0}^1-\frac{1}{8}\cG_{\rho_2,\rho_2,1}^1+\frac{1}{4}\cG_{\rho_2,\rho_2,\rho_2}^1+\frac{1}{2}\zeta_2\cG_0^2\\
 \nonumber &+\frac{1}{2}\zeta_2\cG_{\rho_2}^1-\frac{\zeta_3}{2}-\frac{1}{8}\cG_{0,1,\rho_2}^1-\frac{1}{2}\zeta_2\cG_1^1+\frac{1}{4}\cG_{\rho_2,1,0}^1+\frac{1}{4}\cG_0^2\cG_{\rho_2,\rho_2}^1-\frac{1}{8}\cG_1^2\cG_{\rho_2,\rho_2}^1\\
 \nonumber &-\frac{1}{8}\cG_1^1\cG_{0,0}^2+\frac{1}{8}\cG_{0,0}^2\cG_{\rho_2}^1-\frac{1}{8}\cG_0^2\cG_{0,1}^1+\frac{1}{8}\cG_1^1\cG_{0,1}^2-\frac{1}{8}\cG_{0,1}^2\cG_{\rho_2}^1+\frac{1}{8}\cG_0^2\cG_{0,\rho_2}^1\\
 \nonumber &+\frac{1}{8}\cG_1^2\cG_{1,0}^1+\frac{1}{8}\cG_1^1\cG_{1,0}^2-\frac{1}{8}\cG_{1,0}^2\cG_{\rho_2}^1+\frac{1}{4}\cG_0^2\cG_{1,1}^1+\frac{3}{8}\cG_1^2\cG_{1,1}^1-\frac{1}{2}\cG_1^1\cG_{1,1}^2\\
 \nonumber &+\frac{1}{2}\cG_{1,1}^2\cG_{\rho_2}^1-\frac{3}{8}\cG_0^2\cG_{1,\rho_2}^1+\frac{1}{8}\cG_1^2\cG_{1,\rho_2}^1-\frac{1}{8}\cG_1^2\cG_{\rho_2,0}^1-\frac{1}{8}\cG_0^2\cG_{\rho_2,1}^1-\frac{3}{8}\cG_1^2\cG_{\rho_2,1}^1
\end{align}
\begin{align}
\fc_{1,-+-}^{\text{(0,0)}}\left(\rho_1,\rho_2\right) = &\frac{3}{8}\cG_{0,0,0}^1-\frac{9}{8}\cG_{0,0,0}^2-\frac{1}{4}\cG_{0,0,1}^1+\frac{5}{8}\cG_{0,0,1}^2-\frac{3}{8}\cG_{0,0,\rho_2}^1-\frac{1}{8}\cG_{0,1,0}^1\\
 \nonumber &+\frac{1}{2}\cG_{0,1,1}^1-\frac{1}{2}\cG_{0,1,1}^2+\frac{3}{8}\cG_{0,1,\rho_2}^1+\frac{1}{8}\cG_{0,\rho_2,0}^1+\frac{1}{4}\cG_{0,\rho_2,1}^1+\frac{1}{8}\cG_{0,\rho_2,\rho_2}^1-\frac{1}{4}\cG_{1,0,0}^1\\
 \nonumber &+\frac{5}{8}\cG_{1,0,0}^2+\frac{3}{8}\cG_{1,0,1}^1-\frac{3}{8}\cG_{1,0,1}^2+\frac{3}{8}\cG_{1,0,\rho_2}^1+\frac{1}{2}\cG_{1,1,0}^1-\frac{1}{2}\cG_{1,1,0}^2-\frac{1}{2}\cG_{1,1,\rho_2}^1\\
 \nonumber &-\frac{1}{4}\cG_{1,\rho_2,0}^1-\frac{3}{8}\cG_{1,\rho_2,1}^1+\frac{1}{8}\cG_{1,\rho_2,\rho_2}^1-\frac{1}{8}\cG_{\rho_2,0,0}^1-\frac{1}{8}\cG_{\rho_2,0,1}^1-\frac{3}{8}\cG_{\rho_2,1,0}^1\\
 \nonumber &+\frac{1}{8}\cG_{\rho_2,1,\rho_2}^1+\frac{1}{8}\cG_{\rho_2,\rho_2,0}^1+\frac{1}{8}\cG_{\rho_2,\rho_2,1}^1-\frac{1}{4}\cG_{\rho_2,\rho_2,\rho_2}^1+\frac{1}{2}\zeta_2\cG_0^1-\frac{1}{2}\zeta_2\cG_0^2\\
 \nonumber &+\frac{7}{8}\cG_{0,1,0}^2-\frac{1}{2}\zeta_2\cG_{\rho_2}^1-\frac{1}{2}\cG_{\rho_2,1,1}^1-\frac{\zeta_3}{2}+\frac{3}{8}\cG_1^2\cG_{\rho_2,1}^1-\frac{1}{4}\cG_0^2\cG_{\rho_2,\rho_2}^1+\frac{1}{8}\cG_1^2\cG_{\rho_2,\rho_2}^1\\
 \nonumber &-\frac{3}{8}\cG_0^2\cG_{0,0}^1+\frac{1}{8}\cG_1^2\cG_{0,0}^1+\frac{5}{8}\cG_0^1\cG_{0,0}^2-\frac{1}{2}\cG_1^1\cG_{0,0}^2-\frac{1}{8}\cG_{0,0}^2\cG_{\rho_2}^1+\frac{3}{8}\cG_0^2\cG_{0,1}^1\\
 \nonumber &-\frac{3}{8}\cG_1^2\cG_{0,1}^1-\frac{3}{8}\cG_0^1\cG_{0,1}^2+\frac{1}{4}\cG_1^1\cG_{0,1}^2+\frac{1}{8}\cG_{0,1}^2\cG_{\rho_2}^1+\frac{1}{8}\cG_0^2\cG_{0,\rho_2}^1-\frac{3}{8}\cG_1^2\cG_{0,\rho_2}^1\\
 \nonumber &+\frac{3}{8}\cG_0^2\cG_{1,0}^1-\frac{1}{4}\cG_1^2\cG_{1,0}^1-\frac{1}{2}\cG_0^1\cG_{1,0}^2+\frac{3}{8}\cG_1^1\cG_{1,0}^2+\frac{1}{8}\cG_{1,0}^2\cG_{\rho_2}^1-\frac{1}{2}\cG_0^2\cG_{1,1}^1\\
 \nonumber &+\frac{1}{2}\cG_0^1\cG_{1,1}^2-\frac{1}{2}\cG_{1,1}^2\cG_{\rho_2}^1+\frac{1}{8}\cG_0^2\cG_{1,\rho_2}^1+\frac{1}{4}\cG_1^2\cG_{1,\rho_2}^1+\frac{1}{8}\cG_1^2\cG_{\rho_2,0}^1+\frac{1}{8}\cG_0^2\cG_{\rho_2,1}^1
\end{align}
\beq\bsp
\tilde{h}_{-+-}^{(0,0)} \left(\rho_1,\rho_2\right) = &\fd^{\text{(0,0)}}_{-+-}\left(\rho_1,\rho_2\right)+R_{234}\fe_{1,-+-}^{\text{(0,0)}}\left(\rho_1,\rho_2\right)+\bar{R}_{345}\fe_{2,-+-}^{\text{(0,0)}}\left(\rho_1,\rho_2\right) \\
\nonumber & +R_{234}\bar{R}_{345}\ff_{1,-+-}^{\text{(0,0)}}\left(\rho_1,\rho_2\right)
\esp\eeq
\begin{align}
\fd^{\text{(0,0)}}_{-+-}\left(\rho_1,\rho_2\right) = &\frac{1}{16}\cG_{0,0}^1-\frac{1}{16}\cG_{0,1}^1-\frac{1}{8}\cG_{0,1}^2-\frac{1}{16}\cG_{1,0}^1+\frac{1}{16}\cG_{1,1}^1 \\
\nonumber &+\frac{1}{16}\cG_{1,1}^2+\frac{1}{16}\cG_0^1\cG_1^2-\frac{1}{16}\cG_1^1\cG_1^2
\end{align}
\begin{align}
\fe_{1,-+-}^{\text{(0,0)}}\left(\rho_1,\rho_2\right) = &\frac{1}{8}\cG_{0,1}^2-\frac{1}{8}\cG_0^1\cG_1^2+\frac{1}{8}\cG_1^1\cG_1^2
\end{align}
\begin{align}
\fe_{2,-+-}^{\text{(0,0)}}\left(\rho_1,\rho_2\right) = &\frac{1}{8}\cG_{0,0}^2+\frac{1}{8}\cG_{0,1}^1+\frac{1}{8}\cG_{0,1}^2-\frac{1}{8}\cG_{0,\rho_2}^1-\frac{1}{8}\cG_{1,0}^2+\frac{1}{8}\cG_{1,\rho_2}^1-\frac{1}{8}\cG_{\rho_2,1}^1\\
 \nonumber &-\frac{1}{8}\cG_0^1\cG_0^2+\frac{1}{8}\cG_0^2\cG_1^1-\frac{1}{8}\cG_1^1\cG_1^2+\frac{1}{8}\cG_1^2\cG_{\rho_2}^1
\end{align}
\begin{align}
\ff_{1,-+-}^{\text{(0,0)}}\left(\rho_1,\rho_2\right) = &-\frac{1}{8}\cG_{0,0}^1-\frac{1}{8}\cG_{0,0}^2-\frac{1}{8}\cG_{0,1}^1-\frac{1}{8}\cG_{0,1}^2+\frac{1}{8}\cG_{0,\rho_2}^1+\frac{1}{8}\cG_{1,0}^1+\frac{1}{8}\cG_{1,0}^2\\
 \nonumber &-\frac{1}{8}\cG_{1,\rho_2}^1+\frac{1}{8}\cG_{\rho_2,1}^1+\frac{1}{8}\cG_0^1\cG_0^2-\frac{1}{8}\cG_0^2\cG_1^1+\frac{1}{8}\cG_0^1\cG_1^2-\frac{1}{8}\cG_1^2\cG_{\rho_2}^1
\end{align}
%
%

\bibliographystyle{JHEP}
\bibliography{refs}

\providecommand{\href}[2]{#2}\begingroup\raggedright\begin{thebibliography}{10}

\bibitem{DiVecchia:2007vd}
P.~Di~Vecchia, {\it {The Birth of string theory}},  {\em Lect. Notes Phys.}
  {\bf 737} (2008) 59--118, [\href{http://xxx.lanl.gov/abs/0704.0101}{{\tt
  arXiv:0704.0101}}].

\bibitem{Kuraev:1976ge}
E.~A. Kuraev, L.~N. Lipatov, and V.~S. Fadin, {\it {Multi-Reggeon processes in
  the Yang-Mills theory}},  {\em Sov. Phys. JETP} {\bf 44} (1976) 443.

\bibitem{Kuraev:1977fs}
E.~A. Kuraev, L.~N. Lipatov, and V.~S. Fadin, {\it {The Pomeranchuk singularity
  in nonabelian gauge theories}},  {\em Sov. Phys. JETP} {\bf 45} (1977) 199.

\bibitem{Balitsky:1978ic}
I.~I. Balitsky and L.~N. Lipatov, {\it {The Pomeranchuk singularity in quantum
  chromodynamics}},  {\em Sov. J. Nucl. Phys.} {\bf 28} (1978) 822.

\bibitem{BROWER1974257}
R.~Brower, C.~DeTar, and J.~Weis, {\it Regge theory for multiparticle
  amplitudes},  {\em Physics Reports} {\bf 14} (1974), no.~6 257 -- 367.

\bibitem{Collins:1977jy}
P.~D.~B. Collins, {\em {An Introduction to Regge Theory and High-Energy
  Physics}}.
\newblock Cambridge Monographs on Mathematical Physics. Cambridge Univ. Press,
  Cambridge, UK, 2009.

\bibitem{Forshaw:1997dc}
J.~R. Forshaw and D.~A. Ross, {\it {Quantum chromodynamics and the pomeron}},
  {\em Cambridge Lect. Notes Phys.} {\bf 9} (1997) 1--248.

\bibitem{DelDuca:1995hf}
V.~Del~Duca, {\it {An introduction to the perturbative QCD pomeron and to jet
  physics at large rapidities}},
  \href{http://xxx.lanl.gov/abs/hep-ph/9503226}{{\tt hep-ph/9503226}}.

\bibitem{Brink:1976bc}
L.~Brink, J.~H. Schwarz, and J.~Scherk, {\it {Supersymmetric Yang-Mills
  Theories}},  {\em Nucl. Phys.} {\bf B121} (1977) 77--92.

\bibitem{Gliozzi:1976qd}
F.~Gliozzi, J.~Scherk, and D.~I. Olive, {\it {Supersymmetry, Supergravity
  Theories and the Dual Spinor Model}},  {\em Nucl. Phys.} {\bf B122} (1977)
  253--290.

\bibitem{tHooft:1973alw}
G.~'t~Hooft, {\it {A Planar Diagram Theory for Strong Interactions}},  {\em
  Nucl. Phys.} {\bf B72} (1974) 461.

\bibitem{Anastasiou:2003kj}
C.~Anastasiou, Z.~Bern, L.~J. Dixon, and D.~A. Kosower, {\it {Planar amplitudes
  in maximally supersymmetric Yang-Mills theory}},  {\em Phys. Rev. Lett.} {\bf
  91} (2003) 251602, [\href{http://xxx.lanl.gov/abs/hep-th/0309040}{{\tt
  hep-th/0309040}}].

\bibitem{Bern:2005iz}
Z.~Bern, L.~J. Dixon, and V.~A. Smirnov, {\it {Iteration of planar amplitudes
  in maximally supersymmetric Yang-Mills theory at three loops and beyond}},
  {\em Phys. Rev.} {\bf D72} (2005) 085001,
  [\href{http://xxx.lanl.gov/abs/hep-th/0505205}{{\tt hep-th/0505205}}].

\bibitem{Bartels:2008ce}
J.~Bartels, L.~N. Lipatov, and A.~Sabio~Vera, {\it {BFKL Pomeron, Reggeized
  gluons and Bern-Dixon-Smirnov amplitudes}},  {\em Phys. Rev.} {\bf D80}
  (2009) 045002, [\href{http://xxx.lanl.gov/abs/0802.2065}{{\tt
  arXiv:0802.2065}}].

\bibitem{Alday:2007hr}
L.~F. Alday and J.~M. Maldacena, {\it {Gluon scattering amplitudes at strong
  coupling}},  {\em JHEP} {\bf 06} (2007) 064,
  [\href{http://xxx.lanl.gov/abs/0705.0303}{{\tt arXiv:0705.0303}}].

\bibitem{Drummond:2007aua}
J.~M. Drummond, G.~P. Korchemsky, and E.~Sokatchev, {\it {Conformal properties
  of four-gluon planar amplitudes and Wilson loops}},  {\em Nucl. Phys.} {\bf
  B795} (2008) 385--408, [\href{http://xxx.lanl.gov/abs/0707.0243}{{\tt
  arXiv:0707.0243}}].

\bibitem{Brandhuber:2007yx}
A.~Brandhuber, P.~Heslop, and G.~Travaglini, {\it {MHV amplitudes in N=4 super
  Yang-Mills and Wilson loops}},  {\em Nucl. Phys.} {\bf B794} (2008) 231--243,
  [\href{http://xxx.lanl.gov/abs/0707.1153}{{\tt arXiv:0707.1153}}].

\bibitem{Drummond:2007cf}
J.~M. Drummond, J.~Henn, G.~P. Korchemsky, and E.~Sokatchev, {\it {On planar
  gluon amplitudes/Wilson loops duality}},  {\em Nucl. Phys.} {\bf B795} (2008)
  52--68, [\href{http://xxx.lanl.gov/abs/0709.2368}{{\tt arXiv:0709.2368}}].

\bibitem{Drummond:2007au}
J.~M. Drummond, J.~Henn, G.~P. Korchemsky, and E.~Sokatchev, {\it {Conformal
  Ward identities for Wilson loops and a test of the duality with gluon
  amplitudes}},  {\em Nucl. Phys.} {\bf B826} (2010) 337--364,
  [\href{http://xxx.lanl.gov/abs/0712.1223}{{\tt arXiv:0712.1223}}].

\bibitem{Bern:2008ap}
Z.~Bern, L.~J. Dixon, D.~A. Kosower, R.~Roiban, M.~Spradlin, C.~Vergu, and
  A.~Volovich, {\it {The Two-Loop Six-Gluon MHV Amplitude in Maximally
  Supersymmetric Yang-Mills Theory}},  {\em Phys. Rev.} {\bf D78} (2008)
  045007, [\href{http://xxx.lanl.gov/abs/0803.1465}{{\tt arXiv:0803.1465}}].

\bibitem{Drummond:2008aq}
J.~M. Drummond, J.~Henn, G.~P. Korchemsky, and E.~Sokatchev, {\it {Hexagon
  Wilson loop = six-gluon MHV amplitude}},  {\em Nucl. Phys.} {\bf B815} (2009)
  142--173, [\href{http://xxx.lanl.gov/abs/0803.1466}{{\tt arXiv:0803.1466}}].

\bibitem{Drummond:2006rz}
J.~M. Drummond, J.~Henn, V.~A. Smirnov, and E.~Sokatchev, {\it {Magic
  identities for conformal four-point integrals}},  {\em JHEP} {\bf 01} (2007)
  064, [\href{http://xxx.lanl.gov/abs/hep-th/0607160}{{\tt hep-th/0607160}}].

\bibitem{Bern:2006ew}
Z.~Bern, M.~Czakon, L.~J. Dixon, D.~A. Kosower, and V.~A. Smirnov, {\it {The
  Four-Loop Planar Amplitude and Cusp Anomalous Dimension in Maximally
  Supersymmetric Yang-Mills Theory}},  {\em Phys. Rev.} {\bf D75} (2007)
  085010, [\href{http://xxx.lanl.gov/abs/hep-th/0610248}{{\tt
  hep-th/0610248}}].

\bibitem{Bern:2007ct}
Z.~Bern, J.~J.~M. Carrasco, H.~Johansson, and D.~A. Kosower, {\it {Maximally
  supersymmetric planar Yang-Mills amplitudes at five loops}},  {\em Phys.
  Rev.} {\bf D76} (2007) 125020, [\href{http://xxx.lanl.gov/abs/0705.1864}{{\tt
  arXiv:0705.1864}}].

\bibitem{Alday:2007he}
L.~F. Alday and J.~Maldacena, {\it {Comments on gluon scattering amplitudes via
  AdS/CFT}},  {\em JHEP} {\bf 11} (2007) 068,
  [\href{http://xxx.lanl.gov/abs/0710.1060}{{\tt arXiv:0710.1060}}].

\bibitem{Bartels:2008sc}
J.~Bartels, L.~N. Lipatov, and A.~Sabio~Vera, {\it {N=4 supersymmetric Yang
  Mills scattering amplitudes at high energies: The Regge cut contribution}},
  {\em Eur. Phys. J.} {\bf C65} (2010) 587--605,
  [\href{http://xxx.lanl.gov/abs/0807.0894}{{\tt arXiv:0807.0894}}].

\bibitem{Lipatov:2010qf}
L.~N. Lipatov, {\it {Analytic properties of high energy production amplitudes
  in N=4 SUSY}},  {\em Theor. Math. Phys.} {\bf 170} (2012) 166--180,
  [\href{http://xxx.lanl.gov/abs/1008.1015}{{\tt arXiv:1008.1015}}].

\bibitem{Fadin:2011we}
V.~S. Fadin and L.~N. Lipatov, {\it {BFKL equation for the adjoint
  representation of the gauge group in the next-to-leading approximation at N=4
  SUSY}},  {\em Phys. Lett.} {\bf B706} (2012) 470--476,
  [\href{http://xxx.lanl.gov/abs/1111.0782}{{\tt arXiv:1111.0782}}].

\bibitem{Lipatov:2012gk}
L.~Lipatov, A.~Prygarin, and H.~J. Schnitzer, {\it {The Multi-Regge limit of
  NMHV Amplitudes in N=4 SYM Theory}},  {\em JHEP} {\bf 01} (2013) 068,
  [\href{http://xxx.lanl.gov/abs/1205.0186}{{\tt arXiv:1205.0186}}].

\bibitem{Dixon:2014iba}
L.~J. Dixon and M.~von Hippel, {\it {Bootstrapping an NMHV amplitude through
  three loops}},  {\em JHEP} {\bf 10} (2014) 065,
  [\href{http://xxx.lanl.gov/abs/1408.1505}{{\tt arXiv:1408.1505}}].

\bibitem{Lipatov:2010qg}
L.~N. Lipatov and A.~Prygarin, {\it {Mandelstam cuts and light-like Wilson
  loops in N=4 SUSY}},  {\em Phys. Rev.} {\bf D83} (2011) 045020,
  [\href{http://xxx.lanl.gov/abs/1008.1016}{{\tt arXiv:1008.1016}}].

\bibitem{Lipatov:2010ad}
L.~N. Lipatov and A.~Prygarin, {\it {BFKL approach and six-particle MHV
  amplitude in N=4 super Yang-Mills}},  {\em Phys. Rev.} {\bf D83} (2011)
  125001, [\href{http://xxx.lanl.gov/abs/1011.2673}{{\tt arXiv:1011.2673}}].

\bibitem{Bartels:2010tx}
J.~Bartels, L.~N. Lipatov, and A.~Prygarin, {\it {MHV amplitude for $3 \to 3$
  gluon scattering in Regge limit}},  {\em Phys. Lett.} {\bf B705} (2011)
  507--512, [\href{http://xxx.lanl.gov/abs/1012.3178}{{\tt arXiv:1012.3178}}].

\bibitem{DelDuca:2009au}
V.~Del~Duca, C.~Duhr, and V.~A. Smirnov, {\it {An Analytic Result for the
  Two-Loop Hexagon Wilson Loop in N = 4 SYM}},  {\em JHEP} {\bf 03} (2010) 099,
  [\href{http://xxx.lanl.gov/abs/0911.5332}{{\tt arXiv:0911.5332}}].

\bibitem{DelDuca:2010zg}
V.~Del~Duca, C.~Duhr, and V.~A. Smirnov, {\it {The Two-Loop Hexagon Wilson Loop
  in N = 4 SYM}},  {\em JHEP} {\bf 05} (2010) 084,
  [\href{http://xxx.lanl.gov/abs/1003.1702}{{\tt arXiv:1003.1702}}].

\bibitem{Goncharov:2010jf}
A.~B. Goncharov, M.~Spradlin, C.~Vergu, and A.~Volovich, {\it {Classical
  polylogarithms for amplitudes and Wilson loops}},  {\em Phys.Rev.Lett.} {\bf
  105} (2010) 151605, [\href{http://xxx.lanl.gov/abs/1006.5703}{{\tt
  arXiv:1006.5703}}].

\bibitem{Bartels:2010ej}
J.~Bartels, J.~Kotanski, and V.~Schomerus, {\it {Excited Hexagon Wilson Loops
  for Strongly Coupled N=4 SYM}},  {\em JHEP} {\bf 01} (2011) 096,
  [\href{http://xxx.lanl.gov/abs/1009.3938}{{\tt arXiv:1009.3938}}].

\bibitem{Bartels:2013dja}
J.~Bartels, J.~Kotanski, V.~Schomerus, and M.~Sprenger, {\it {The Excited
  Hexagon Reloaded}},  {\em {}} (2013)
  [\href{http://xxx.lanl.gov/abs/1311.1512}{{\tt arXiv:1311.1512}}].

\bibitem{Basso:2014pla}
B.~Basso, S.~Caron-Huot, and A.~Sever, {\it {Adjoint BFKL at finite coupling: a
  short-cut from the collinear limit}},  {\em JHEP} {\bf 01} (2015) 027,
  [\href{http://xxx.lanl.gov/abs/1407.3766}{{\tt arXiv:1407.3766}}].

\bibitem{Alday:2010ku}
L.~F. Alday, D.~Gaiotto, J.~Maldacena, A.~Sever, and P.~Vieira, {\it {An
  Operator Product Expansion for Polygonal null Wilson Loops}},  {\em JHEP}
  {\bf 04} (2011) 088, [\href{http://xxx.lanl.gov/abs/1006.2788}{{\tt
  arXiv:1006.2788}}].

\bibitem{Gaiotto:2010fk}
D.~Gaiotto, J.~Maldacena, A.~Sever, and P.~Vieira, {\it {Bootstrapping Null
  Polygon Wilson Loops}},  {\em JHEP} {\bf 03} (2011) 092,
  [\href{http://xxx.lanl.gov/abs/1010.5009}{{\tt arXiv:1010.5009}}].

\bibitem{Gaiotto:2011dt}
D.~Gaiotto, J.~Maldacena, A.~Sever, and P.~Vieira, {\it {Pulling the straps of
  polygons}},  {\em JHEP} {\bf 12} (2011) 011,
  [\href{http://xxx.lanl.gov/abs/1102.0062}{{\tt arXiv:1102.0062}}].

\bibitem{Sever:2011da}
A.~Sever, P.~Vieira, and T.~Wang, {\it {OPE for Super Loops}},  {\em JHEP} {\bf
  11} (2011) 051, [\href{http://xxx.lanl.gov/abs/1108.1575}{{\tt
  arXiv:1108.1575}}].

\bibitem{Basso:2013vsa}
B.~Basso, A.~Sever, and P.~Vieira, {\it {Spacetime and Flux Tube S-Matrices at
  Finite Coupling for N=4 Supersymmetric Yang-Mills Theory}},  {\em Phys. Rev.
  Lett.} {\bf 111} (2013), no.~9 091602,
  [\href{http://xxx.lanl.gov/abs/1303.1396}{{\tt arXiv:1303.1396}}].

\bibitem{Basso:2013aha}
B.~Basso, A.~Sever, and P.~Vieira, {\it {Space-time S-matrix and Flux tube
  S-matrix II. Extracting and Matching Data}},  {\em JHEP} {\bf 01} (2014) 008,
  [\href{http://xxx.lanl.gov/abs/1306.2058}{{\tt arXiv:1306.2058}}].

\bibitem{Basso:2014koa}
B.~Basso, A.~Sever, and P.~Vieira, {\it {Space-time S-matrix and Flux-tube
  S-matrix III. The two-particle contributions}},  {\em JHEP} {\bf 08} (2014)
  085, [\href{http://xxx.lanl.gov/abs/1402.3307}{{\tt arXiv:1402.3307}}].

\bibitem{Basso:2014jfa}
B.~Basso, A.~Sever, and P.~Vieira, {\it {Collinear Limit of Scattering
  Amplitudes at Strong Coupling}},  {\em Phys. Rev. Lett.} {\bf 113} (2014),
  no.~26 261604, [\href{http://xxx.lanl.gov/abs/1405.6350}{{\tt
  arXiv:1405.6350}}].

\bibitem{Basso:2014nra}
B.~Basso, A.~Sever, and P.~Vieira, {\it {Space-time S-matrix and Flux-tube
  S-matrix IV. Gluons and Fusion}},  {\em JHEP} {\bf 09} (2014) 149,
  [\href{http://xxx.lanl.gov/abs/1407.1736}{{\tt arXiv:1407.1736}}].

\bibitem{Basso:2014hfa}
B.~Basso, J.~Caetano, L.~Cordova, A.~Sever, and P.~Vieira, {\it {OPE for all
  Helicity Amplitudes}},  {\em JHEP} {\bf 08} (2015) 018,
  [\href{http://xxx.lanl.gov/abs/1412.1132}{{\tt arXiv:1412.1132}}].

\bibitem{Basso:2015rta}
B.~Basso, J.~Caetano, L.~Cordova, A.~Sever, and P.~Vieira, {\it {OPE for all
  Helicity Amplitudes II. Form Factors and Data analysis}},  {\em JHEP} {\bf
  12} (2015) 088, [\href{http://xxx.lanl.gov/abs/1508.02987}{{\tt
  arXiv:1508.02987}}].

\bibitem{Basso:2015uxa}
B.~Basso, A.~Sever, and P.~Vieira, {\it {Hexagonal Wilson loops in planar ${
  \mathcal N }=4$ SYM theory at finite coupling}},  {\em J. Phys.} {\bf A49}
  (2016), no.~41 41LT01, [\href{http://xxx.lanl.gov/abs/1508.03045}{{\tt
  arXiv:1508.03045}}].

\bibitem{Bartels:2011xy}
J.~Bartels, L.~N. Lipatov, and A.~Prygarin, {\it {Collinear and Regge behavior
  of $2 \to 4$ MHV amplitude in N = 4 super Yang-Mills theory}},
  \href{http://xxx.lanl.gov/abs/1104.4709}{{\tt arXiv:1104.4709}}.

\bibitem{Hatsuda:2014oza}
Y.~Hatsuda, {\it {Wilson loop OPE, analytic continuation and multi-Regge
  limit}},  {\em JHEP} {\bf 10} (2014) 38,
  [\href{http://xxx.lanl.gov/abs/1404.6506}{{\tt arXiv:1404.6506}}].

\bibitem{Drummond:2015jea}
J.~M. Drummond and G.~Papathanasiou, {\it {Hexagon OPE Resummation and
  Multi-Regge Kinematics}},  {\em JHEP} {\bf 02} (2016) 185,
  [\href{http://xxx.lanl.gov/abs/1507.08982}{{\tt arXiv:1507.08982}}].

\bibitem{Dixon:2012yy}
L.~J. Dixon, C.~Duhr, and J.~Pennington, {\it {Single-valued harmonic
  polylogarithms and the multi-Regge limit}},  {\em JHEP} {\bf 1210} (2012)
  074, [\href{http://xxx.lanl.gov/abs/1207.0186}{{\tt arXiv:1207.0186}}].

\bibitem{Pennington:2012zj}
J.~Pennington, {\it {The six-point remainder function to all loop orders in the
  multi-Regge limit}},  {\em JHEP} {\bf 1301} (2013) 059,
  [\href{http://xxx.lanl.gov/abs/1209.5357}{{\tt arXiv:1209.5357}}].

\bibitem{Broedel:2015nfp}
J.~Broedel and M.~Sprenger, {\it {Six-point remainder function in
  multi-Regge-kinematics: an efficient approach in momentum space}},  {\em
  JHEP} {\bf 05} (2016) 055, [\href{http://xxx.lanl.gov/abs/1512.04963}{{\tt
  arXiv:1512.04963}}].

\bibitem{Dixon:2011pw}
L.~J. Dixon, J.~M. Drummond, and J.~M. Henn, {\it {Bootstrapping the three-loop
  hexagon}},  {\em JHEP} {\bf 11} (2011) 023,
  [\href{http://xxx.lanl.gov/abs/1108.4461}{{\tt arXiv:1108.4461}}].

\bibitem{Dixon:2013eka}
L.~J. Dixon, J.~M. Drummond, M.~von Hippel, and J.~Pennington, {\it {Hexagon
  functions and the three-loop remainder function}},  {\em JHEP} {\bf 12}
  (2013) 049, [\href{http://xxx.lanl.gov/abs/1308.2276}{{\tt
  arXiv:1308.2276}}].

\bibitem{Dixon:2014voa}
L.~J. Dixon, J.~M. Drummond, C.~Duhr, and J.~Pennington, {\it {The four-loop
  remainder function and multi-Regge behavior at NNLLA in planar N = 4
  super-Yang-Mills theory}},  {\em JHEP} {\bf 06} (2014) 116,
  [\href{http://xxx.lanl.gov/abs/1402.3300}{{\tt arXiv:1402.3300}}].

\bibitem{Dixon:2015iva}
L.~J. Dixon, M.~von Hippel, and A.~J. McLeod, {\it {The four-loop six-gluon
  NMHV ratio function}},  {\em JHEP} {\bf 01} (2016) 053,
  [\href{http://xxx.lanl.gov/abs/1509.08127}{{\tt arXiv:1509.08127}}].

\bibitem{Caron-Huot:2016owq}
S.~Caron-Huot, L.~J. Dixon, A.~McLeod, and M.~von Hippel, {\it {Bootstrapping a
  Five-Loop Amplitude Using Steinmann Relations}},  {\em Phys. Rev. Lett.} {\bf
  117} (2016), no.~24 241601, [\href{http://xxx.lanl.gov/abs/1609.00669}{{\tt
  arXiv:1609.00669}}].

\bibitem{Drummond:2014ffa}
J.~M. Drummond, G.~Papathanasiou, and M.~Spradlin, {\it {A Symbol of
  Uniqueness: The Cluster Bootstrap for the 3-Loop MHV Heptagon}},  {\em JHEP}
  {\bf 03} (2015) 072, [\href{http://xxx.lanl.gov/abs/1412.3763}{{\tt
  arXiv:1412.3763}}].

\bibitem{Dixon:2016nkn}
L.~J. Dixon, J.~Drummond, T.~Harrington, A.~J. McLeod, G.~Papathanasiou, and
  M.~Spradlin, {\it {Heptagons from the Steinmann Cluster Bootstrap}},  {\em
  JHEP} {\bf 02} (2017) 137, [\href{http://xxx.lanl.gov/abs/1612.08976}{{\tt
  arXiv:1612.08976}}].

\bibitem{ArkaniHamed:2012nw}
N.~Arkani-Hamed, J.~L. Bourjaily, F.~Cachazo, A.~B. Goncharov, A.~Postnikov,
  and J.~Trnka, {\em {Scattering Amplitudes and the Positive Grassmannian}}.
\newblock Cambridge University Press, 2012.

\bibitem{Golden:2013xva}
J.~Golden, A.~B. Goncharov, M.~Spradlin, C.~Vergu, and A.~Volovich, {\it
  {Motivic Amplitudes and Cluster Coordinates}},  {\em JHEP} {\bf 01} (2014)
  091, [\href{http://xxx.lanl.gov/abs/1305.1617}{{\tt arXiv:1305.1617}}].

\bibitem{CaronHuot:2012ab}
S.~Caron-Huot and K.~J. Larsen, {\it {Uniqueness of two-loop master contours}},
   {\em JHEP} {\bf 10} (2012) 026,
  [\href{http://xxx.lanl.gov/abs/1205.0801}{{\tt arXiv:1205.0801}}].

\bibitem{Prlina:2017tvx}
I.~Prlina, M.~Spradlin, J.~Stankowicz, and S.~Stanojevic, {\it {Boundaries of
  Amplituhedra and NMHV Symbol Alphabets at Two Loops}},
  \href{http://xxx.lanl.gov/abs/1712.08049}{{\tt arXiv:1712.08049}}.

\bibitem{DelDuca:2016lad}
V.~Del~Duca, S.~Druc, J.~Drummond, C.~Duhr, F.~Dulat, R.~Marzucca,
  G.~Papathanasiou, and B.~Verbeek, {\it {Multi-Regge kinematics and the moduli
  space of Riemann spheres with marked points}},  {\em JHEP} {\bf 08} (2016)
  152, [\href{http://xxx.lanl.gov/abs/1606.08807}{{\tt arXiv:1606.08807}}].

\bibitem{Bartels:2011ge}
J.~Bartels, A.~Kormilitzin, L.~N. Lipatov, and A.~Prygarin, {\it {BFKL approach
  and $2 \to 5$ maximally helicity violating amplitude in ${\cal N}=4$
  super-Yang-Mills theory}},  {\em Phys. Rev.} {\bf D86} (2012) 065026,
  [\href{http://xxx.lanl.gov/abs/1112.6366}{{\tt arXiv:1112.6366}}].

\bibitem{Bartels:2013jna}
J.~Bartels, A.~Kormilitzin, and L.~Lipatov, {\it {Analytic structure of the
  $n=7$ scattering amplitude in $\mathcal{N}=4$ SYM theory in the multi-Regge
  kinematics: Conformal Regge pole contribution}},  {\em Phys. Rev.} {\bf D89}
  (2014), no.~6 065002, [\href{http://xxx.lanl.gov/abs/1311.2061}{{\tt
  arXiv:1311.2061}}].

\bibitem{Bartels:2014jya}
J.~Bartels, A.~Kormilitzin, and L.~N. Lipatov, {\it {Analytic structure of the
  $n=7$ scattering amplitude in $\mathcal{N}=4$ theory in multi-Regge
  kinematics: Conformal Regge cut contribution}},  {\em Phys. Rev.} {\bf D91}
  (2015), no.~4 045005, [\href{http://xxx.lanl.gov/abs/1411.2294}{{\tt
  arXiv:1411.2294}}].

\bibitem{Caron-Huot:2013fea}
S.~Caron-Huot, {\it {When does the gluon reggeize?}},  {\em JHEP} {\bf 05}
  (2015) 093, [\href{http://xxx.lanl.gov/abs/1309.6521}{{\tt
  arXiv:1309.6521}}].

\bibitem{Bargheer:2015djt}
T.~Bargheer, G.~Papathanasiou, and V.~Schomerus, {\it {The Two-Loop Symbol of
  all Multi-Regge Regions}},  {\em JHEP} {\bf 05} (2016) 012,
  [\href{http://xxx.lanl.gov/abs/1512.07620}{{\tt arXiv:1512.07620}}].

\bibitem{Prygarin:2011gd}
A.~Prygarin, M.~Spradlin, C.~Vergu, and A.~Volovich, {\it {All Two-Loop MHV
  Amplitudes in Multi-Regge Kinematics From Applied Symbology}},  {\em Phys.
  Rev.} {\bf D85} (2012) 085019, [\href{http://xxx.lanl.gov/abs/1112.6365}{{\tt
  arXiv:1112.6365}}].

\bibitem{Bartels:2012gq}
J.~Bartels, V.~Schomerus, and M.~Sprenger, {\it {Multi-Regge Limit of the
  n-Gluon Bubble Ansatz}},  {\em JHEP} {\bf 11} (2012) 145,
  [\href{http://xxx.lanl.gov/abs/1207.4204}{{\tt arXiv:1207.4204}}].

\bibitem{Bartels:2014mka}
J.~Bartels, V.~Schomerus, and M.~Sprenger, {\it {The Bethe roots of Regge cuts
  in strongly coupled $ \mathcal{N}=4 $ SYM theory}},  {\em JHEP} {\bf 07}
  (2015) 098, [\href{http://xxx.lanl.gov/abs/1411.2594}{{\tt
  arXiv:1411.2594}}].

\bibitem{DelDuca:2008jg}
V.~Del~Duca, C.~Duhr, and E.~W.~N. Glover, {\it {Iterated amplitudes in the
  high-energy limit}},  {\em JHEP} {\bf 12} (2008) 097,
  [\href{http://xxx.lanl.gov/abs/0809.1822}{{\tt arXiv:0809.1822}}].

\bibitem{BrownSVMPLs}
F.~C.~S. Brown, {\it {Single-valued hyperlogarithms and unipotent differential
  equations}},  {\em \verb+http://www.ihes.fr/~brown/RHpaper5.pdf+}.

\bibitem{Brown_Notes}
F.~C.~S. Brown, {\it {Notes on motivic periods}},  {\em {}} (2015)
  [\href{http://xxx.lanl.gov/abs/1512.06410}{{\tt arXiv:1512.06410}}].

\bibitem{Freyhult:2010kc}
L.~Freyhult, {\it {Review of AdS/CFT Integrability, Chapter III.4: Twist States
  and the cusp Anomalous Dimension}},  {\em Lett. Math. Phys.} {\bf 99} (2012)
  255--276, [\href{http://xxx.lanl.gov/abs/1012.3993}{{\tt arXiv:1012.3993}}].

\bibitem{BassoTalk}
B.~Basso.
\newblock Talk at Amplitudes 2016,
  \href{http://agenda.albanova.se/contributionDisplay.py?contribId=272&confId=5285}{\texttt{http://agenda.albanova.se/getFile.py/access?contribId=272\&amp;resId=250\&amp;materialId=slides\&amp;confId=5285}}.

\bibitem{Brown:2009qja}
F.~C. Brown, {\it {Multiple zeta values and periods of moduli spaces
  $\mathcal{M}_{0,n}(\mathbb{R})$}},  {\em Annales Sci.Ecole Norm.Sup.} {\bf
  42} (2009) 371, [\href{http://xxx.lanl.gov/abs/math/0606419}{{\tt
  math/0606419}}].

\bibitem{Papathanasiou:2013uoa}
G.~Papathanasiou, {\it {Hexagon Wilson Loop OPE and Harmonic Polylogarithms}},
  {\em JHEP} {\bf 11} (2013) 150,
  [\href{http://xxx.lanl.gov/abs/1310.5735}{{\tt arXiv:1310.5735}}].

\bibitem{HeptLimits}
L.~J. Dixon, J.~Drummond, A.~J. McLeod, G.~Papathanasiou, and M.~Spradlin.
\newblock To appear.

\bibitem{Moch:2001zr}
S.~Moch, P.~Uwer, and S.~Weinzierl, {\it {Nested sums, expansion of
  transcendental functions and multiscale multiloop integrals}},  {\em J. Math.
  Phys.} {\bf 43} (2002) 3363--3386,
  [\href{http://xxx.lanl.gov/abs/hep-ph/0110083}{{\tt hep-ph/0110083}}].

\bibitem{Bauer20021}
C.~Bauer, A.~Frink, and R.~Kreckel, {\it Introduction to the ginac framework
  for symbolic computation within the c++ programming language},  {\em
  J.Symb.Comput.} {\bf 33} (2002) 1--12,
  [\href{http://xxx.lanl.gov/abs/cs/0004015}{{\tt cs/0004015}}].

\bibitem{Vermaseren:2000nd}
J.~A.~M. Vermaseren, {\it {New features of FORM}},
  \href{http://xxx.lanl.gov/abs/math-ph/0010025}{{\tt math-ph/0010025}}.

\bibitem{Weinzierl:2002hv}
S.~Weinzierl, {\it {Symbolic expansion of transcendental functions}},  {\em
  Comput. Phys. Commun.} {\bf 145} (2002) 357--370,
  [\href{http://xxx.lanl.gov/abs/math-ph/0201011}{{\tt math-ph/0201011}}].

\bibitem{Moch2006759}
S.~Moch and P.~Uwer, {\it -xsummer- transcendental functions and symbolic
  summation in form},  {\em Computer Physics Communications} {\bf 174} (2006),
  no.~9 759 -- 770.

\bibitem{Schnetz:2013hqa}
O.~Schnetz, {\it {Graphical functions and single-valued multiple
  polylogarithms}},  {\em Commun. Num. Theor. Phys.} {\bf 08} (2014) 589--675,
  [\href{http://xxx.lanl.gov/abs/1302.6445}{{\tt arXiv:1302.6445}}].

\bibitem{Bargheer:2016eyp}
T.~Bargheer, {\it {Systematics of the Multi-Regge Three-Loop Symbol}},  {\em
  {}} (2016) [\href{http://xxx.lanl.gov/abs/1606.07640}{{\tt
  arXiv:1606.07640}}].

\bibitem{Lipatov:2009nt}
L.~N. Lipatov, {\it {Integrability of scattering amplitudes in N=4 SUSY}},
  {\em J. Phys.} {\bf A42} (2009) 304020,
  [\href{http://xxx.lanl.gov/abs/0902.1444}{{\tt arXiv:0902.1444}}].

\bibitem{Chachamis:2018bys}
G.~Chachamis and A.~Sabio~Vera, {\it {Open Spin Chains and Complexity in the
  High Energy Limit}},  \href{http://xxx.lanl.gov/abs/1801.04872}{{\tt
  arXiv:1801.04872}}.

\bibitem{symbolsC}
K.~T. Chen, {\it Iterated path integrals},  {\em Bull.\ Amer.\ Math.\ Soc.}
  {\bf 83} (1977) 831.

\bibitem{Bogner:2012dn}
C.~Bogner and F.~Brown, {\it {Symbolic integration and multiple
  polylogarithms}},  \href{http://xxx.lanl.gov/abs/1209.6524}{{\tt
  arXiv:1209.6524}}. [PoSLL2012,053(2012)].

\bibitem{Duhr:2012fh}
C.~Duhr, {\it {Hopf algebras, coproducts and symbols: an application to Higgs
  boson amplitudes}},  {\em JHEP} {\bf 08} (2012) 043,
  [\href{http://xxx.lanl.gov/abs/1203.0454}{{\tt arXiv:1203.0454}}].

\bibitem{Anastasiou:2013srw}
C.~Anastasiou, C.~Duhr, F.~Dulat, and B.~Mistlberger, {\it {Soft triple-real
  radiation for Higgs production at N3LO}},  {\em JHEP} {\bf 07} (2013) 003,
  [\href{http://xxx.lanl.gov/abs/1302.4379}{{\tt arXiv:1302.4379}}].

\bibitem{Duhr:2014woa}
C.~Duhr, {\it {Mathematical aspects of scattering amplitudes}},  in {\em
  {Theoretical Advanced Study Institute in Elementary Particle Physics:
  Journeys Through the Precision Frontier: Amplitudes for Colliders (TASI 2014)
  Boulder, Colorado, June 2-27, 2014}}, 2014.
\newblock \href{http://xxx.lanl.gov/abs/1411.7538}{{\tt arXiv:1411.7538}}.

\end{thebibliography}\endgroup

\end{document}